\documentclass[a4paper,11pt]{article}
\pdfoutput=1 

\usepackage{geometry}
\usepackage{fullpage}

\usepackage{mathtools} 
	\numberwithin{equation}{section} 

\usepackage[small,skip=0pt]{caption} 
\usepackage{amssymb}

\usepackage{bbm} 

\usepackage{bm} 
	\newcommand{\vect}[1]{\bm{{#1}}}

\usepackage[british]{babel} 

\usepackage[dvipsnames]{xcolor} 
\usepackage{graphicx}
\usepackage{tikz}
	\usetikzlibrary{matrix,arrows,decorations.pathmorphing} 
	\usetikzlibrary{calc} 
\usepackage{xifthen} 

\usepackage{marginnote}

\usepackage[
	ocgcolorlinks, 
	linkcolor=BlueViolet, citecolor=OliveGreen, urlcolor=RawSienna, filecolor=Sepia,
	hyperfootnotes=false,
	]{hyperref} 

\usepackage{booktabs} 

\usepackage{enumitem}
\usepackage[utf8]{inputenc}
\usepackage[T1]{fontenc} 
\usepackage{lmodern}
\AtBeginDocument{
	\DeclareFontShape{T1}{lmr}{m}{scit}{<->ssub*lmr/m/scsl}{}%
}

\usepackage[alphabetic]{amsrefs} 

\newcommand{\ev}{\mathrm{ev}}
\newcommand{\I}{\mathrm{i}}
\newcommand{\E}{\mathrm{e}}
\newcommand{\rk}{r}

\DeclareMathOperator*{\res}{Res}

\DeclarePairedDelimiter{\bra}{\langle}{\rvert}
\DeclarePairedDelimiter{\ket}{\lvert}{\rangle}
\DeclarePairedDelimiterX{\braket}[2]{\langle}{\rangle}{#1\vert#2}

\DeclarePairedDelimiter{\cbra}{\langle\!\langle}{\rvert}
\DeclarePairedDelimiter{\cket}{\lvert}{\rangle\!\rangle}
\DeclarePairedDelimiterX{\cbraket}[2]{\langle\!\langle}{\rangle}{#1\vert#2}
\DeclarePairedDelimiterX{\bracket}[2]{\langle}{\rangle\!\rangle}{#1\vert#2}
\DeclarePairedDelimiterX{\cbracket}[2]{\langle\!\langle}{\rangle\!\rangle}{#1\vert#2}

\newcommand{\choice}[1]{\vect{#1}}
\renewcommand{\underbar}[1]{\underline{#1\mspace{-3mu}}\mspace{3mu}}

\title{\textbf{From fermionic spin-Calogero--Sutherland models \\ to the Haldane--Shastry chain by freezing}}

\author{\normalsize Jules Lamers and Didina Serban \\ \\ \normalsize Université Paris--Saclay, CNRS, CEA \\ \normalsize Institut de Physique Théorique \\ \normalsize 91191 Gif-sur-Yvette, France \\ \\ \normalsize \texttt{jules.lamers@ipht.fr} {\small\textbullet} \texttt{didina.serban@ipht.fr}}

\date{\normalsize\today}

\begin{document}

\maketitle	

\begin{abstract}
\noindent The Haldane--Shastry spin chain has a myriad of remarkable properties, including Yangian symmetry and, for spin $1/2$, explicit highest-weight eigenvectors featuring (the case $\alpha = 1/2$ of) Jack polynomials. This stems from the spin-Calogero--Sutherland model, which reduces to Haldane--Shastry in a special `freezing' limit. 

In this work we clarify various points that, to the best of our knowledge, were missing in the literature. We have two main results. First, we show that freezing the \emph{fermionic} spin-1/2 Calogero--Sutherland model naturally accounts for the precise form of the Haldane--Shastry wave functions, including the Vandermonde factor squared. Second, we use the fermionic framework to prove the claim of Bernard--Gaudin--Haldane--Pasquier that the Yangian highest-weight eigenvectors of the $\textit{SU}\mspace{1mu}(\rk)$-version of the Haldane--Shastry chain arise by freezing $\textit{SU}\mspace{1mu}(\rk-1)$ spin-Calogero--Sutherland eigenvectors at $\alpha = 1/2$.
\end{abstract}

\tableofcontents

\section{Introduction}

Long-range interacting (quantum) integrable spin chains are of considerable interest in theoretical physics, due to their beautiful and intricate mathematical structure as well as their range of applications. In particular, in the last decades, long-range spin chains naturally appeared in AdS/CFT integrability~\cite{Bei_12} and as solvable models that can be engineered using cold atoms in optical lattices. From a theoretical point of view, progress in solving these models was slow because the standard tools for studying integrable nearest-neighbour Heisenberg spin chains, like the algebraic Bethe Ansatz, do not suffice. A few isolated models were solved, and some partially, through connections with `dynamical' models (quantum many-body systems) that come with their own techniques. We shall focus on the best understood example: the Haldane--Shastry spin chain~\cite{Hal_88,Sha_88} 
\begin{equation} \label{eq:HS_intro}
	H^\textsc{hs} = \sum_{i<j}^N \frac{1-P_{ij}}{4\sin^2[\pi(i-j)/N]} \, , 
	\qquad\quad
	\tikz[baseline={([yshift=-.5*11pt*.15]current bounding box.center)},scale=.5,font=\scriptsize]{
		\draw[very thin,gray!70] (0,0) ellipse (3 and 1);
		\foreach \n in {0,...,11}{
			\fill [black] ($(30*\n:3 and 1)$) circle (3pt);
			\ifthenelse%
				{\expandafter\isin\n{0,1,3,7,8,11}}
				{\draw [thick,->] ($(30*\n:3 and 1)+(0,.3)$) -- ($(30*\n:3 and 1)-(0,.4)$);}%
				{\draw [thick,->] ($(30*\n:3 and 1)-(0,.3)$) -- ($(30*\n:3 and 1)+(0,.4)$);%
			};
		};
		\draw[dotted] ($(90:3 and 1)$) node[xshift=4,yshift=7] {$i$} -- ($(240:3 and 1)$) node [xshift=-5,yshift=-7] {$j$};
		\node at ($(0:3 and 1)$) [right=2pt] {1};
		\node at($(330:3 and 1)$) [below right] {$N$};
		%
	}
\end{equation}
where $P_{ij}$ is the spin permutation operator, and the sites may be taken to have spin~1/2. While looking more complicated than the Heisenberg \textsc{xxx} chain, \eqref{eq:HS_intro} has a lot more spin \emph{symmetries} (namely the Yangian) and a much simpler, \emph{explicit} description of its energies and eigenvectors \cite{Hal_91a,BG+_93}. In particular, in the antiferromagnetic regime it has a neat CFT interpretation in terms of the level-one $\textit{SU}(2)$ Kac--Moody algebra~\cite{Hal_91a,BPS_94}. 
These special features arise from connections to the (trigonometric) Calogero--Sutherland model, where the particles move around on a circle with positions $x_j$ and may have spins \cite{HH_92, MP_93, HW_93}:
\begin{equation} \label{eq:CS_spin_intro}
	\widetilde{H}^\textsc{cs}_\pm = -\frac{1}{2}\sum_{j=1}^N \frac{\partial^2}{\partial x_{\mspace{-1mu}j}^2} + \beta \sum_{i < j}^N \frac{\beta\mp P_{ij}}{4 \sin^2[(x_i - x_j)/2]} \, , 
	\qquad\quad
	\tikz[baseline={([yshift=-.5*11pt*.15]current bounding box.center)},scale=.5,font=\scriptsize]{
		\draw[very thin,gray!50] (0,0) ellipse (3 and 1);
		\foreach \n/\thi/\thii/\thf in {0/5/8/22, 1/15/-3/-13, 2/-1/2.5/8, 3/7/-2.5/-25, 4/-2/2.5/10, 5/4/-4.5/-13, 6/1/0/0, 7/7/-3.5/-12, 8/8/3/12, 9/0/-2.5/-7, 10/-22/2.5/10, 11/0/-4.5/-25}{
			\ifthenelse{\NOT \n = 6}{\draw[gray!70,-stealth] (30*\n+\thi + \thii:3.02 and 1.02) arc [start angle = 30*\n+\thi + \thii, end angle=30*\n+\thi+\thf, x radius = 3.02, y radius = 1.02]}{};
			\fill [black] ($(30*\n+\thi:3 and 1)$) circle (3pt);
			\ifthenelse%
				{\expandafter\isin\n{0,1,3,7,8,11}}
				{\draw [thick,->] ($(30*\n+\thi:3 and 1)+(0,.3)$) -- ($(30*\n+\thi:3 and 1)-(0,.4)$);}%
				{\draw [thick,->] ($(30*\n+\thi:3 and 1)-(0,.3)$) -- ($(30*\n+\thi:3 and 1)+(0,.4)$);%
			};
		};
		\draw[dotted] ($(3*30+7:3 and 1)$) node[xshift=4,yshift=7] {$i$} -- ($(8*30+8:3 and 1)$) node[xshift=-5,yshift=-7] {$j$};
		\node at ($(0*30+5:3 and 1)$) [right=2pt] {1};
		\node at($(11*30:3 and 1)$) [below right] {$N$};
		%
	}
\end{equation}
with $\beta$ a (`reduced') coupling constant. The sign in \eqref{eq:CS_spin_intro} determines the symmetry (exchange statistics) of the eigenvectors: bosons (upper sign) or fermions (lower sign). For scalar particles (no spins) we recover~\cite{Sut_71, Sut_72},
\begin{equation} \label{eq:CS_scalar_intro}
	H^\textsc{cs}_\pm = -\frac{1}{2}\sum_{j=1}^N \frac{\partial^2}{\partial x_{\mspace{-1mu}j}^2} + \beta(\beta\mp1) \sum_{i < j}^N \frac{1}{4 \sin^2[(x_i - x_j)/2]} \, . 
	\qquad\quad
	\tikz[baseline={([yshift=-.5*11pt*.15]current bounding box.center)},scale=.5,font=\scriptsize]{
		\draw[very thin,gray!50] (0,0) ellipse (3 and 1);
		\foreach \n/\thi/\thii/\thf in {0/5/8/22, 1/15/-3/-13, 2/-1/2.5/8, 3/7/-2.5/-25, 4/-2/2.5/10, 5/4/-4.5/-13, 6/1/0/0, 7/7/-3.5/-12, 8/8/3/12, 9/0/-2.5/-7, 10/-22/2.5/10, 11/0/-4.5/-25}{
			\ifthenelse{\NOT \n = 6}{\draw[gray!70,-stealth] (30*\n+\thi + \thii:3.02 and 1.02) arc [start angle = 30*\n+\thi + \thii, end angle=30*\n+\thi+\thf, x radius = 3.02, y radius = 1.02]}{};
			\fill [black] ($(30*\n+\thi:3 and 1)$) circle (3pt);
		};
		\draw[dotted] ($(3*30+7:3 and 1)$) node[xshift=4,yshift=7] {$i$} -- ($(8*30+8:3 and 1)$) node[xshift=-5,yshift=-7] {$j$};
		\node at ($(0*30+5:3 and 1)$) [right=2pt] {1};
		\node at($(11*30:3 and 1)$) [below right] {$N$};
		%
	}
\end{equation}

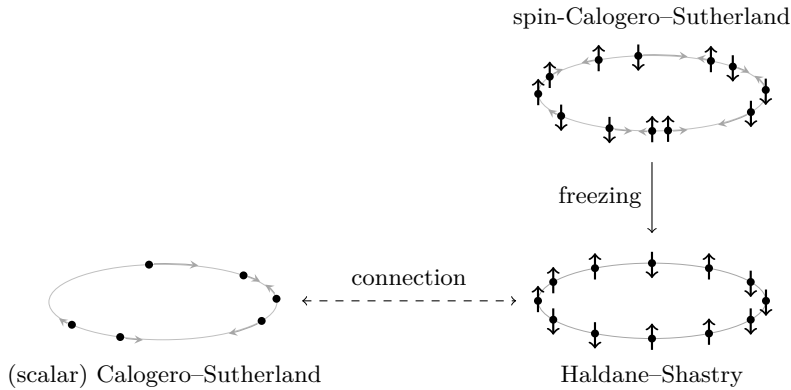
\begin{figure}[h]
	\centering
	\begin{tikzpicture}
		\matrix (m) [matrix of nodes,row sep=3em,column sep=8em]{
			& 
			\tikz[baseline={([yshift=-.5*11pt*.15]current bounding box.center)},scale=.5]{
				\draw[very thin,gray!50] (0,0) ellipse (3 and 1);
				\foreach \n/\thi/\thii/\thf in {0/5/8/22, 1/15/-3/-13, 2/-1/2.5/8, 3/7/-2.5/-25, 4/-2/2.5/10, 5/4/-4.5/-13, 6/1/0/0, 7/7/-3.5/-12, 8/8/3/12, 9/0/-2.5/-7, 10/-22/2.5/10, 11/0/-4.5/-25}{
					\ifthenelse{\NOT \n = 6}{\draw[gray!70,-stealth] (30*\n+\thi + \thii:3.02 and 1.02) arc [start angle = 30*\n+\thi + \thii, end angle=30*\n+\thi+\thf, x radius = 3.02, y radius = 1.02]}{};
					\fill [black] ($(30*\n+\thi:3 and 1)$) circle (3pt);
					\ifthenelse%
						{\expandafter\isin\n{0,1,3,7,8,11}}
						{\draw [thick,->] ($(30*\n+\thi:3 and 1)+(0,.3)$) -- ($(30*\n+\thi:3 and 1)-(0,.4)$);}%
						{\draw [thick,->] ($(30*\n+\thi:3 and 1)-(0,.3)$) -- ($(30*\n+\thi:3 and 1)+(0,.4)$);%
					};
				};
			}
			\\
			\tikz[baseline={([yshift=-.5*11pt*.15]current bounding box.center)},scale=.5]{
				\draw[very thin,gray!50] (0,0) ellipse (3 and 1);
				\foreach \n/\thi/\thii/\thf in {0/5/8/22, 1/15/-3/-13, 2/-1/2.5/8, 3/7/-2.5/-25, 4/-2/2.5/10, 5/4/-4.5/-13, 6/1/0/0, 7/7/-3.5/-12, 8/8/3/12, 9/0/-2.5/-7, 10/-22/2.5/10, 11/0/-4.5/-25}{
					\ifthenelse%
						{\expandafter\isin\n{0,1,3,7,8,11}}
						{\fill [black] ($(30*\n+\thi:3 and 1)$) circle (3pt);
						\draw[gray!70,-stealth] (30*\n+\thi + \thii:3.02 and 1.02) arc [start angle = 30*\n+\thi + \thii, end angle=30*\n+\thi+\thf, x radius = 3.02, y radius = 1.02];
						}%
						{%
					};
				};
			} 
			& 
			\tikz[baseline={([yshift=-.5*11pt*.15]current bounding box.center)},scale=.5]{
				\draw[very thin,gray!70] (0,0) ellipse (3 and 1);
				\foreach \n in {0,...,11}{
					\fill [black] ($(30*\n:3 and 1)$) circle (3pt);
					\ifthenelse%
						{\expandafter\isin\n{0,1,3,7,8,11}}
						{\draw [thick,->] ($(30*\n:3 and 1)+(0,.3)$) -- ($(30*\n:3 and 1)-(0,.4)$);}%
						{\draw [thick,->] ($(30*\n:3 and 1)-(0,.3)$) -- ($(30*\n:3 and 1)+(0,.4)$);%
					};
				};
			}
			\\
		};
		\path[->] ([yshift=-.1cm]m-1-2.south) edge node [left,font=\footnotesize] {freezing} ([yshift=.1cm]m-2-2.north);
		\path[dashed,<->] ([xshift=.15cm]m-2-1.east) edge node[above,yshift=2pt,align=center,font=\footnotesize] {connection} ([xshift=-.1cm]m-2-2.west);	
		\node at (m-1-2) [yshift=1cm,font=\footnotesize] {spin-Calogero--Sutherland};
		\node at (m-2-1) [yshift=-1cm,font=\footnotesize] {(scalar) Calogero--Sutherland};
		\node at (m-2-2) [yshift=-1cm,font=\footnotesize] {Haldane--Shastry};
		\node at (m-2-2) [yshift=-1.3cm] {};
	\end{tikzpicture}
	\caption{Schematic relation of the Haldane--Shastry spin chain to the spin- and scalar Calogero--Sutherland models for the case of spin~1/2. See Figure~\ref{fig:diagram} for details.} \label{fig:diagram_cartoon}
\end{figure}

The \textbf{goal} of this paper is to fill in several gaps in the literature regarding the connections between \eqref{eq:HS_intro} and \eqref{eq:CS_spin_intro}--\eqref{eq:CS_scalar_intro}. Broadly speaking, the spin-1/2 Haldane--Shastry chain is related to \emph{both} of these versions of the Calogero--Sutherland model (Figure~\ref{fig:diagram_cartoon}):
\begin{itemize}
	\item \eqref{eq:HS_intro} is a special `freezing' limit of \eqref{eq:CS_spin_intro}, and inherits various of its properties;
	\item per magnon sector, \eqref{eq:HS_intro} is related to a special case of \eqref{eq:CS_scalar_intro}, yielding the wave functions.
\end{itemize}
To explain our precise goals let us examine both connections in more detail. 

\textbf{Background.} The Haldane--Shastry hamiltonian~\eqref{eq:HS_intro} can be seen as the part of \eqref{eq:CS_spin_intro} linear in $\beta$ with positions fixed to equispaced values $x_j^\star = 2\pi j/N$. More precisely, Polychronakos \cite{Pol_93} identified \eqref{eq:HS_intro} as a properly regularised limit $\beta\to \infty$ of \eqref{eq:CS_spin_intro}, with $x_j^\star$ the (classical) equilibrium positions of \eqref{eq:CS_scalar_intro}. This \emph{freezing} limit will be reviewed in more detail in Section~\ref{sec:freezing}. An important feature is the decoupling between the spins and the positions~$x_j$ (`spin-charge separation'). Indeed, the charge degrees of freedom can be stripped away in such a way that the remaining excitations are purely magnetic: magnons (respectively spinons) in the (anti)ferromagnetic regime. In the resulting Haldane--Shastry spin chain, the magnons do not form bound states, and both magnons and spinons have purely (exclusion) statistical interactions with fractional statistics~\cite{Hal_91b}. 
Amongst others, freezing explains the appearance of fractional statistics. Moreover, since the eigenvalues of the spin-Calogero--Sutherland model are known explicitly, it allows one to directly read off the energies that may occur for the Haldane--Shastry spin chain. In addition, the Yangian spin symmetry~\cite{HH+_92} of \eqref{eq:HS_intro} originates from freezing~\cite{BG+_93}. This last point, which lies outside our focus here, is reviewed in Appendix~\ref{sec:nonabelian}, where we also link the descriptions of the Yangian actions from \cite{HH+_92} and \cite{BG+_93}; while straightforward, this link seems to be lacking in the literature. 

The exact solution of \eqref{eq:HS_intro} includes very simple exact wave functions. It turns out that for $M$ magnons $\downarrow$ located at sites $j_1,\dots,j_M$, any Haldane--Shastry wave function can be written as
\begin{equation} \label{eq:HS_wave_fn_ev_intro}
	\Psi(j_1,\dots,j_M) \, \equiv \bra{\uparrow \cdots \uparrow} \, \sigma^-_{j_1} \cdots \sigma^-_{j_1} \, \ket{\Psi} = \widetilde{\Psi}\bigl(\mathrm{e}^{2\pi \mathrm{i} \mspace{2mu} j_1/N},\dots,\mathrm{e}^{2\pi \mathrm{i} \mspace{2mu} j_M/N}\bigr)
\end{equation}
for some symmetric polynomial $\widetilde{\Psi}(z_1,\dots,z_M)$~\cite{BG+_93,LPS_22}. Here $z_m^\star = \E^{\I x_m^\star} = \E^{2\pi\I j_m/N}$ are just multiplicative versions of $x_m^\star$. An important class of states has explicit wave functions \cite{Hal_91a,BG+_93}
\begin{equation} \label{eq:HS_wave_fn_hw_intro}
	\widetilde{\Psi}(z_1,\dots,z_M) \equiv \biggl(\, \prod_{n<m}^M \! (z_n - z_m) \biggr)^{\!\!2} P_{\choice{\nu}}^\star(z_1,\dots,z_M) \, , \qquad P_{\choice{\nu}}^\star \equiv P_{\choice{\nu}}^{(1/2)} \, ,
\end{equation}
containing a Laughlin-like square of the Vandermonde polynomial along with a Jack polynomial $P_{\choice{\nu}}^{(\alpha)}$ at the (zonal spherical) point $\alpha^\star = 1/2$ of the Jack parameter. Not \emph{all} Haldane--Shastry wave functions are given by~\eqref{eq:HS_wave_fn_hw_intro}~\cite{Hal_91a}, but partitions $\choice{\nu} = (\nu_1 \geqslant \cdots \geqslant \nu_M)$ with
\begin{equation} \label{eq:hw_cond_intro}
	\nu_1 \leqslant N-2M+1 \, , \qquad \nu_M \geqslant 1 \, , 
\end{equation}
precisely yield all $M$-particle wave functions with \emph{Yangian highest weight} \cite{BG+_93,BPS_95a}. They have known momentum, energy~\cite{Hal_91a}, and Drinfeld polynomial~\cite{BG+_93}, and allow one to compute e.g.\ the dynamical spin-spin correlation function exactly \cite{LPS_94, Ha_94}.

The focus of this paper is the origin of the exact functions~\eqref{eq:HS_wave_fn_hw_intro}. We are aware of two derivations in the literature. The first of these uses Lagrange interpolation to show that the spin-chain hamiltonian~\eqref{eq:HS_intro} acts on $M$-magnon wave functions as a discretised version of the scalar hamiltonian~$H^\textsc{cs}_+$ from \eqref{eq:CS_scalar_intro} with $N^\star = M$ \emph{bosons} (required by the symmetry of spin-chain wave functions) and reduced coupling $\beta^\star = 2$. This leads to the eigenfunctions \eqref{eq:HS_wave_fn_hw_intro} with $\alpha^\star = 1/\beta^\star$ evaluated at $z_m = z_m^\star$ as in \eqref{eq:HS_wave_fn_ev_intro}. We believe that this is how Haldane \cite{Hal_91a} originally found the eigenfunctions~\eqref{eq:HS_wave_fn_hw_intro}. A proof, which seems to be missing in the literature, is given in Appendix~\ref{sec:lagrange}. Although direct, this derivation is not very enlightening: it is not clear \emph{why} it works, and does not explain any of the other properties of the Haldane--Shastry spin chain. In particular, it gives no understanding of which wave functions are obtained in this way.

The second derivation of~\eqref{eq:HS_wave_fn_hw_intro} is less direct but more insightful \cite{BG+_93,LPS_22}. It exploits the freezing limit of \eqref{eq:CS_spin_intro}, which provides an underlying algebraic framework given by \mbox{(Cherednik--)}Dunkl operators. These commuting operators are simultaneously diagonalised by so-called nonsymmetric Jack polynomials. The eigenvectors of the spin-Calogero--Sutherland model are determined by partially symmetrised versions of nonsymmetric Jacks~\cite{Ugl_96,TU_97}. However, the freezing limit is quite brutal at the level of the eigenvectors: it is not obvious which wave functions survive, or what the survivors look like explicitly. We intend to investigate this in a separate publication. Thus, one has to take a detour to arrive at \eqref{eq:HS_wave_fn_hw_intro}~\cite{BG+_93}, cf.~\cite{LPS_22}.%
\footnote{\ \label{fn:second_derivation} This goes as follows. The spin-chain hamiltonian~\eqref{eq:HS_intro} can be simultaneously diagonalised with the `(partially) frozen' Dunkl operators $\lim_{\beta\to\infty} d_j$ (no derivative). For $M$-magnon eigenvectors the first $M$ of those operators are, on a suitable space of polynomials, related by a `gauge transformation' (conjugation by the Vandermonde) to new Dunkl operators $d_m^\star$ in only $N^\star = M$ variables and with $\beta^\star =1$. Symmetric combinations of $d_1^\star,\dots,d_M^\star$ generate a scalar Calogero--Sutherland model with $N^\star$ particles and coupling $\beta^\star =1$. It must be a \emph{fermionic} model: due to the symmetry of spin-chain wave functions, the Vandermonde factor that we already used requires the wave functions of this scalar Calogero--Sutherland model to be antisymmetric. Its exact wave functions equal another Vandermonde factor times a Jack polynomial with $\alpha^\star = 1/(\beta^\star + 1)$, yielding \eqref{eq:HS_wave_fn_hw_intro}.}
This derivation has the benefits that it exploits the algebraic structure (quantum integrability), but a drawback is that it is somewhat indirect; for instance, the energies of the wave functions constructed in this way has to be found by other means.

\begin{figure}[h]
	\centering
	\begin{tikzpicture}
		\matrix (m) [matrix of nodes,row sep=5em,column sep=9.9em,nodes={text width=14em,align=left},font=\footnotesize]{
			{nonsymmetric theory \\ \textbullet\,commuting (Dunkl) operators \\ \textbullet\,joint eigenfunctions $E_{\choice{\lambda}}$} 
			& {spin-Calogero--Sutherland \textbullet\,commuting hamiltonians incl.~\eqref{eq:CS_spin_intro} \\ \textbullet\,joint eigenvectors \\ \textbullet\,Yangian symmetry} \\
			{(scalar) Calogero--Sutherland \\ \textbullet\,commuting hamiltonians incl.~\eqref{eq:CS_scalar_intro} \\ \textbullet\,joint eigenfunctions $P_{\choice{\lambda}}$} 
			& {Haldane--Shastry \\ \textbullet\,commuting hamiltonians incl.~\eqref{eq:HS_intro} \\ \textbullet\,joint eigenvectors with $P_{\choice{\nu}}^\star$ \\ \textbullet\,Yangian symmetry} \\
		};
		\path[densely dotted,->] ([xshift=-.8cm,yshift=-.9cm]m-1-1.north east) edge node [below,yshift=-2pt,align=left,font=\footnotesize] {$N$ spin-1/2 particles \\[.2em] \scriptsize reduced coupling $\beta$} ([yshift=-.9cm]m-1-2.north west);
		\path[densely dotted,->] ([xshift=-1cm]m-1-1.south) edge node [right,xshift=1pt,align=left,font=\footnotesize] {$N$ (spinless) particles \\[.2em] \scriptsize reduced coupling $\beta$} ([xshift=-1cm]m-2-1.north);
		\path[->] ([xshift=-.9cm]m-1-2.south) edge node [left,align=left,font=\scriptsize] { \footnotesize freezing \\[.2em] $\beta \to \infty$ \\ $x_j \to 2\pi \mspace{1mu} j/N$} ([xshift=-.9cm]m-2-2.north);
		\path[dashed,<->] ([xshift=-.07cm,yshift=-.89cm]m-2-1.north east) edge node[above,yshift=2pt,align=left,font=\scriptsize] {$N^\star = M$ \\ $\alpha^\star = 1/2$ \\ $x_m^\star = 2\pi \mspace{1mu} j_m/N$ \\[.2em] \footnotesize connection} ([yshift=-.87cm]m-2-2.north west);
	\end{tikzpicture}
	\caption{The nonsymmetric theory (degenerate affine Hecke algebra) gives rise to both the scalar and spin-Calogero--Sutherland model, here for spin~1/2. By freezing, the latter yields the Haldane--Shastry spin chain with its commuting hamiltonians and Yangian symmetry. The spin-chain eigenvectors with $M$ magnons and Yangian highest weight arise from a connection to an $M$-particle scalar Calogero--Sutherland model at the (zonal spherical) point $\alpha^\star = 1/2$. (For the generalisation to $\mathfrak{sl}_\rk$-spins, the scalar model is upgraded to rank~$\rk^\star = \rk-1$.)} \label{fig:diagram}
\end{figure}
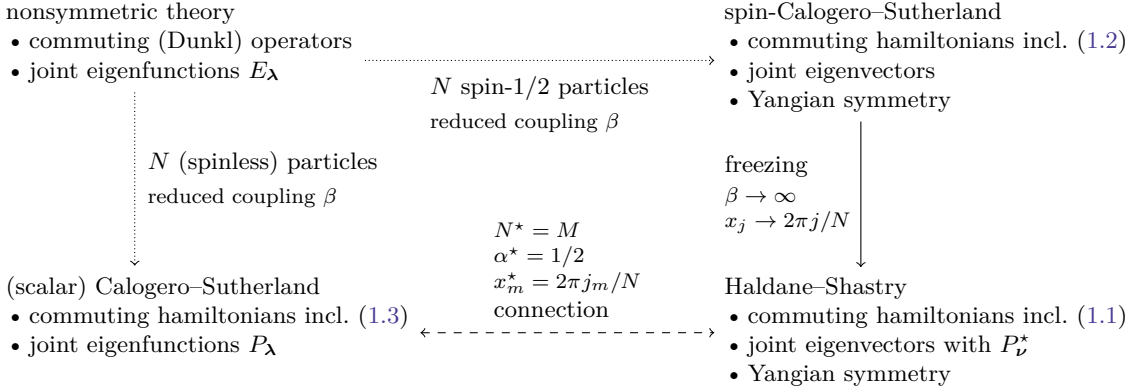

The situation is summarised in Figure~\ref{fig:diagram}. Since magnons cannot occupy the same site, both methods of diagonalising the Haldane--Shastry spin chain assume that the wave functions vanish whenever any two arguments coincide, $z_i = z_j$; in the second derivation, this is the origin of the gauge transformation by the Vandermonde factor. Although physically reasonable, this assumption concerns values of the wave function that do not appear in the eigenvectors and, therefore, are strictly speaking non physical. Moreover, the analogous derivation for the $q$-deformed Haldane--Shastry spin chain \cite{Ugl_95u, Lam_18} instead assumes that the wave function vanishes at $z_i = q^2 \, z_j$, yielding a factor of (the `symmetric square' of) the $q$-deformed Vandermonde polynomial \cite{LPS_22}. This suggest that even at $q=1$ the true origin of the Vandermonde factor might be different. 

\textbf{In this paper} we propose a \emph{third} derivation of the exact Yangian highest-weight wave functions~\eqref{eq:HS_wave_fn_hw_intro} that passes through the \emph{fermionic} spin-Calogero--Sutherland model.%
\footnote{\ The astute reader will note that the derivation described in Footnote~\ref{fn:second_derivation} also uses the fermionic model. This observation is retrospective: the fermionic model only appeared implicitly in the derivations of~\cite{BG+_93,LPS_22}.}
To do so we use the action of (the gauge transformed version of) the fermionic case of \eqref{eq:CS_spin_intro} on the so-called \emph{wedge} basis of the fermionic Fock space~\cite{Ste_95,KMS_95,Ugl_96}. We show that, upon freezing, the fermionic route naturally gives rise to the square of the Vandermonde factor in \eqref{eq:HS_wave_fn_hw_intro}. We moreover establish the connection between the $M$-magnon sector of the Haldane--Shastry spin chain and the spinless \emph{fermionic} Calogero--Sutherland model, i.e.\ $H^\textsc{cs}_-$ from \eqref{eq:CS_scalar_intro}, with $N^\star=M$ and $\beta^\star =1$. That is, the fermionic setting fits with the algebraic framework and freezing \emph{and} allows us to directly connect the spin-chain wave functions to those of the scalar Calogero--Sutherland model at special values of its parameters. Moreover, we show that this generalises to higher rank: the eigenvectors of the $\mathit{SU}(\rk)$ Haldane--Shastry spin chain \cite{BG+_93} are similarly connected to those of the fermionic spin-Calogero--Sutherland model with $\mathit{SU}(\rk^\star)$ spins for $\rk^\star = \rk - 1$ and reduced coupling $\beta^\star = 1$ for \emph{any} rank~$\rk$, as was conjectured in \cite{BG+_93}. This `one-step nesting' contains the spin-1/2 case as $r=2$, whence $\rk^\star = \rk-1$ yields the scalar Calogero--Sutherland model.%
\footnote{\ We stress that, unlike for higher-rank Heisenberg spin chains, the nesting for Haldane--Shastry and Calogero--Sutherland only goes one step: the model of rank $\rk^\star = \rk-1$ has \emph{fixed} coupling $\beta^\star =1$ and cannot be frozen.}
While these results are not new, to the best of our knowledge we give the first proof of this connection for higher rank ($\rk>2$). Moreover, we think that our method is more transparent and will extend to more general situations, including the \textit{q}-deformed (\textsc{xxz}-type) Haldane--Shastry spin chain \cite{Ugl_95u,Lam_18,LPS_22} (using \textit{q}-wedges~\cite{KMS_95}), the Inozemtsev spin chain~\cite{Ino_90,Ino_03,KL_22}, as well as its \textit{q}-deformation~\cite{KL_23u}.
\\

\noindent
The \textbf{outline} of this paper is as follows. Section~\ref{sec:spin_CS} gives the general background on the scalar and spin-versions of the (fermionic) Calogero--Sutherland model, the structure of its wave functions, and the definition of the wedge basis. Sections~\ref{sec:HS}--\ref{sec:higer_rank} concentrate on the freezing limit and the action of the Haldane--Shastry spin chain on the frozen wedges, as well as the relation to the scalar Calogero--Sutherland model. They contain the main results of the paper. To avoid cumbersome notations, we focus on spin~1/2 in Section~\ref{sec:HS}, and repeat the main steps of the derivation for general $\mathit{SU}(\rk)$-spins in Section~\ref{sec:higer_rank}. 

There are three appendices. Appendix~\ref{sec:lagrange} contains the direct derivation (via Lagrange interpolation) of the connection between Haldane--Shastry and the scalar Calogero--Sutherland model. Appendix~\ref{sec:bosonic} provides an overview of the differences between the bosonic and fermionic versions of the spin Calogero--Sutherland model. In Appendix~\ref{sec:nonabelian} we review the Yangian invariance of these models, and show the equivalence between the Yangian symmetries of the Haldane--Shastry chain postulated in \cite{HH+_92} and those derived in \cite{BG+_93}.

\subsection*{Acknowledgements} 
We thank G.~Ferrando, F.~Levkovich-Maslyuk and V.~Pasquier for discussions about related topics. We are grateful to A.~Ben Moussa for detailed feedback on an earlier version.

The work of JL was funded by Labex Mathématique Hadamard (LMH). We acknowledge hospitality and support from  the Galileo Galilei Institute, and from the scientific program \textit{Randomness, Integrability, and Universality} (Spring 2022), where part of this work was done. DS also acknowledges hospitality from the \textsc{kitp} Santa Barbara and partial support  by the National Science Foundation under Grant No.\ \textsc{nsf} \textsc{phy}-1748958.

\section{Spin-Calogero--Sutherland model} \label{sec:spin_CS}

The (trigonometric, quantum) Calogero--Sutherland model describes $N$ point particles moving on a circle while interacting in pairs with a periodic inverse-square potential
\begin{equation} \label{eq:potential}
	\frac{1}{4 \sin^2[(x_i - x_j)/2]} = \frac{z_i \, z_j}{z_{ij}\,z_{ji}} = \frac{-z_i\,z_j}{(z_i-z_j)^2} = \frac{1}{|z_i - z_j|^2} \; , \qquad z_j = \mathrm{e}^{\mathrm{i}\mspace{1mu}x_j} \, ,
\end{equation}
where we use the short-hand $z_{ij} \equiv z_i - z_j$, and the multiplicative coordinates are unimodular, $|z_j| = 1$. In this notation, the hamiltonian~\eqref{eq:CS_scalar_intro} reads
\begin{equation} \label{eq:CS_scalar}
	\text{scalar case:} \qquad
	H^\textsc{cs}_\pm = \frac{1}{2}\sum_{j=1}^N \, (z_j \, \partial_{z_j})^2 +  \beta\,(\beta\mp1) \sum_{i < j}^N \frac{z_i \, z_j}{z_{ij}\,z_{ji}} \, .
\end{equation}
Here the `reduced coupling parameter'~$\beta$ parametrises the coupling constant $\beta \, (\beta \mp 1)$ in such a way that the energy and wave function will have a simple dependence on $\beta$. The choice of sign distinguishes between bosons (upper sign) and fermions (lower sign), which is possible since we are in $1+1$ dimensions. 

In addition, the particles may have an internal degree of freedom taking values in $\mathbb{C}^\rk$: they can be scalars ($\rk=1$), have spin~1/2 ($\rk=2$), or come in $\rk$ `colours' or `species'. As our terminology anticipates, for $\rk>2$ this $\mathbb{C}^\rk$ should be interpreted as the vector (first fundamental) representation of $\mathfrak{sl}_\rk$, rather than the spin-$(\rk-1)/2$ representation of $\mathfrak{sl}_2$.%
\footnote{\ As we work at the Lie-algebra level and over $\mathbb{C}$ we deal with $\mathfrak{sl}_\rk = \mathfrak{su}_\rk^\mathbb{C}$ (or $\mathfrak{gl}_\rk = \mathfrak{sl}_\rk \oplus \mathbb{C}$) rather than $\mathit{SU}(\rk)$.}
We write $P_{ij}$ for the operator permuting this internal degree of freedom for particles $i$ and~$j$, e.g.~if $r=2$ then $P_{ij} = (1+\vec{\sigma}_i \cdot \vec{\sigma}_j)/2$ in terms of Pauli spin matrices. As we will see, the appropriate generalisation \cite{HH_92, MP_93, HW_93} of \eqref{eq:CS_scalar} is \eqref{eq:CS_spin_intro}, i.e.\
\begin{equation} \label{eq:CS_spin}
	\widetilde{H}^\textsc{cs}_\pm = \frac{1}{2}\sum_{j=1}^N \, (z_j \, \partial_{z_j})^2 +  \beta \sum_{i < j}^N \frac{z_i \, z_j}{z_{ij} \, z_{ji}} \, (\beta \mp P_{ij})\, .
\end{equation}
This \emph{spin-Calogero--Sutherland model} is quantum integrable: the hamiltonian \eqref{eq:CS_spin}
\begin{itemize}
	\item belongs to a hierarchy of commuting hamiltonians \cite{BG+_93, Che_94b, Res_17}, each of which
	\item has enhanced (Yangian) $\mathfrak{sl}_\rk$ symmetry for $\rk>1$ \cite{HH+_92, BG+_93}, and
	\item admits an explicit description of its (Yangian) highest-weight vectors in terms of partially symmetrised nonsymmetric Jack polynomials, cf.\ \cite{TU_97, Ugl_98}. 
\end{itemize}
We refer to the first property as the \emph{abelian symmetries} and the second as the \emph{nonabelian symmetries}. In this section we recall the underlying algebraic structure and how it gives rise to the abelian symmetries and eigenvectors of the scalar and spin-Calogero--Sutherland model, aiming for the case of fermions with spin. See Appendix~\ref{sec:bosonic} for more about bosons with spin, and Appendix~\ref{sec:nonabelian} for the nonabelian symmetries.

\subsection{Formalism: nonsymmetric theory} \label{sec:nonsymmetric}

We start with $N$~distinguishable particles. The algebraic framework underlying Calogero--Sutherland models is centred around the \emph{(Cherednik--)Dunkl operators} \cite{Dun_89, Che_91}
\begin{equation} \label{eq:Dunkl}
	\begin{aligned}
		d_j & \equiv \frac{1}{\beta} \, z_j \,\partial_{z_j} - z_j \sum_{i=1}^{j-1} \frac{1}{z_{ij}} \, (1-s_{ij}) + \sum_{k=j+1}^N \frac{z_k}{z_{jk}} \, (1-s_{jk}) +\frac{1}{2}\,(N-2j+1) \\
		& = \frac{1}{\beta} \, z_j \,\partial_{z_j} - \frac{1}{2}\sum_{i(\neq j)}^N \frac{z_i + z_j}{z_i - z_j} \, (1-s_{ij}) - \frac{1}{2} \sum_{i(<j)} \! s_{ij} + \frac{1}{2} \sum_{k(>j)}^N \! s_{jk} \, ,
	\end{aligned}
\end{equation}
where $\beta > 0$ is a parameter, and $s_{ij}$ denotes the coordinate permutation $z_i \leftrightarrow z_j$. These operators act on the space $\mathbb{C}[\vect{z}, \vect{z}^{-1}]$ of Laurent polynomials in $z_1,\dots,z_N$, and obey the commutation relations
\begin{equation} \label{eq:dAHA}
	d_i \, d_j = d_j \, d_i \, , \qquad d_i \, s_{i,i+1} = s_{i,i+1} \, d_{i+1} + 1 \, , \qquad d_i \, s_{jk} = s_{jk} \, d_i \quad (i\neq j,k) \, ,
\end{equation}
which are the relations of the degenerate affine Hecke algebra~\cite{Dri_86, Lus_89}.
Note that shifts of all $d_j$ by a common constant do not affect \eqref{eq:dAHA}, and such shifts occur in the literature. In \cite{BG+_93} the $d_j$ were called `gauge transformed' Dunkl operators. See also the closely related `exchange operator formalism' of Polychronakos~\cite{Pol_99}.
\medskip

\textbf{Nonsymmetric Jack polynomials.}
Since Dunkl operators commute they can be diagonalised simultaneously, with joint eigenvectors that can be described using a few combinatorial ingredients. Dunkl operators are triangular with respect to the basis of monomials, 
\begin{equation} \label{eq:Dunkl_monomial}
	d_i \, z_1^{\lambda_1} \cdots z_N^{\lambda_N} = \delta_i(\choice{\lambda}) \, z_1^{\lambda_1} \cdots z_N^{\lambda_N}  + \text{lower monomials} \, , 
\end{equation}
where `lower' refers to a (partial) ordering on the set of all monomials called the dominance order,%
\footnote{\ \label{fn:dominance} Namely, for~$\choice{\lambda} \in \mathbb{Z}^N$ denote the corresponding partition by $\choice{\lambda}^{\mspace{-1mu}+}$, i.e.\ $\lambda^+_i \geqslant \lambda^+_{i+1}$ for all $i$. For partitions,
\begin{equation*}
	\choice{\lambda} \geqslant \choice{\nu} \qquad\quad \text{if{f}} \qquad\quad \lambda_1 + \dots + \lambda_j \geqslant \nu_1 + \dots + \nu_j \quad \text{for all $1\leqslant j\leqslant N$} \, ,
\end{equation*}
and $\choice{\lambda} > \choice{\nu}$ means $\choice{\lambda} \geqslant \choice{\nu}$ but $\choice{\lambda} \neq \choice{\nu}$. This is refined to compositions $\choice{\lambda} \in \mathbb{Z}^N$ as~\cite{BG+_93,Opd_95} 
\begin{equation*}
		\choice{\lambda} \succ \choice{\nu} \qquad\quad \text{if{f}} \qquad\quad \text{either} \quad \choice{\lambda}^{\mspace{-1mu}+} > \choice{\nu}^+ \qquad \text{or} \qquad \choice{\lambda}^{\mspace{-1mu}+} = \choice{\nu}^+ \ \text{and} \ \choice{\lambda} > \choice{\nu} \, .
\end{equation*}
Thus, $\choice{\lambda}^{\mspace{-1mu}+}$ is the highest among all permutations of $\choice{\lambda}$. For example, $(3,0,0) \succ (0,3,0) \succ (0,0,3) \succ (2,1,0) \succ [(2,0,1) \text{ and } (1,2,0)] \succ [(1,0,2) \text{ and } (0,2,1)] \succ (0,1,2) \succ (1,1,1)$, while $(2,0,1)$ and~$(1,2,0)$ are incomparable, as are $(1,0,2)$ and~$(0,2,1)$. The monomial $z_1^{\lambda_1} \cdots z_N^{\lambda_N}$ is \emph{lower} than $z_1^{\nu_1} \cdots z_N^{\nu_N}$ if $\choice{\nu} \succ \choice{\lambda}$. Incomparability is a feature rather than a bug, restricting which (lower) monomials appear in \eqref{eq:nonsymm_Jack}. \\ Note: in the mathematical literature the set of all permutations of $\choice{\lambda}$ is sometimes sorted in the reverse order.}
and
\begin{equation} \label{eq:Dunkl_eigvals}
	\begin{aligned}
	\delta_i(\choice{\lambda}) & \equiv \frac{1}{\beta} \, \lambda_i + \frac12 \bigl(N-2 \, \sigma^{\choice{\lambda}}(i) + 1\bigr) \, , \\
	\sigma^{\choice{\lambda}}(i) & \equiv \# \Bigl\{k \, (< i) : \lambda_k \geqslant \lambda_i \Bigr\} + 1 + \, \# \Bigl\{(i<)\,k\,(\leqslant N) : \lambda_k > \lambda_i \Bigr\} \, .
	\end{aligned}
\end{equation}
The diagonal entries \eqref{eq:Dunkl_eigvals} of $d_i$ together encode $\choice{\lambda} \in \mathbb{Z}^N$, so Dunkl operators are `separating' in the sense that their joint spectrum is simple (multiplicity free). Since the triangularity~\eqref{eq:Dunkl_monomial} only features a partial order, it does not immediately follow that Dunkl operators are diagonalisable; but this can be shown.
Hence their joint eigenfunctions form a basis for $\mathbb{C}[\vect{z},\vect{z}^{-1}]$ indexed by \emph{compositions} $\choice{\lambda} \in \mathbb{Z}^N$.
These eigenfunctions are called \emph{non-symmetric Jack polynomials} (with parameter~$\alpha \equiv 1/\beta$), and are uniquely characterised by the conditions \cite{Opd_95}
\begin{equation} \label{eq:nonsymm_Jack}
	E_{\choice{\lambda}}(\vect{z}) \equiv E_\lambda^{(1/\beta)}(\vect{z}) 
	= z_1^{\lambda_1} \cdots z_N^{\lambda_N} + \text{lower monomials} \, , \qquad
	d_i \, E_{\choice{\lambda}}(\vect{z}) = \delta_i(\choice{\lambda}) \, E_{\choice{\lambda}}(\vect{z}) \, , 
\end{equation}
with the same meaning of `lower' as in \eqref{eq:Dunkl_monomial}, and eigenvalues \eqref{eq:Dunkl_eigvals}.
Nonsymmetric Jack polynomials have various useful properties. For example,
\begin{itemize}
	\item The definition~\eqref{eq:nonsymm_Jack} implies that $E_{\choice{\lambda}}(\vect{z})$ is a homogeneous polynomial of (total) degree equal to the \emph{weight} $|\choice{\lambda}| \equiv \sum_i \lambda_i$. 
	\item Since $\choice{1} \equiv (1^N) = (1,\dots,1)$ is the lowest composition of weight~$N$ we have the \emph{stability} property
	\begin{equation} \label{eq:nonsym_jack_stability}
		E_{\choice{\lambda} + \choice{1}}(\vect{z}) = z_1 \cdots z_N \, E_{\choice{\lambda}}(\vect{z}) \, ,
	\end{equation}
	where addition of compositions is defined pointwise, so $\choice{\lambda} + \choice{1}$ has parts $\lambda_i +1$. 
	\item Permutations act by
	\begin{subequations}
		\begin{gather}
			s_{i,i+1} \, E_{\choice{\lambda}}(\vect{z}) = a_i(\choice{\lambda}) \, E_{\choice{\lambda}}(\vect{z})  + b_i(\choice{\lambda}) \, E_{(i,i+1) \mspace{1mu} \choice{\lambda}}(\vect{z})\, ,
			\intertext{where $(i,i+1) \, \choice{\lambda} \equiv (\lambda_1,\dots,\lambda_{i+1},\lambda_i,\dots, \lambda_N)$, and the coefficients are}
			\begin{aligned}
				a_i(\choice{\lambda}) \equiv \frac{1}{\delta_i(\choice{\lambda}) - \delta_{i+1}(\choice{\lambda})} \, , \qquad 
				b_i(\choice{\lambda}) \equiv 
				\begin{cases}
					1 - a_i(\choice{\lambda})^2 & \lambda_i > \lambda_{i+1} \, , \\
					0 & \lambda_i = \lambda_{i+1} \, , \\
					1 & \lambda_i < \lambda_{i+1} \, .
				\end{cases}
			\end{aligned}
		\end{gather}
	\end{subequations}
	\item In particular, 
	\begin{equation} \label{eq:nonsym_jack_symm}
		s_{i,i+1} \, E_{\choice{\lambda}}(\vect{z}) = E_{\choice{\lambda}}(\vect{z}) \quad \text{if{f}} \quad \lambda_i = \lambda_{i+1} \, .
	\end{equation}
\end{itemize}
Nonsymmetric Jack polynomials arise as the limit $t\to1$ with $t = q^\beta$ of nonsymmetric Macdonald polynomials. This provides combinatorial formulas in terms of non-attacking fillings \cite{HHL_08, Ale_19}, alcove walks \cite{RY_17, GR_21a, GR_21b} and integrable higher-rank vertex models \cite{BW_22}. 
\medskip

\textbf{Effective hamiltonians.}
To make contact with Calogero--Sutherland-type models consider the hierarchy of commuting operators that are generated by symmetric combinations of Dunkl operators, e.g.\ power sums $\sum_j d_j^{\mspace{2mu}k}$. In particular these include the first-order differential operator
\begin{equation} \label{eq:momentum_operator_eff}
	P^{\prime\mspace{2mu}\textsc{cs}} \equiv \beta \sum_{j=1}^N d_j = \sum_{j=1}^N z_j \,\partial_{z_j} \, ,
\end{equation}
and the second-order differential and permutation operator
\begin{align} \label{eq:CS_eff}
	H^{\prime\mspace{2mu}\textsc{cs}} \equiv {} & \frac{\beta^2}{2} \Biggl(\,\sum_{j=1}^N d_j^{\mspace{2mu}2} \mspace{7mu} - E^0 \Biggr) \nonumber \\
	= {} & \frac{1}{2}\sum_{j=1}^N (z_j \, \partial_{z_j})^2 - \frac{\beta}{2} \sum_{j=1}^N \, (N-2j+1) \,z_j \, \partial_{z_j} + \beta \sum_{i<j}^N \frac{z_i}{z_{ij}} \, \Bigl(z_i \, \partial_{z_i} - z_j \, \partial_{z_j} + \frac{z_j}{z_{ji}}\, (1-s_{ij})\Bigr) \nonumber \\
	= {} & \frac{1}{2} \sum_{j=1}^N (z_j \, \partial_{z_j})^2 + \frac{\beta}{2} \sum_{i<j}^N \frac{z_i + z_j}{z_i - z_j} \, \bigl(z_i \, \partial_{z_i} - z_j \, \partial_{z_j}\bigr) + \beta \sum_{i<j}^N \frac{z_i \, z_j}{z_{ij} \, z_{ji}} \, (1-s_{ij}) \, ,
\end{align}
where we find it convenient to remove the contribution of the constant part of the Dunkls~\eqref{eq:Dunkl},
\begin{equation} \label{eq:E0}
	E^0 \equiv \frac{1}{4} \sum_{j=1}^N  \, (N-2j+1)^2 = \frac{1}{12} \, N \, (N^2-1) \, .
\end{equation} 
Nonsymmetric Jack polynomials remain eigenvectors for the family of symmetric combinations of Dunkls, with eigenvalues that follow from \eqref{eq:Dunkl_eigvals}: for example, \eqref{eq:momentum_operator_eff}--\eqref{eq:CS_eff} have eigenvalues
\begin{equation} \label{eq:E_cs_eff}
	P^{\prime\mspace{2mu}\textsc{cs}}(\choice{\lambda}) = \beta \sum_{i=1}^N \delta_i(\choice{\lambda}) = \sum_{i=1}^N \lambda_i = |\choice{\lambda}| \, , \qquad
	E^{\prime\mspace{2mu}\textsc{cs}}(\choice{\lambda}) = \frac{\beta^2}{2} \Biggl( \, \sum_{i=1}^N \delta_i(\choice{\lambda})^2 \mspace{7mu} - E^0 \Biggr) \, .
\end{equation}

To interpret \eqref{eq:momentum_operator_eff}--\eqref{eq:CS_eff} we set $z_j = \E^{\I x_j}$ as in \eqref{eq:potential}. Since $z_j \,\partial_{z_j} = -\mathrm{i}\,\partial_{x_j}$, \eqref{eq:momentum_operator_eff} is just the (continuum) total momentum operator, and the first term of \eqref{eq:CS_eff} is a kinetic term. We further recognise the potential function \eqref{eq:potential} in \eqref{eq:CS_eff}. We will call \eqref{eq:CS_eff} the \emph{effective} hamiltonian. To understand the origin of the remaining parts of \eqref{eq:CS_eff} consider the function
\begin{equation} \label{eq:psi0}
	\Phi_0(\vect{z}) \equiv \prod_{i\neq j}^N (1-z_i/z_j)^{\beta/2} \, .
\end{equation}
This symmetric rational function is a Laurent polynomial if $\beta \in 2\,\mathbb{Z}_{\geqslant 0}$, and equals $|\Delta(\vect{z})|^\beta$ if the $z_j^* = 1/z_j$ are unimodular, where we define the Vandermonde polynomial
\begin{equation} \label{eq:Vandermonde}
	\Delta(\vect{z}) \equiv \prod_{i<j}^N (z_i - z_j) = \det\Bigl(z_i^{j-1}\Bigr)_{i,j=1}^N \, .
\end{equation}
On functions divisible by (i.e.\ containing a factor of) $\Phi_0(\vect{z})$ the operators \eqref{eq:momentum_operator_eff}--\eqref{eq:CS_eff} act as follows. The momentum operator~\eqref{eq:momentum_operator_eff} is invariant under the `gauge transformation' of conjugating by $\Phi_0$ (viewed as a multiplication operator),
\begin{equation} \label{eq:momentum_operator}
	\Phi_0 \, P^{\prime\mspace{2mu}\textsc{cs}} \; \Phi_0^{-1}
	= \sum_{j=1}^N z_j \,\partial_{z_j}  \, ,
\end{equation}
while the effective hamiltonian \eqref{eq:CS_eff} can be recognised as a `gauge transformed' Calogero--Sutherland-type operator:
\begin{equation} \label{eq:Ham_gauge}
	\Phi_0 \, \Biggl( H^{\prime\mspace{2mu}\textsc{cs}} + \frac{\beta^2}{2} \, E^0 \Biggr) \, \Phi_0^{-1} = \frac{1}{2} \sum_{j=1}^N (z_j \, \partial_{z_j})^2 + \sum_{i<j}^N \frac{z_i \, z_j}{z_{ij}\,z_{ji}}\, \beta \, (\beta - s_{ij}) \, .
\end{equation}
The identities \eqref{eq:momentum_operator_eff}--\eqref{eq:CS_eff} \& \eqref{eq:momentum_operator}--\eqref{eq:Ham_gauge} hold on arbitrary (sufficiently regular) functions. 

The physical picture is as follows. The hamiltonian \eqref{eq:Ham_gauge} governs $N$ particles moving on a circle with coordinates $z_j = \E^{\I x_j}$ and inverse-square exchange interaction. Higher symmetric polynomials in Dunkl operators provide a hierarchy of abelian symmetries. If $\beta>0$ then $\Phi_0(\vect{z})$ is the ground-state wave function, with zero momentum and energy $\beta^2 E_0/2$ due to \eqref{eq:momentum_operator}--\eqref{eq:Ham_gauge} and $H^{\prime\mspace{2mu}\textsc{cs}} \, 1 =0$.
Compositions $\choice{\lambda} \in \mathbb{Z}^N$ are the (quasi)momenta of excited states, with wave functions $\Phi_0(\vect{z}) \, E_{\choice{\lambda}}(\vect{z})$ and eigenvalues following from \eqref{eq:E_cs_eff}. The relation \eqref{eq:nonsym_jack_stability} describes a simultaneous shift (Galilean boost) of the momenta.
So far we have treated these particles as distinguishable. 
For \emph{in}distinguishable particles we have to prescribe the symmetry of the wave functions under particle exchange. This will allow us to reinterpret the coordinate permutations $s_{ij}$ in \eqref{eq:Ham_gauge} and retrieve the Calogero--Sutherland models from \eqref{eq:CS_scalar} and~\eqref{eq:CS_spin}, as we show next.

\subsection{Warm-up: scalar Calogero--Sutherland model} \label{sec:spinless}

We begin by reviewing the case of spinless particles, which will allow us to set up our notation and prepare for the case with spins in the next subsection.
Consider the subspaces of Laurent polynomials that are totally (anti)symmetric,
\begin{equation} \label{eq:pol_phys_space}
	\Bigl\{ f(\vect{z}) \in \mathbb{C}[\vect{z}, \vect{z}^{-1}] : s_{ij} \, f(\vect{z}) = \pm f(\vect{z}) , \ 1\leqslant i<j\leqslant N \Bigl\} \; = 
	\begin{cases} 
		\, \mathbb{C}[\vect{z}, \vect{z}^{-1}]^{S_N} \, , \\ 
		\, \Delta(\vect{z}) \, \mathbb{C}[\vect{z}, \vect{z}^{-1}]^{S_N} \, ,
	\end{cases} 
\end{equation}
where we use that any antisymmetric Laurent polynomial is divisible by the Vandermonde polynomial~\eqref{eq:Vandermonde} and the ratio is symmetric. Note that these spaces arise as the images of the coordinate (anti)symmetrisers on $\mathbb{C}[\vect{z}, \vect{z}^{-1}]$,
\begin{equation} \label{eq:anti/symm}
	\Pi^\text{pol}_\pm \equiv \! \sum_{\sigma \in S_N} \! (\pm1)^{\ell(\sigma)} \, s_\sigma \, , \quad (-1)^{\ell(\sigma)} \equiv \mathrm{sgn}(\sigma) \, , \qquad \bigl(\Pi^\text{pol}_\pm\bigr)^2 = N! \; \Pi^\text{pol}_\pm \, .
\end{equation} 
Note that symmetric combinations of Dunkl operators, such as \eqref{eq:momentum_operator_eff}--\eqref{eq:CS_eff}, commute with any coordinate permutation $s_{ij}$ by the relations \eqref{eq:dAHA}. They therefore also commute with the projectors \eqref{eq:anti/symm}, so that they preserve (leave invariant) the subspaces~\eqref{eq:pol_phys_space}. This is where \eqref{eq:CS_eff} becomes the effective form of the scalar Calogero--Sutherland hamiltonian~\eqref{eq:CS_scalar}.
Indeed, using the `physical condition' $s_{ij} = \pm1$ in \eqref{eq:Ham_gauge} gives coupling $\beta \, (\beta \mp 1)$ from \eqref{eq:CS_scalar}, with the upper (lower) sign for the bosonic (fermonic) case of totally (anti)symmetric Laurent polynomials. In this way, the restriction to \eqref{eq:pol_phys_space} of symmetric polynomials in Dunkl operators provide the abelian symmetries of the spinless Calogero--Sutherland model. 

The nonsymmetric theory provides the exact eigenvectors too via the projector~\eqref{eq:anti/symm}. At the nonsymmetric level, the hierarchy of commuting operators was not separating. Indeed, while the nonsymmetric Jacks are eigenvectors of the $\sum_j d_j^{\mspace{2mu}k}$, the eigenvalues, including \eqref{eq:E_cs_eff}, are invariant under rearranging the parts of $\choice{\lambda}$. The multiplicities in the joint spectrum are lifted by the restriction to the `physical space'~\eqref{eq:pol_phys_space}. The (anti)symmetrisation the joint Dunkl-eigenbasis $E_{\choice{\lambda}}(\vect{z})$ provides eigenvectors of the scalar Calogero--Sutherland hamiltonians, which are again separating, so that we obtain a basis of eigenvectors for the subspaces~\eqref{eq:pol_phys_space}.
\medskip

\textbf{Spinless bosons.} Let us consider the most commonly studied case in more detail: spinless bosons, with upper sign in \eqref{eq:pol_phys_space}. On this subspace the effective hamiltonian~\eqref{eq:CS_eff} becomes
\begin{equation} \label{CSgauge_scalar}
	H_+^{\prime\mspace{2mu}\textsc{cs}} \equiv \frac{1}{2} \sum_{j=1}^N (z_j \, \partial_{z_j})^2 + \frac{\beta}{2} \sum_{i<j}^N \frac{z_i + z_j}{z_i - z_j} \, \bigl(z_i \, \partial_{z_i} - z_j \, \partial_{z_j}\bigr) \, ,
\end{equation}
where the remarkable disappearance of one sum in \eqref{eq:CS_eff} is coincidental and not important. This is the effective form of the hamiltonian (with upper sign) from \eqref{eq:CS_scalar}. Together with the momentum operator~\eqref{eq:momentum_operator_eff} and higher hamiltonians constructed from other symmetric polynomials in Dunkl operators, this provides the abelian symmetries of the spinless bosonic Calogero--Sutherland model. To obtain their joint eigenvectors we symmetrise the nonsymmetric Jack polynomials. For compositions $\choice{\lambda} \in \mathbb{Z}^N$ differing only in ordering symmetrisation gives the same symmetric polynomial up to normalisation:
\begin{equation} \label{eq:nonsymm_to_symm_Jack}
	\text{cst}_+(\choice{\lambda}) \; \Pi^\text{pol}_+ \, E_{\choice{\lambda}}(\vect{z}) = P_{\choice{\lambda}^{\mspace{-1mu}+}}(\vect{z}) \, ,
\end{equation}
where $\choice{\lambda}^{\mspace{-1mu}+}$ is the partition corresponding to $\choice{\lambda}$, i.e.\ with parts sorted in weakly decreasing order. We may thus take $\choice{\lambda}  = (\lambda_1 \geqslant \lambda_2 \geqslant \dots)$ to be a \emph{partition}. We use the constant $\text{cst}_+(\choice{\lambda})$ to absorb the coefficient of the highest term (in the dominance order\,\textsuperscript{\ref{fn:dominance} (p.\,\pageref{fn:dominance})}), so that the polynomial \eqref{eq:nonsymm_to_symm_Jack} is monic.
It is an eigenfunction of the restrictions of the abelian symmetries since the latter commute with the projector~\eqref{eq:anti/symm}.
We thus conclude that the \emph{(symmetric) Jack polynomial} \cite{Jac_70} (with parameter~$\alpha \equiv 1/\beta$) obeys
\begin{subequations} \label{eq:symm_Jack_conditions}
	\begin{gather} 
	P_{\choice{\lambda}}(\vect{z}) 
	= m_{\choice{\lambda}}(\vect{z}) + \text{lower symmetric monomials} \, , \qquad m_{\choice{\lambda}}(\vect{z}) \equiv \sum_{\choice{\nu} \in S_N \cdot \choice{\lambda}} \!\!\!\! z_1^{\nu_1} \cdots z_N^{\nu_N} \, , \label{eq:symm_Jack} \\
	H_+^{\prime\mspace{2mu}\textsc{cs}} \, P_{\choice{\lambda}}(\vect{z}) = E^{\prime\mspace{2mu}\textsc{cs}}(\choice{\lambda}) \, P_{\choice{\lambda}}(\vect{z}) \, . \label{eq:symm_Jack_eigval_eqn}
	\end{gather}
\end{subequations}
Since $\choice{\lambda}$ is a partition, the eigenvalue~\eqref{eq:E_cs_eff} can be simplified to
\begin{equation} \label{eq:E_cs_eff_bosonic}
	E^{\prime\mspace{2mu}\textsc{cs}}(\choice{\lambda}) = \frac{1}{2}\sum_{i=1}^N \lambda_i^2 + \frac{\beta}{2} \sum_{i=1}^N (N-2\mspace{1mu}i+1)\,\lambda_i \, .
\end{equation}
If $\choice{\lambda},\choice{\nu}$ are comparable, then for generic~$\beta$ we have $E^{\prime\mspace{2mu}\textsc{cs}}(\choice{\lambda}) = E^{\prime\mspace{2mu}\textsc{cs}}(\choice{\nu})$ if{f} $\choice{\lambda} = \choice{\nu}$.
We will get back to various useful properties of Jack polynomials in a moment.

In physical terms we have the following. The scalar bosonic Calogero--Sutherland model with hamiltonian $H^\textsc{cs}_+$ given in \eqref{eq:CS_scalar} has quantum numbers consisting of partitions $\choice{\lambda} = \choice{\lambda}^+ \in \mathbb{Z}^N$, labelling (symmetric) wave functions 
\begin{equation} \label{eq:bosonic_wave_fns}
	\Phi_{\choice{\lambda}}(\vect{z}) \equiv \Phi_0(\vect{z}) \, P_{\choice{\lambda}}(\vect{z}) \, ,
\end{equation}
with $\Phi_0$ defined in \eqref{eq:psi0}. The momentum is $|\choice{\lambda}| = \sum_i \lambda_i$ and the energy reads
\begin{equation} \label{eq:E_cs}
	E^\textsc{cs}(\choice{\lambda}) \equiv E^{\prime\mspace{2mu}\textsc{cs}}(\choice{\lambda}) + \frac{\beta^2}{2} \, E^0 \, .
\end{equation}
As in the nonsymmetric case, \eqref{eq:sym_jack_stability} is a Galilean boost of the momenta.
When $\beta=0,1$ the potential term vanishes, yielding free bosons. If $\beta \to 0$ the bosonic wave function $\Phi_{\choice{\lambda}}(\vect{z})|_{\beta = 0} = m_{\choice{\lambda}}(\vect{z})$ is just a symmetrised plane wave in multiplicative coordinates. For $\beta=1$ we instead get $\Phi_{\choice{\lambda}}(\vect{z})|_{\beta = 1} = |\Delta(\vect{z})| \, s_{\choice{\lambda}}(\vect{z}) = |a_{\choice{k}}(\vect{z})|$ by \eqref{eq:schur}. The factor $\Phi_0(\vect{z})|_{\beta = 1} = |\Delta(\vect{z})|$ forbids particles at coinciding positions, i.e.\ free hard-core bosons.
For $\beta>0$ the lowest energy is attained for $\choice{\lambda} = \choice{0}$ the empty partition: the bosonic ground state has wave function $\Phi_{\choice{0}} = \Phi_0$ given by \eqref{eq:psi0}, vanishing momentum, and energy $E^\textsc{cs}(\choice{0}) = \beta^2 E^0\!/2$.
\medskip

\textbf{(Symmetric) Jack polynomials.}
The polynomial $P_{\choice{\lambda}}(\vect{z})$ is uniquely characterised by the conditions \eqref{eq:symm_Jack_conditions}, analogously to \eqref{eq:nonsymm_Jack} from the nonsymmetric case. Jack polynomials form a basis for $\mathbb{C}[\vect{z},\vect{z}^{-1}]^{S_N}$ labelled by partitions $\choice{\lambda}$. They have various useful properties, see e.g.\ \cite{Sta_89,Mac_95,BF_99}. We record some of them for later use.
\begin{itemize}
	\item By \eqref{eq:symm_Jack}, $P_{\choice{\lambda}}(\vect{z})$ is a symmetric homogeneous polynomial of (total) degree $|\choice{\lambda}|$. 
	\item In particular, $P_{\choice{0}}(\vect{z})=1$ if $\choice{\lambda} = \choice{0} \equiv (0^N)$ is the empty partition.
	\item Jack polynomials inherit the \emph{stability} property \eqref{eq:nonsym_jack_stability}, i.e.\
\begin{equation} \label{eq:sym_jack_stability}
	P_{\choice{\lambda} + \choice{1}}(\vect{z}) = z_1 \cdots z_N \, P_{\choice{\lambda}}(\vect{z}) \, .
\end{equation}
	\item Special cases of Jacks include monomial symmetric polynomials~$m_{\choice{\lambda}}$ ($\beta\to 0$), zonal polynomials ($\beta = 1/2$), Schur polynomials $s_{\choice{\lambda}}$ ($\beta=1$), zonal spherical polynomials ($\beta=2$), and elementary symmetric polynomials $e_{\choice{\lambda}'}$ ($\beta\to \infty$). Here $\choice{\lambda}'$ denotes the partition  \emph{conjugate} to $\choice{\lambda}$, and 
\begin{equation} \label{eq:e_n}
	e_{\choice{\lambda}}(\vect{z}) \equiv \prod_{n \in  \choice{\lambda}} e_n(\vect{z}) \, , \qquad e_n(\vect{z}) \equiv \sum_{i_1 < \cdots < i_n}^N \!\!\! z_{i_1} \cdots z_{i_n} \, .
\end{equation}
We further recall that Schur polynomials are given by Cauchy's bialternant formula
\begin{equation} \label{eq:schur}
	s_{\choice{\lambda}}(\vect{z}) \equiv \frac{a_{\choice{k}}(\vect{z})}{\Delta(\vect{z})} \, , \qquad k_j \equiv \lambda_j + N - j \, ,
\end{equation}
where the alternant determinant
\begin{equation} \label{eq:asym_schur}
	a_{\choice{k}}(\vect{z}) \equiv \det\Bigl(z_i^{k_j}\Bigr)_{i,j=1}^N 
	= \sum_{\sigma \in S_N} \! \mathrm{sgn}(\sigma) \, z_1^{k_{\sigma(1)}} \cdots z_N^{k_{\sigma(N)}} = \Pi^\mathrm{pol}_- \, z_1^{k_1} \cdots z_N^{k_N}
\end{equation}
reduces to the Vandermonde polynomial~\eqref{eq:Vandermonde} for $\choice{\lambda} = \choice{0}$. In that case $\choice{k}$ is the \emph{staircase partition} $\choice{\delta_N} \equiv (N-1,\dots,1,0)$, i.e.\ the highest exponent of $\Delta(\vect{z}) = \prod_j z_j^{N-j} + \text{lower}$.
\end{itemize}
\medskip

\textbf{Spinless fermions.} 
If we instead \emph{anti}symmetrise we arrive at the fermionic case, with lower sign in \eqref{eq:pol_phys_space}. The restriction of the effective Hamiltonian~\eqref{eq:CS_eff} reads
\begin{equation} \label{eq:CS_eff_ferm_scalar}
	H_-^{\prime\mspace{2mu}\textsc{cs}} = \frac{1}{2} \sum_{j=1}^N (z_j \, \partial_{z_j})^2 + \frac{\beta}{2} \sum_{i<j}^N \frac{z_i + z_j}{z_i - z_j} \, \bigl(z_i \, \partial_{z_i} - z_j \, \partial_{z_j}\bigr) + 2\,\beta \sum_{i<j}^N \frac{z_i \, z_j}{z_{ij} \, z_{ji}} \, .
\end{equation}
This time the potential explicitly appears in the the effective Hamiltonian.

To obtain the eigenfunctions we use the antisymmetriser~\eqref{eq:anti/symm}.
By \eqref{eq:nonsym_jack_symm} the result of $\Pi^\text{pol}_- E_{\choice{k}}$ is nonzero if{f} $\choice{k}$ has pairwise distinct parts, i.e.\ if{f} $\choice{k}^+ = (k_1 > k_2 > \dots)$ is a \emph{strict} partition. This is precisely true when $\choice{k}^+ = \choice{\lambda} + \choice{\delta_N}$ for some (not necessarily strict) partition $\choice{\lambda}$ as in \eqref{eq:schur}. The result of the antisymmetrisation is divisible by the Vandermonde polynomial, with symmetric quotient. Remarkably, this quotient is again a Jack polynomial, with partition $\choice{\lambda}$ and reduced coupling parameter shifted as $\beta \mapsto \beta +1$ \cite{BF_99, Mac_00}:\,%
\footnote{\ Macdonald \cite{Mac_00} writes $Q_\lambda(\vect{z})$ for the antisymmetric Jack polynomial $\Delta(\vect{z}) \, [P_{\choice{\lambda}}(\vect{z})]_{\beta\mapsto \beta + 1}$.}
\begin{equation} \label{eq:nonsymm_to_asymm_Jack}
	\mathrm{cst}_-(\choice{k}) \; \Pi^\text{pol}_- \, E_{\choice{k}}(\vect{z}) = 
	\begin{cases}
		\, \Delta(\vect{z}) \, [P_{\choice{\lambda}}(\vect{z})]_{\beta \mapsto \beta +1} & \text{if{f}} \ \choice{k}^+ = \choice{\lambda} + \choice{\delta_N} \ \text{has distinct parts} \, , \\
		\, 0 & \text{else} \, .
	\end{cases}
\end{equation}
Note that $\Delta(\vect{z}) \, [P_{\choice{\lambda}}(\vect{z})]_{\beta \mapsto \beta +1} = z^{\choice{\lambda} + \choice{\delta_N}} + \text{lower monomials}$. 
We now have
\begin{equation} \label{eq:asymm_Jack_eigval_eqn}
	H_-^{\prime\mspace{2mu}\textsc{cs}} \; \Delta(\vect{z}) \, [P_{\choice{\lambda}}(\vect{z})]_{\beta \mapsto \beta +1} = E^{\prime\mspace{2mu}\textsc{cs}}(\choice{\lambda} + \choice{\delta_N}) \; \Delta(\vect{z}) \, [P_{\choice{\lambda}}(\vect{z})]_{\beta \mapsto \beta +1}\, . 
\end{equation}
This yields a basis for $\Delta(\vect{z}) \, \mathbb{C}[\vect{z},\vect{z}^-]^{S_N}$ indexed by \emph{strict partitions} $\choice{k} = \choice{\lambda} + \choice{\delta_N}$, i.e.\ the quantum numbers for spinless fermions. By a gauge transformation it follows that the original fermionic hamiltonian $H^\textsc{cs}_-$ has (antisymmetric) wave functions 
\begin{equation} \label{eq:fermionic_wave_fns}
	\Psi_{\choice{\lambda}}(\vect{z}) \equiv \Delta(\vect{z}) \, \Phi_0(\vect{z}) \, [P_{\choice{\lambda}}(\vect{z})]_{\beta \mapsto \beta+1} \, ,
\end{equation}
with momentum $|\choice{k}| = |\choice{\lambda}| + N(N-1)/2$ and energy $E^\textsc{cs}(\choice{\lambda} + \choice{\delta_N})$. Assuming $\beta\geqslant 0$ the free case occurs at $\beta = 0$ (fermions). Here the wave functions $\Psi_{\choice{\lambda}}(\vect{z})|_{\beta = 0} = a_{\choice{k}}(\vect{z})$ are just the antisymmetrised plane wave~\eqref{eq:asym_schur}.

The bosonic and fermionic wave functions are related by $\Psi_{\choice{\lambda}}(\vect{z}) = \mathrm{sgn}(\vect{z}) \, [\Phi_{\choice{\lambda}}(\vect{z})]|_{\beta \mapsto \beta+1}$ for particles on the unit circle $|z_j| = 1$. At the level of the hamiltonians~\eqref{eq:CS_scalar} the fermionic coupling constant is likewise obtained from the bosonic case by $\beta\,(\beta+1) = [\beta\,(\beta-1)]_{\beta \mapsto \beta +1}$.

\subsection{Fermionic spin-1/2 Calogero--Sutherland model} \label{sec:rank_one}

Now consider $N$ particles with coordinates and spins. For simplicity we focus on $SU(2)$ with spin~1/2, but this can be generalised to higher rank as we will show in Section~\ref{sec:higer_rank}. Here we concentrate on fermions; the bosonic case is outlined in Appendix~\ref{sec:bosonic}. 
The fermionic \emph{physical space} consists of totally antisymmetric vectors
\begin{equation} \label{eq:phys_space_fermionic}
	\widetilde{\mathcal{F}} \equiv \Bigl\{ \ket{\widetilde{\Psi}} \in (\mathbb{C}^2)^{\otimes N} \otimes \mathbb{C}[\vect{z}, \vect{z}^{-1}] : P_{ij} \, s_{ij}\, \ket{\widetilde{\Psi}} = -\ket{\widetilde{\Psi}} \Bigr\} \, .
\end{equation} 
It is the image in the ambient space%
\,\footnote{\ The fermionic physical space can be viewed either as a quotient, cf.\ the exterior algebra $\Lambda(V) = T(V)/\text{(relations enforcing antisymmetry)}$ of a vector space~$V$, or as a subspace, as we do here following~\cite{Ugl_96}.} 
\begin{equation} \label{eq:big_space}
	(\mathbb{C}^2)^{\otimes N} \otimes \mathbb{C}[\vect{z}, \vect{z}^{-1}]
\end{equation}
of the total (spin and coordinate) antisymmetriser
\begin{align} \label{eq:antisymmetriser}
	\Pi^\mathrm{tot}_- \equiv \sum_{\sigma \in S_N} \! \mathrm{sgn}(\sigma) \, P_{\sigma} \, s_{\sigma} \, , \qquad \bigl(\Pi^\mathrm{tot}_-\bigr)^2 = N! \; \Pi^\mathrm{tot}_- \, .
\end{align}
We extend the Dunkl operators to \eqref{eq:big_space} by letting them act trivially on the spin part, and keep the notation $d_j$ for $\mathbbm{1} \otimes d_j$. Like in the scalar case, symmetric combinations of these Dunkl operators commute with the permutations $P_{ij}$ and $s_{ij}$ and thus preserve the fermionic space~$\widetilde{\mathcal{F}}$. This again furnishes a hierarchy of abelian symmetries. The momentum operator is $\mathbbm{1} \otimes$ \eqref{eq:momentum_operator} like in the scalar case. Using the fermionic condition the effective hamiltonian~\eqref{eq:CS_eff} becomes
\begin{equation} \label{eq:CS_eff_ferm_spin}
	\widetilde{H}_-^{\prime\mspace{2mu}\textsc{cs}} = \frac{1}{2} \sum_{j=1}^N (z_j \, \partial_{z_j})^2 + \frac{\beta}{2} \sum_{i<j}^N \frac{z_i + z_j}{z_i - z_j} \, \bigl(z_i \, \partial_{z_i} - z_j \, \partial_{z_j}\bigr) + \beta \sum_{i<j}^N \frac{z_i \, z_j}{z_{ij} \, z_{ji}} \, (1+P_{ij}) \, .
\end{equation}
The scalar case~\eqref{eq:CS_eff_ferm_scalar} is recovered by replacing $P_{ij} \rightsquigarrow 1$. Upon gauge transforming as in \eqref{eq:Ham_gauge} we get the fermionic spin-Calogero--Sutherland hamiltonian $\widetilde{H}_-^\textsc{cs}$ from~\eqref{eq:CS_spin}.

A new feature of the spin case is the presence of further symmetries, which enhance the obvious $\mathfrak{sl}_2$ spin symmetry (isotropy) of the abelian symmetries. The extra spin symmetries form a representation of an infinite-dimensional algebra called the Yangian, see Appendix~\ref{sec:nonabelian}. The presence of these \emph{nonabelian symmetries} causes the spectrum of the spin-Calogero--Sutherland model to exhibit (high) degeneracies. Like in the scalar case the eigenvectors can be described exactly. We outline two approaches.

\subsubsection{Coordinate basis} \label{sec:coord_basis}

Following \cite{LPS_22}, elements of the fermionic space admit an explicit description, as follows. Recall that the spin space $(\mathbb{C}^2)^{\otimes N}$ decomposes into magnon sectors ($S^z$-eigenspaces, weight spaces). Anticipating the higher-rank case in Section~\ref{sec:higer_rank}, we will write $(\mathbb{C}^2)^{\otimes N}[N-2M]$ for the $M$-magnon sector.
It has a basis labelled by the positions of the $\downarrow$s on the chain,\,%
\footnote{\ This basis for spin chains is usually called the \emph{coordinate basis}, not related to our `coordinates' $z_j$.}
\begin{equation} \label{eq:coord_basis}
	\cket{i_1,\dots,i_M} \equiv \sigma^-_{i_1} \cdots \sigma^-_{i_M} \, \ket{\uparrow\cdots \uparrow} \, , \qquad 1\leqslant i_1 < \dots < i_M \leqslant N \, .
\end{equation}
The fermionic space $\widetilde{\mathcal{F}}$ similarly decomposes into magnon sectors\,%
\footnote{\ More precisely: the ambient space~\eqref{eq:big_space} clearly decomposes in this way. Its $M$-particle sectors are preserved by the spin permutations $P_{ij}$, so make sense at the level of the \eqref{eq:phys_space_fermionic} space too.} 
with fixed $S^z$-eigenvalue:
\begin{equation} \label{eq:magnon_sectors}
	\widetilde{\mathcal{F}} = \bigoplus_{M=0}^N \widetilde{\mathcal{F}}[N-2M] \, , \qquad \widetilde{\mathcal{F}}[N-2M]  \equiv \ker\bigl[S^z - \tfrac12(N-2M) \bigr] \, .
\end{equation}
This time the $M$-magnons sectors $\widetilde{\mathcal{F}}[N-2M]$ are infinite dimensional because of the polynomial part, but their elements can still be described explicitly. Namely, any fermionic $M$-magnon vector $\ket{\widetilde{\Psi}} \in \widetilde{\mathcal{F}}[N-2M]$ is completely determined by a single spin component, which we will take to be the \emph{simple spin component} $\cbraket{1,\dots,M}{\widetilde{\Psi}}$ where all $\downarrow$s are on the left. Moreover, this simple component is a Laurent polynomial in the $z_j$ of definite (anti)symmetry, cf.~\cite{LPS_22}. The reason for the latter is that, in order to compensate for the symmetry of $\cbra{1,\dots,M}$ in the $\downarrow$s (and in the $\uparrow$s), the fermionic condition (total antisymmetry) requires $\cbraket{1,\dots,M}{\widetilde{\Psi}}$ to be \emph{anti}symmetric in $z_1,\dots,z_M$ (and in $z_{M+1},\dots,z_N$). It follows that the simple component contains two Vandermonde factors:
\begin{equation} \label{eq:simple_component}
	\cbraket{1,\dots,M}{\widetilde{\Psi}} = \Delta(z_1,\dots,z_M) \, \Delta(z_{M+1},\dots,z_N) \, \widetilde{\Psi}(z_1,\dots,z_M;z_{M+1},\dots,z_N) \, ,
\end{equation}
where 
\begin{equation} \label{eq:psi_tilde}
	\widetilde{\Psi}(\vect{z}) \in \mathbb{C}[\vect{z}, \vect{z}^{-1}]^{S_M \times S_{N-M}}
\end{equation}
is a Laurent polynomial that is \emph{symmetric} in $z_1,\dots,z_M$ and $z_{M+1},\dots,z_N$ separately. The fermionic condition furthermore determines any spin component $\cbraket{i_1,\dots,i_M}{\widetilde{\Psi}}$ in terms of the simple spin component~\eqref{eq:simple_component}.\,%
\footnote{\ Sketch of the proof. For $M=1$ compute $\cbraket{i}{\widetilde{\Psi}} = \cbra{i-1} \, P_{i-1,i} \, \ket{\widetilde{\Psi}} = \cbra{i-1} \,(-s_{i-1,i}) \,\ket{\widetilde{\Psi}} = -s_{i-1,i} \, \cbraket{i-1}{\widetilde{\Psi}}$. Iteration gives $\cbraket{i}{\widetilde{\Psi}} = (-1)^{i-1} s_{i-1,i} \cdots s_{12} \, \cbraket{1}{\widetilde{\Psi}} = (-1)^{i-1} s_{1i} \, \cbraket{1}{\widetilde{\Psi}}$. For $M>1$ use this to push the $\downarrow$ at $i_1$ to $1$, then the $\downarrow$ from $i_2$ to $2$, and so on, to get \eqref{eq:physvec}. See \cite{LPS_22} for the full argument in the $q$-bosonic case.} 

For example, for $M=0$ the only spin component is a skew-symmetric Laurent polynomial,
\begin{align}
	\label{eq:physvecM=0}
	& M=0\colon \quad && \ket{\widetilde{\Psi}} = \Delta(z_1,\dots,z_N) \, \widetilde{\Psi}(z_1,\dots,z_N) \, \ket{\uparrow\cdots\uparrow} \, ,
\intertext{where $\widetilde{\Psi}(\vect{z}) \in \mathbb{C}[\vect{z}, \vect{z}^{-1}]^{S_N}$ is a (totally) symmetric Laurent polynomial. Next,}
	\label{eq:physvecM=1}
	& M=1\colon \quad && \ket{\widetilde{\Psi}} = \sum_{i=1}^N \, (-1)^{i-1} \Delta(z_1,\dots\widehat{z_i}\dots,z_N) \, \widetilde{\Psi}(z_i;z_1,\dots\widehat{z_i}\dots,z_N) \, \cket{i} \, ,
\end{align}
where the caret indicates that $z_i$ is omitted. In this case, $\widetilde{\Psi}(\vect{z}) \in \mathbb{C}[\vect{z}, \vect{z}^{-1}]^{S_1 \times S_{N-1}}$ is symmetric in $z_2,\dots,z_N$.

To write down the form of fermionic vectors in an arbitrary $M$-magnon sector we just need to set up some notation. We will use the shorthand
\begin{equation}
	\cket{I} \equiv \cket{i_1,\dots,i_M} \, , \quad \vect{z}_I \equiv \{z_{i_1},\dots,z_{i_M}\} , \qquad \text{for} \quad I = \{i_1,\dots,i_M \} \subset \{1,\dots,N\} \, .
\end{equation}
We further denote the complement of $I$ in $\{1,\dots,N\}$ by
\begin{equation}
	I^\text{c} \equiv \{1,\dots,N\} \setminus I \, ,
\end{equation}
and define
\begin{equation}
	\mathrm{sgn}(I) \equiv (-1)^{i_1 + \cdots + i_M - M(M+1)/2} \, .
\end{equation}
Then any $\ket{\widetilde{\Psi}} \in \widetilde{\mathcal{F}}[N-2M]$ can be written as
\begin{equation} \label{eq:physvec}
	\ket{\widetilde{\Psi}} = \sum_{\substack{ I \subset \{1,\dots,N\} \\ \# I = M }}^N \!\!\!\! \mathrm{sgn}(I) \, \Delta(\vect{z}_I) \, \Delta(\vect{z}_{I^\text{c}}) \, \widetilde{\Psi}(\vect{z}_I \, ; \vect{z}_{I^\text{c}}) \, \cket{I\mspace{1mu}} \, ,
\end{equation}
where the sum is over all $M$-element subsets of $\{1,\dots,N\}$ and the Laurent polynomial $\widetilde{\Psi}$ is doubly symmetric as in \eqref{eq:psi_tilde}. Note that the latter is indeed obtained from the simple spin component as in \eqref{eq:simple_component}.\,%
\footnote{\ This result, that physical $M$-particle vectors are completely determined by Laurent polynomials with appropriate (anti)symmetry, may be stated mathematically as follows: we have a bijection
\begin{equation*} 
	\begin{aligned}
		& \Delta(z_1,\dots,z_M) \, \Delta(z_{M+1},\dots,z_N) \, \mathbb{C}[\vect{z}, \vect{z}^{-1}]^{S_M \times S_{N-M}} \!\!\!\!\! && \xrightarrow{\ \sim\ } \widetilde{\mathcal{F}}[N-2M] \\
		& \Delta(z_1,\dots,z_M) \, \Delta(z_{M+1},\dots,z_N) \, \widetilde{\Psi}(\vect{z}) && \xmapsto{\ \hphantom{\sim} \ } \eqref{eq:physvec} \, , \\
		& \qquad\qquad\qquad\qquad\qquad \eqref{eq:simple_component} && \ \raisebox{.2cm}{\rotatebox{180}{$\longmapsto$}} \ \, \ket{\widetilde{\Psi}} \, ,
	\end{aligned}
\end{equation*}
where the space on the left is also in bijection with \eqref{eq:psi_tilde}. To be precise, these are isomorphisms of $\mathbb{C}[\vect{z}, \vect{z}^{-1}]^{S_N}$-modules: totally symmetric Laurent polynomials remain overall (pre)factors throughout these bijections. Cf.~e.g.\ the discussion about $\mathbb{C}[\chi]$-modules in \cite{CLV_22} and references therein.} 
The preceding provides a basis of $\widetilde{\mathcal{F}}[N-2M]$ by picking a basis of partially symmetric polynomials in $\mathbb{C}[\vect{z}, \vect{z}^{-1}]^{S_M \times S_{N-M}}$.
\bigskip

\noindent
We can now construct the eigenvectors of the spin-Calogero--Sutherland model per $M$-particle sector by proceeding analogously to the scalar case. Namely, for the $\widetilde{\mathcal{F}}[N-2M]$ we can apply the total antisymmetriser~\eqref{eq:antisymmetriser} to $E_{\choice{\lambda}}(\vect{z}) \, \cket{1\cdots M}$. As in the scalar case, see \eqref{eq:nonsymm_to_asymm_Jack}, the resulting vectors are not all linearly independent. To see this, consider the simple spin component~\eqref{eq:simple_component}. It is obtained using the partial coordinate antisymmetrisers
\begin{equation} \label{eq:partial_antisym}
	\Pi_{M,-}^\mathrm{pol} \equiv \!\!\! \sum_{\sigma \in S_M \times S_{N-M}} \!\!\!\!\!\!\!\!\! \mathrm{sgn}(\sigma) \, s_\sigma \, , \qquad \bigl(\Pi_{M,-}^\mathrm{pol}\bigr)^2  = \binom{N}{M} \, \Pi_{M,-}^\mathrm{pol} \, .
\end{equation}
For example, $\Pi_{0,-}^\mathrm{pol} = \Pi_{N,-}^\mathrm{pol} = \Pi_-^\mathrm{pol}$ are the coordinate antisymmetriser from the case of spinless fermions.
The nonsymmetric Jack polynomial $E_{\choice{\mu}}(\vect{z})$ with $\choice{\mu} \in \mathbb{Z}^N$ is thus sent to
\begin{equation} \label{eq:partial_symm}
	\Pi_{M,-}^\mathrm{pol} \, E_{\choice{\mu}}(\vect{z}) = \Delta(z_1,\dots,z_M) \, \Delta(z_{M+1},\dots,z_N) \; \widetilde{\Psi}(\vect{z}) \, .
\end{equation} 
The partially symmetric quotient $\widetilde{\Psi}(\vect{z}) \in \mathbb{C}[\vect{z}, \vect{z}^{-1}]^{S_M \times S_{N-M}}$ is nonzero provided $(\mu_1,\dots,\mu_M)$ are pairwise distinct and the $(\mu_{M+1},\dots,\mu_N)$ are so too. Moreover, up to normalisation, the results depend only $(\mu_1,\dots,\mu_M)^+$ and $(\mu_{M+1},\dots,\mu_N)^+$. We thus obtain a basis for the space $\widetilde{\mathcal{F}}[N-2M]$ of $M$-magnon fermions indexed by two strict partitions $(\mu_1>\dots>\mu_M)$ and $(\mu_{M+1}>\dots>\mu_N)$. These are the quantum numbers of spin-1/2 (really: $\mathfrak{sl}_2$) fermions. See \cite{TU_97} and \cite{FLLS_24} for an explicit description of the resulting eigenspaces. 
Since we will not need the details we suffice with a simple but instructive example. If $M=0$ then by \eqref{eq:nonsymm_to_asymm_Jack} we get $\widetilde{\Psi}(\vect{z}) = [P_{\choice{\lambda}}(\vect{z})]_{\beta \mapsto \beta+1}$, yielding a vector in $\widetilde{\mathcal{F}}[N]$ via \eqref{eq:physvecM=0}. Thus, the zero-magnon sector of the spin-Calogero--Sutherland model is already equivalent to the complete scalar fermionic space! Each of these $M=0$ vectors is the highest-weight vector for a Yangian irrep of dimension~$2^N$: the spin case exhibits up to exponentially large degeneracies because of the nonabelian symmetries. We will see these degeneracies in terms of eigenvalues towards the end of the next section.

\subsubsection{Wedge basis} \label{sec:wedges}

Uglov~\cite{Ugl_96, Ugl_98} proposed to diagonalise the fermionic spin-Calogero--Sutherland model on another basis of $\widetilde{\mathcal{F}}$, obtained in the following way. For a single particle ($N=1$) define the vectors\,%
\footnote{\ Note that our conventions differ from Uglov~\cite{Ugl_98} in that for us $k$ \emph{in}creases with $\bar{k}$.}
\begin{equation} \label{eq:wedge_single_particle}
	u_k \equiv z^{\bar{k}} \, \ket{\underbar{k}} \,, \quad k \in \mathbb{Z} \, , \quad \text{with} \quad \bar{k} \equiv \lfloor k/2 \rfloor = 
	\begin{cases*}
		k/2 \\
		(k-1)/2 
	\end{cases*}
	\ \ \ \text{and} \quad \underbar{k} \equiv 
	\begin{cases*}
		\uparrow \quad & $k$ even, \\
		\downarrow & $k$ odd,
	\end{cases*}
\end{equation}
so that
\begin{equation} \label{eq:wedge_diagram}
	\begin{aligned}
		& \ \, \vdots && \qquad \ \, \vdots \\[-.3\baselineskip]
		u_4 & = z^{2\hphantom{-}} \ket{\uparrow} \qquad && \ \; u_5 = z^{2\hphantom{-}} \mspace{1mu} \ket{\downarrow} \\
		u_2 & = z^{\hphantom{-1}} \ket{\uparrow} && \ \; u_3 = z^{\hphantom{-1}} \mspace{1mu} \ket{\downarrow} \\
		u_0 & = \hphantom{z^{-1}} \ket{\uparrow} && \ \; u_1 = \hphantom{z^{-1}} \mspace{1mu} \ket{\downarrow} \\
		u_{-2} & = z^{-1} \ket{\uparrow} && u_{-1} = z^{-1} \ket{\downarrow} \\[-.3\baselineskip]
		& \ \, \vdots && \qquad \ \, \vdots 
	\end{aligned}
\end{equation}
This is just a reordering of the (standard) basis $z^k$ ($k \in \mathbb{Z}$) of $\mathbb{C}[z,z^{-1}]$ and $\ket{\uparrow},\ket{\downarrow}$ of $\mathbb{C}^2$.
From these vectors one builds a basis of $\widetilde{\mathcal{F}}$ by applying the total antisymmetriser~\eqref{eq:antisymmetriser} to $N$-fold tensor products in order to get skew-symmetric vectors called \emph{wedges} (cf.~\cite{KMS_95}), 
\begin{align} \label{eq:wedge}
	\widehat{u}_{\choice{k}} \equiv u_{k_1} \wedge \ldots \wedge u_{k_N} \equiv \sum_{\sigma\in S_N} \! {\rm sgn}(\sigma)\, u_{k_{\sigma(1)}}\otimes \cdots \otimes u_{k_{\sigma(N)}} \, .
\end{align}
Here the coordinate $z$ in the $i$th factor of a wedge or tensor product is identified with $z_i$. By antisymmetry we may assume that $\choice{k}=(k_1,\ldots,k_N)$ is ordered as $k_1>\ldots>k_N$. In particular, $\widehat{u}_{\choice{k}} = 0$ if any two $k_j$ are equal. Following \cite{Ugl_96} will write $\bar{\choice{k}} \equiv (\bar{k}_1,\dots,\bar{k}_N)$ for the `polynomial part' of $\choice{k}$ and $\underbar{\choice{k}} \equiv (\underbar{k}_1,\dots,\underbar{k}_N)$ for its `spin part'.

Before we work out the form of a general wedge let us give some simple examples of wedges. For $N=2$ we have
\begin{equation}
	\begin{aligned}
	\widehat{u}_{(1,0)} = 
	\hphantom{(z \,} \ket{\downarrow} \wedge \ket{\uparrow}  \hphantom{)} & = \ket{\downarrow\uparrow} - \ket{\uparrow\downarrow} \, , \\
	\widehat{u}_{(2,0)} = (z\,\ket{\uparrow}) \wedge \ket{\uparrow} & = (z_1-z_2)\,\ket{\uparrow\uparrow} \, , \\
	\widehat{u}_{(3,0)} = (z\,\ket{\downarrow}) \wedge \ket{\uparrow} & = z_1 \,\ket{\downarrow\uparrow} - z_2 \,\ket{\uparrow\downarrow} \\
		& = \frac{1}{2} (z_1 - z_2) \, \bigl( \ket{\downarrow\uparrow} + \ket{\uparrow\downarrow} \bigr) + \frac{1}{2}(z_1 + z_2) \, \bigl( \ket{\downarrow\uparrow} - \ket{\uparrow\downarrow} \bigr) \,.
	\end{aligned}
\end{equation}
If $N=4$ the simplest ferromagnetic vector at $M=0$ is 
\begin{equation} \label{eq:N=4_FM_wedge}
	\widehat{u}_{(6,4,2,0)} = \bigl(z^3\,\ket{\uparrow}\bigl) \,\wedge\, \bigl(z^2\,\ket{\uparrow}\bigr) \wedge (z\,\ket{\uparrow}) \wedge \ket{\uparrow} = \Delta(z_1,z_2,z_3,z_4) \,\ket{\uparrow\uparrow\uparrow\uparrow} \, ,
\end{equation}
while the antiferromagnetic ground state at $M=2$ is 
\begin{equation} \label{eq:N=4_AFM_wedge}
	\begin{aligned}
	\widehat{u}_{(3,2,1,0)} & = (z\,\ket{\downarrow}) \wedge (z\,\ket{\uparrow})\wedge \ket{\downarrow}\wedge \ket{\uparrow}) \\
	& = z_{13} \, z_{24} \, \bigl( \ket{\downarrow\uparrow\downarrow\uparrow} + \ket{\uparrow\downarrow\uparrow\downarrow} \bigr) - z_{14} \, z_{23} \, \bigl( \ket{\uparrow\downarrow\downarrow\uparrow}  + \ket{\downarrow\uparrow\uparrow\downarrow} \bigr) \\
	& \hphantom{\ = z_{13} \, z_{24} \, \bigl( \ket{\downarrow\uparrow\downarrow\uparrow} + \ket{\uparrow\downarrow\uparrow\downarrow} \bigr) } - z_{12}\,z_{34} \, \bigl( \ket{\uparrow\uparrow\downarrow\downarrow} + \ket{\downarrow\downarrow\uparrow\uparrow} \bigr)\,.
	\end{aligned}
\end{equation}
These patterns persist to general $N$: the ferromagnetic pseudovacuum at $M=0$ is $\widehat{u}_{2 \mspace{1mu} \choice{\delta_N}} = \Delta(\vect{z})\,\ket{\uparrow\!\cdots\!\uparrow}$ where $2\,\choice{\delta_N} \equiv (2N-2,\dots,2,0)$, and, if $N$ is even, then the antiferromagnetic ground state at the equator $M=N/2$ corresponds to $\choice{k} = \choice{\delta_N}$ \cite{Ugl_96}. 
Back to $N=4$, an example of a wedge with $M=1$ is
\begin{equation}
	\begin{aligned}
	\widehat{u}_{(4,2,1,0)} = {} & \bigl(z^2\,\ket{\uparrow}\bigr) \wedge (z\,\ket{\uparrow})\wedge \ket{\downarrow}\wedge \ket{\uparrow} \\
	= {} & z_{12} \, z_{14} \, z_{24} \, \ket{\uparrow\uparrow\downarrow\uparrow} - z_{12} \, z_{13} \, z_{23} \, \ket{\uparrow\uparrow\uparrow\downarrow} \\
	& + z_{23} \, z_{24} \, z_{34} \, \ket{\downarrow\uparrow\uparrow\uparrow} - z_{13} \, z_{14} \, z_{34}\ket{\uparrow\downarrow\uparrow\uparrow} \, ,
	\end{aligned}
\end{equation}
with one magnon, corresponding to $(1,0,0,0)$, above the ferromagnetic ground state labelled by $\choice{k} = \choice{\delta_4} = (3,2,1,0)$. 
Finally, for general $N$ when $M=0$ we have $\choice{k} = 2\,\bar{\choice{k}}$. The symmetry of $\ket{\uparrow\cdots \uparrow}$ forces the polynomial part to be totally antisymmetric:
\begin{equation} \label{eq:wedge_M=0}
	\widehat{u}_{2\,\bar{\choice{k}}}^{(M=0)} = \bigl( z^{\bar{k}_1} \, \ket{\uparrow} \bigr) \wedge \ldots \wedge \bigl( z^{\bar{k}_N} \, \ket{\uparrow} \bigr) = a_{\bar{\choice{k}}}(\vect{z}) \, \ket{\uparrow \cdots \uparrow} = \Delta(\vect{z}) \, s_{\choice{\lambda}}(\vect{z}) \, \ket{\uparrow \cdots \uparrow} \, ,
\end{equation}
where $\bar{k}_j = \lambda_j + N-j$ now plays the role of $k_j$ in the definition~\eqref{eq:schur} of the Schur polynomial. That is, wedges can be viewed as spin-generalisations of the antisymmetric Schur polynomials~\eqref{eq:asym_schur}. (Unlike polynomials, however, wedges cannot be multiplied at a fixed~$N$.)

To understand the explicit form of an arbitrary $M$-magnon wedge we divide the `coordinate integers'~$\bar{k}_j$ into two groups, one for magnons ($\downarrow\mspace{1mu}$, `particles') and one for holes ($\uparrow$), and relabel them as
\begin{equation} \label{eq:splitk}
	\underbar{k}_j = 1 \colon \quad \bar{k}_1^{(\downarrow)} >\ldots>\bar{k}_{M}^{(\downarrow)} \, , \qquad \underbar{k}_j = 0 \colon \quad \bar{k}_1^{(\uparrow)} >\ldots>\bar{k}_{N-M}^{(\uparrow)} \, .
\end{equation}
In this way we split $\choice{k}$ into two strict partitions
\begin{equation}
	\choice{k}^{(\downarrow)} \equiv \bigl( 2\,\bar{k}_1^{(\downarrow)} + 1 , \ldots , 2\,\bar{k}_{M}^{(\downarrow)} + 1 \bigr) \, , \qquad \choice{k}^{(\uparrow)} \equiv \bigl( 2\,\bar{k}_1^{(\uparrow)} , \ldots , 2\,\bar{k}_{N-M}^{(\uparrow)} \bigr) \, .
\end{equation}
Let us accordingly reorder the factors of the $M$-magnon wedge as
\begin{equation}
	\widehat{u}_{\choice{k}}^{(M)} = \pm\widehat{u}_{\choice{k}^{(\downarrow)}} \wedge \widehat{u}_{\choice{k}^{(\uparrow)}} \, ,
\end{equation}
where the sign comes from reordering the factors in the wedge product.
Using \eqref{eq:wedge_M=0} for $\widehat{u}_{\choice{k}^{(\downarrow)}}$ and $\widehat{u}_{\choice{k}^{(\uparrow)}}$ separately we compute the (simple) component
\begin{equation} \label{eq:hwwedge}
	\begin{aligned}
	\cbra{1,\dots,M} \, \widehat{u}_{\choice{k}}^{(M)} \cket{1,\dots,M} = {} \pm & a_{\bar{\choice{k}}^{(\uparrow)}}(z_1,\dots,z_M) \, a_{\bar{\choice{k}}^{(\downarrow)}}(z_{M+1},\dots,z_N) \\ 
	= {} \pm & \Delta(z_1,\dots,z_M)\, s_{\choice{\lambda}^{(\downarrow)}} (z_1,\ldots, z_M) \\
	& \! \times  \Delta(z_{M+1},\dots,z_N) \, s_{\choice{\lambda}^{(\uparrow)}} (z_{M+1},\dots,z_N) \, ,
	\end{aligned}
\end{equation}
where, like in \eqref{eq:schur},
\begin{equation} \label{eq:ktolambda}
	\lambda^{(\downarrow)}_m = \bar{k}_m^{(\downarrow)} - M+m \,, \qquad \lambda^{(\uparrow)}_j = \bar{k}_j^{(\uparrow)} - N + M + j \, .
\end{equation}
Working out the remaining antisymmetrisation that moves around the $\downarrow$s, we conclude that
\begin{equation} \label{eq:wedgeM}
	\begin{aligned}
	\widehat{u}_{\choice{k}}^{(M)} & = \,\pm\!\!\!\!\!\! \sum_{\substack{I \subset \{1,\dots,N\} \\ \# I = M}} \!\!\!\! \mathrm{sgn}(I) \, a_{\bar{\choice{k}}^{(\downarrow)}}(\vect{z}_I)  \, a_{\bar{\choice{k}}^{(\uparrow)}}(\vect{z}_{I^\mathrm{c}}) \, \cket{I} \\
	& = \,\pm\!\!\!\!\!\! \sum_{\substack{I \subset \{1,\dots,N\} \\ \# I = M}} \!\!\!\! \mathrm{sgn}(I) \, \Delta(\vect{z}_I) \, \Delta(\vect{z}_{I^\mathrm{c}}) \, s_{\choice{\lambda}^{(\downarrow)}}(\vect{z}_I)  \, s_{\choice{\lambda}^{(\uparrow)}}(\vect{z}_{I^\mathrm{c}}) \, \cket{I} \, .
	\end{aligned}
\end{equation}
This expresses the wedge basis in the coordinate basis~\eqref{eq:physvec}.

On the wedge basis the effective hamiltonian~\eqref{eq:CS_eff_ferm_spin} of the fermionic spin-Calogero--Sutherland model acts triangularly (with respect to the dominance order\,\textsuperscript{\ref{fn:dominance} (p.\,\pageref{fn:dominance})}),
\begin{equation} \label{eq:hamwedge}
	\widetilde{H}_-^{\prime\mspace{2mu}\textsc{cs}} \,\widehat{u}_{\choice{k}} = E^{\prime\mspace{2mu}\textsc{cs}}\bigl(\bar{\choice{k}}\bigr) \, \widehat{u}_{\choice{k}} + \beta \sum_{i<j}^N \widetilde{h}_{ij}\,\widehat{u}_{\choice{k}} \, .
\end{equation}
Here the diagonal elements (and therefore eigenvalues) are as in the scalar case, see \eqref{eq:E_cs_eff}; in particular they do not see the spin part $\underbar{\choice{k}}$ of $\choice{k}$. This reflects the Yangian symmetry: for most choices of $\bar{\choice{k}}$ there are large degeneracies, with eigenspaces of dimension up to $2^N$. The off-diagonal part in \eqref{eq:hamwedge} is given by the `squeezing' operator
\begin{subequations} \label{eq:offdiag}
	\begin{gather}
	\widetilde{h}_{ij} \, \widehat{u}_{\choice{k}} \equiv \!\!\! \sum_{p=1}^{\bar{k}_i -\bar{k}_j-1} \!\!\! \bigl( \bar{k}_i-\bar{k}_j-p \bigr) \, u_{k_1}\wedge\ldots \wedge u_{k_i-2\,p}\wedge\ldots \wedge u_{k_j+2\,p}\wedge\ldots  \wedge u_{k_N} \, .
\intertext{Writing $\choice{\varepsilon_i}$ for the vector with entries $(\varepsilon_i)_j \equiv \delta_{ij}$ this can be recast in the more compact form}
 	\widetilde{h}_{ij} \, \widehat{u}_{\choice{k}} = \!\!\! \sum_{p=1}^{k_i -k_j -1} \!\!\! \bigl( k_i - k_j - p \bigr) \, \widehat{u}_{\choice{k}- 2 \, p \, (\choice{\varepsilon_i} - \choice{\varepsilon_j})} \, .
	\end{gather}
\end{subequations}
Note that $\choice{k} > \choice{\varepsilon_i} - \choice{\varepsilon_j})$ for $i<j$ and $p\geqslant 1$: the squeezing operator produces \emph{lower} wedges. 
Using antisymmetry the resulting subscripts can be ordered decreasingly to get back to the wedge basis. 
For instance, for $N=3$ the wedge
$\widehat{u}_{(6,1,0)} = \bigl(z^3 \, \ket{\uparrow}\bigr) \wedge \ket{\downarrow} \wedge \ket{\uparrow}$ is sent to 
\begin{equation*}
	\begin{aligned}
	\widetilde{h}_{12} \,  \widehat{u}_{(6,1,0)} & = 2 \, \widehat{u}_{(4,3,0)} + \widehat{u}_{(2,5,0)} = 2 \, \widehat{u}_{(4,3,0)} - \widehat{u}_{(5,2,0)} \\
		& = 2\, \bigl(z^2 \, \ket{\uparrow}\bigr) \wedge (z \, \ket{\downarrow}) \wedge \ket{\uparrow} - \bigl(z^2 \, \ket{\downarrow} \bigr) \wedge (z \, \ket{\uparrow}) \wedge \ket{\uparrow} \, , \\
	\widetilde{h}_{13} \, \widehat{u}_{(6,1,0)} & = 2 \, \widehat{u}_{(4,1,2)} + \widehat{u}_{(2,1,4)} = {-}\widehat{u}_{(4,2,1)} \\
		& = -\bigl(z^2 \, \ket{\uparrow}\bigr) \wedge (z \, \ket{\uparrow}) \wedge \ket{\downarrow} \, , \\
		\widetilde{h}_{23} \, \widehat{u}_{(6,1,0)} & = 0 \, .
	\end{aligned}
\end{equation*}
In the free-fermionic case $\beta=0$ the off-diagonal part disappears from \eqref{eq:hamwedge}, so the hamiltonian is diagonal on the wedges $\widehat{u}_{\choice{k}}$, which can therefore be interpreted as free-fermion eigenvectors (Slater determinants). For arbitrary $\beta$, \eqref{eq:hamwedge} is just the action of the hamiltonian on this basis.

For later use we note that, in particular, wedges have a stability property under simultaneous shifts (Galilean boosts) of their momenta
\begin{equation} \label{eq:wedge_boost}
	\widehat{u}_{\choice{k} + (l^N)} = (z_1\cdots z_N)^l \; \widehat{u}_{\choice{k}} \, , \qquad l \in \mathbb{Z} \, ,
\end{equation}
cf.~\eqref{eq:sym_jack_stability} in the scalar case. Since such factors can simply be pulled out in front of the off-diagonal part~\eqref{eq:offdiag}, eigenvectors are boosted to eigenvectors, with energy $E^{\prime\mspace{2mu}\textsc{cs}}\bigl(\bar{\choice{k}}+(l^{\mspace{-1mu}N})\bigr)$.
Since for generic values of $\beta>0$ and comparable $\choice{k},\choice{l}$ we have $E^{\prime\mspace{2mu}\textsc{cs}}\bigl(\bar{\choice{k}}\bigr) = E^{\prime\mspace{2mu}\textsc{cs}}\bigl(\bar{\choice{l}}\bigr)$ if{f} $\bar{\choice{k}} = \bar{\choice{l}}$, for each $\choice{k}$ we get an eigenvector of the form $\widehat{u}_{\choice{k}}[N-2M] + \text{lower wedges} \in \widetilde{\mathcal{F}}[N-2M]$. Uglov called these (antisymmetric) eigenvectors \emph{$\mathfrak{gl}_2$-Jack polynomials}, and found an alternative description of them as a particular limit of a (spinless!) Macdonald polynomial~\cite{Ugl_98}.

The decomposition of the fermionic space in terms of Yangian representations, Drinfeld polynomials, and norms of the eigenvectors and dynamical correlation functions can be found in \cite{Ugl_96, TU_97, Ugl_98}. Coming back to the original fermionic spin-Calogero--Sutherland hamiltonian~$\widetilde{H}^\textsc{cs}_-$ from \eqref{eq:CS_spin} we gauge transform to obtain eigenvectors $\ket{\widetilde{\Psi}_{\choice{k}}} = \Phi_0 \times (\widehat{u}_{\choice{k}}^{(M)} + \text{lower})$ with energy $E^\textsc{cs}\bigl(\bar{\choice{k}}\bigr)$ taking the same values  \eqref{eq:E_cs_eff}--\eqref{eq:E_cs} as in the scalar case and independent of the spin part $\underbar{\choice{k}}$ of $\choice{k}$, leading to degeneracies controlled by the Yangian.

\section{Haldane--Shastry spin chain} \label{sec:HS}

The spin-1/2 Haldane--Shastry spin chain is quantum integrable in the sense that it has abelian symmetries, nonabelian symmetries, and is exactly solvable: the hamiltonian \eqref{eq:HS_intro}
\begin{itemize}
	\item belongs to a hierarchy of commuting hamiltonians \cite{Ino_90, HH+_92}, each of which
	\item has enhanced (Yangian) $\mathfrak{sl}_2$ spin symmetry \cite{HH+_92}, and
	\item admits an explicit description of its (Yangian \cite{BPS_95a}) highest-weight  eigenvectors in terms of (the zonal spherical special case of) symmetric Jack polynomials~\cite{Hal_91a}. 
\end{itemize}
The striking parallel with the properties of the spin-Calogero--Sutherland model, cf.\ the start of \textsection\ref{sec:spin_CS}, is no coincidence. In \cite{BG+_93, TH_95} it was shown that the above properties can all be derived from the corresponding properties of the spin-Calogero--Sutherland model in a special limit called \emph{freezing}.

\subsection{Freezing} \label{sec:freezing}

Consider the limit $\beta\to\infty$ of the effective spin-Calogero--Sutherland hamiltonian~\eqref{eq:CS_eff_ferm_spin}. The dominant term is the part linear in $\beta$. The term with derivatives,
\begin{equation}
	\frac{\beta}{2} \sum_{i<j}^N \frac{z_i + z_j}{z_i - z_j} \, \bigl(z_i \, \partial_{z_i} - z_j \, \partial_{z_j}\bigr) = \frac{\beta}{2} \sum_{j=1}^N \Biggl( \, \sum_{i (\neq j)}^N \frac{z_i + z_j}{z_i - z_j} \Biggr) z_j \, \partial_{z_j} \, ,
\end{equation}
vanishes when the $z_j$ are fixed to distinct $N$th roots of unity (up to an irrelevant common factor). Indeed, defining the `evaluation' (specialisation)\,%
\footnote{\ More properly \cite{Ugl_96}, evaluation corresponds to taking the quotient of the physical space, e.g.\ \eqref{eq:pol_phys_space} or \eqref{eq:phys_space_fermionic}, by the ideal generated by $p_1(\vect{z}), \dots, p_{N-1}(\vect{z}),p_N(\vect{z}) - N$ where $p_k(\vect{z}) = \sum_i z_i^k$ are the power sums.}
\begin{equation} \label{eq:ev}
	\ev_\omega \colon z_j \mapsto \omega^j \,, \qquad \omega \equiv \mathrm{e}^{2\pi\mathrm{i}/N} \, ,
\end{equation}
we have the identity
\begin{equation} \label{eq:ev_sum_cot}
	\ev_\omega \sum_{i(\neq j)}^N \frac{z_i + z_j}{z_i - z_j} = -\mathrm{i} \sum_{n=1}^{N-1} \cot \frac{\pi\,n}{N} = 0 \, ,
\end{equation}
since the summand is odd, while the sum may be taken over a set symmetric around $n=0$ ($n=N/2$ does not contribute). The values in \eqref{eq:ev} are classical equilibrium positions. 
Thus we are led to linearising~\eqref{eq:CS_eff_ferm_spin} in $\beta$ and evaluating using \eqref{eq:ev}. This yields the \emph{anti}ferromagnetic version of the Haldane--Shastry spin chain~\eqref{eq:HS_intro},
\begin{equation} \label{eq:HS_AFM}
	H^\textsc{hs}_- = \ev_\omega \, \partial_\beta \big|_{\beta=0} \, \widetilde{H}_-^{\prime\mspace{2mu}\textsc{cs}} = \sum_{i < j}^N \ev_\omega \, \frac{z_i \, z_j}{z_{ij} \, z_{ji}} \, (1+P_{ij}) \, .
\end{equation}
Note that freezing the \emph{bosonic} spin-Calogero--Sutherland hamiltonian likewise yields the \emph{ferromagnetic} Haldane--Shastry spin chain (see Appendix~\ref{sec:freezing_symm_vs_asymm}).

By \eqref{eq:hamwedge} we further get possible eigenvalues for free by likewise taking the linear part in $\beta$ of the effective spin-Calogero--Sutherland energy~\eqref{eq:E_cs_eff}, cf.\,
\begin{equation} \label{eq:HSspectrum}
	E^\textsc{hs}_-(\bar{\choice{k}}) = \partial_\beta \big|_{\beta=0} \, E^{\prime\mspace{2mu}\textsc{cs}}(\bar{\choice{k}}) = \frac{1}{2}\sum_{i=1}^N \, (N-2\mspace{1mu}i+1)\, \bar{k}_i \, .
\end{equation}
This result needs to be supplemented by a rule that selects which $\bar{\choice{k}}$ are allowed, and thus which of the infinitely many possible energies of the form \eqref{eq:HSspectrum} actually occur in the spectrum of the Haldane--Shastry spin chain. One observation is that the states with $\bar{k}_1 + l,\dots,\bar k_N+l$ related by a (Galilean) boost for some~$l$ all have the same Haldane--Shastry energy since $\sum_i \, (N-2\mspace{1mu}i+1) = 0$. This is consistent with the property \eqref{eq:wedge_boost} of the wedges, which shows that the wedges related by a boost only differ by the overall constant $\ev_\omega \, e_N(\vect{z})^l \neq 0$ [cf.~\eqref{eq:ev_elementary}]. Thus we may certainly restrict our attention to $\bar{\choice{k}}$ with nonnegative parts and $\bar{k}_N = 0$. However, further conditions on $\bar{\choice{k}}$ are needed to determine the energy spectrum; we will return to this in \textsection\ref{sec:connection}--\ref{sec:descendants}. Given the appropriate conditions one can verify that \emph{all} energies are of the form \eqref{eq:HSspectrum} by counting the eigenvalues taking into account their degeneracy, cf.\ the penultimate paragraph of \textsection\ref{sec:connection}.

The abelian symmetries of the Haldane--Shastry spin chain also arise from freezing. The (total) momentum operator~\eqref{eq:momentum_operator} is traded for the (lattice) shift operator $P_{N-1,N} \cdots P_{12}$, see Proposition~3.11 of \cite{LPS_22} for a proof in the $q$-deformed case. Higher abelian symmetries of the Haldane--Shastry spin chain can be extracted from those of the spin-Calogero--Sutherland model by freezing too, although this is a little more subtle~\cite{TH_95}. The nonabelian symmetries are readily obtained from freezing too, see \textsection\ref{sec:nonabelian}.

\subsubsection{Strong coupling vs classical limit} \label{sec:strong_coupling_vs_class}

To understand freezing in more detail consider the (ordinary, i.e.\ not effective) spin-Calogero--Sutherland hamiltonian~\eqref{eq:CS_spin}. Reinstate Planck's constant by setting $\beta = g/\hbar$ and multiplying $\widetilde{H}^\textsc{cs}_-$ by $\hbar^2$. Hence $\beta \to \infty$ can be interpreted either as a strong coupling limit ($g \to \infty$) or as a classical limit ($\hbar \to 0$). 

In the classical limit the momenta become $\hat{p}_j \equiv -\mathrm{i} \, \hbar \, \partial_{x_j} = \hbar \, z_j \,\partial_{z_j} \to p_j$, while in the potential energy $g\,(g - \hbar \, P_{ij}) \to g^2$ the distinction between bosons and fermions is lost. This yields (the identity operator on spins times) the classical-mechanical \emph{Calogero--Sutherland--Moser} model,
\begin{equation}
	H^\textsc{csm} \equiv \frac{1}{2}\sum_{j=1}^N p_j^2 + g^2 \sum_{i < j}^N \frac{1}{4 \sin^2[(x_i - x_j)/2]} 
	= \frac{1}{2}\sum_{j=1}^N p_j^2 + g^2 \sum_{i < j}^N \frac{z_i \, z_j}{z_{ij} \, z_{ji}}\, .
\end{equation}
The equations of motion read
\begin{equation} \label{eq:classical_eom}
	\begin{aligned}
	\dot{x}_i & = \{x_i, H^\textsc{csm} \} = \hphantom{+}\frac{\partial H^\textsc{csm}}{\partial p_i} = p_i  \, , \\
	\dot{p}_j & = \{p_j,  H^\textsc{csm} \} = -\frac{\partial  H^\textsc{csm}}{\partial x_j} = -g^2 \, \partial_{x_j} \! \sum_{i (\neq j)}^N \frac{1}{4 \sin^2[(x_i - x_j)/2]} \, .
	\end{aligned}
\end{equation}
The classical equilibrium positions, where these Poisson brackets vanish, are $p_j = 0$ and $x_j = j/N$ up to permutations and a common shift of the positions. This leads us to the evaluation~\eqref{eq:ev}. 

The energy at classical equilibrium only receives a contribution from the potential energy. This is also the dominant (most divergent) term for $g \to \infty$. To find it we calculate
\begin{equation} \label{eq:cont_int_1}
	\begin{aligned}
	\frac{1}{2} \, \ev_\omega\sum_{i\neq j}^N \frac{z_i \, z_j}{z_{ij} \, z_{ji}} & = \frac{N}{2}  \sum_{n=1}^{N-1}\frac{1}{(1-\omega^n)(1-\omega^{-n})} \\
	& = \frac{N}{2} \oint_{\bar{C}_1} \! \frac{\mathrm{d}y}{2\pi\mathrm{i}} \; \frac{1}{(1-y)(1-y^{-1})} \, \frac{\mathrm{d}}{\mathrm{d}y}\log \frac{y^N-1}{y-1}\;,
	\end{aligned}
\end{equation}
where $\bar{C}_1$ is a counter-clockwise contour encircling all the poles at the roots of unity $y=\omega^n$ except $y=1$. By deforming the integration contour and taking into account that the integral over the contour at infinity vanishes, we compute 
\begin{equation} \label{eq:cont_int_2}
	\begin{aligned}
		\oint_{\bar{C}_1} \! \frac{\mathrm{d}y}{2\pi\mathrm{i}}
		\;
		\frac{1}{(1-y)(1-y^{-1})} \, \frac{\mathrm{d}}{\mathrm{d}y} \log\frac{y^N-1}{y-1}
		& = -\!\oint_{C_1} \! \frac{\mathrm{d}y}{2\pi\mathrm{i}}
		\; 
		\frac{1}{(1-y)(1-y^{-1})} \, \frac{\mathrm{d}}{\mathrm{d}y} \log \frac{y^N-1}{y-1} \\
		& = \res\limits_{y=1} \, \frac{1}{(1-y^{-1})(y^N-1)} \, \frac{\mathrm{d}}{\mathrm{d}y}\frac{y^N-1}{y-1} \\ 
		& = \frac{N^2-1}{12}\;,
	\end{aligned}
\end{equation}
where $C_1$ is a counter-clockwise contour encircling $y=1$.
Hence the classical equilibrium energy is given by
\begin{equation} \label{eq:class_equilibrium_energy}
	g^2 \, \ev_\omega \sum_{i < j}^N  \frac{z_i \, z_j}{z_{ij} \, z_{ji}} = \frac{g^2}{2} \, E^0 \, ,
\end{equation}
where $E^0$ was defined in \eqref{eq:E0}. This value coincides (up to a factor of $\hbar^2$) with the contribution proportional to $\beta^2$ in the quantum-mechanical energy, cf.\ \eqref{eq:E_cs_eff}--\eqref{eq:E_cs}.

Now expand the spin-Calogero--Sutherland hamiltonian one order further in $\hbar$:
\begin{equation} \label{eq:semiclassical}
	\begin{aligned} 
	\hbar^2 \, \widetilde{H}^\textsc{cs}_- & = \frac{1}{2}\sum_{j=1}^N \, (\hbar\,z_j \, \partial_{z_j})^2 + g \sum_{i < j}^N \frac{z_i \, z_j}{z_{ij} \, z_{ji}} \, (g + \hbar \, P_{ij}) \\
	& = H^\textsc{csm} \, \mathbbm{1} + \hbar \, g \sum_{i < j}^N  \frac{z_i \, z_j}{z_{ij} \, z_{ji}} \, P_{ij} + O(\hbar^2) \, . 
	\end{aligned}
\end{equation}
At classical equilibrium the \emph{semiclassical limit}, i.e.\ the coefficient of $\hbar$, becomes ($g$ times) the Haldane--Shastry hamiltonian~\eqref{eq:HS_AFM} up to a shift by the constant~\eqref{eq:class_equilibrium_energy}.
We conclude that the Haldane--Shastry spin chain can be obtained both from the original or from the effective spin-Calogero--Sutherland hamiltonian by freezing.\,%
\footnote{\ A similar result was recently established for a (not quite directly related) long-range spin chain with elliptic interactions~\cite{MZ_23}.}
The physical picture is that the freezing limit is dominated by the potential energy, so that the particles slow down and come to a halt at their equally spaced classical equilibrium positions to reduce to the Haldane--Shastry spin chain.

\subsection{Haldane--Shastry eigenvectors}

As we have just seen, the Haldane--Shastry hamiltonian and its eigenvalues can be obtained by freezing. The same holds for the abelian symmetries (with a little more effort) and nonabelian symmetries (see \textsection\ref{sec:nonabelian}). Unfortunately it is not as simple to derive the Haldane--Shastry \emph{eigenvectors} directly from those of the spin-Calogero--Sutherland model by freezing. There are several obstacles.
First, one needs to compute limit $\beta\to \infty$ of non-symmetric Jack polynomials, or rather their partial symmetrisation~\eqref{eq:partial_symm}. Second, for many states the result vanishes upon evaluation. Third, the wave functions of the surviving states require further work. The reason for this is that evaluation allows one to rewrite partially symmetric polynomials from $\mathbb{C}[\vect{z}]^{S_M \times S_{N-M}}$ as symmetric polynomials in only $M$ variables; physically the latter will correspond to the wave functions for the excitations~$\downarrow$. Indeed, consider the evaluation of the power sums $p_n(\vect{z}) \equiv \sum_i z_i^n$,
\begin{equation} \label{eq:p_r_ev}
	\ev_\omega \, p_n(\vect{z}) = N \, \delta_{N \mid n} \, , \qquad 
	\delta_{N \mid n} \equiv
	\begin{cases} 
		1 & n \in N \, \mathbb{Z}_{\geqslant 0} \, , \\ 
		0 & \text{else} \, . 
	\end{cases}
\end{equation}
Hence $p_n(z_{M+1},\dots,z_N) = p_n(\vect{z}) - p_n(z_1,\dots,z_M)$ coincides with $N \, \delta_{N \mid n} - p_n(z_1,\dots,z_M)$ upon evaluation. Moreover, any Laurent polynomial can be written as a polynomials times some (negative) power of $z_1 \cdots z_N$, yielding a nonzero prefactor upon evaluation. In principle, any element in $\mathbb{C}[\vect{z},\vect{z}^{-1}]^{S_M \times S_{N-M}}$ can be reduced to a polynomial in $\mathbb{C}[z_1,\dots,z_M]^{S_M}$ in this way. Carrying out this procedure in practice, however, is not straightforward in general.

The exception is formed by the simple $M=0$ eigenstates $\Delta(\vect{z}) \, P_{\choice{\lambda}}(\vect{z}) \, \ket{\uparrow\cdots\uparrow}$ and their spin-flipped counterparts at $M=N$. Here we only need to recall that $P_{\choice{\lambda}}(\vect{z}) \to e_{\choice{\lambda}'}(\vect{z}) \equiv e_{\lambda'_1}(\vect{z}) \, e_{\lambda'_2}(\vect{z}) \cdots$ when $\beta\to \infty$, and note that Newton's identities \eqref{eq:p_r_ev} imply
\begin{equation} \label{eq:ev_elementary}
	\ev_\omega \ e_n(\vect{z})=\delta_{n,0}+(-1)^{N-1}\delta_{n,N} \, .
\end{equation}
Thus the only $M=0$ states that survive evaluation correspond to partitions $\choice{\lambda} = (l^N)$ for some $l = \ell(\choice{\lambda}') \in \{ 0, \dots, N-1\}$. By \eqref{eq:sym_jack_stability} their wave function $P_{(l^N)}(\vect{z}) = e_N(\vect{z})^l$ is independent of $\beta$. Upon evaluation all these states reduce to $\ket{\uparrow\cdots\uparrow}$ up to a phase. 
That is, all states with $\bar{\choice{k}} = (l^N) + \choice{\delta_N}$ are, in the freezing limit, identified with the ferromagnetic state $\ket{\uparrow\cdots\uparrow}$. Their energy is
\begin{equation} \label{eq:HSF}
	E^\textsc{hs}_-(\choice{\delta_N}) =  \frac{1}{2}\sum_{i=1}^N  \, (N - 2\mspace{1mu}i +1)\,(N-i) = \frac{N(N^2-1)}{12} = E^0 \, ,
\end{equation}
where we used the boost invariance mentioned following \eqref{eq:HSspectrum} to set the last momentum to zero. This eigenvalue is consistent with the evaluation of the Haldane--Shastry hamiltonian~\eqref{eq:HS_AFM} at $P_{ij} = 1$ using \eqref{eq:class_equilibrium_energy} as appropriate for symmetric spin vectors.

For arbitrary $M$ the spin-Calogero--Sutherland eigenvectors are not as simple, and it is much less straightforward to follow them in the freezing limit. We do not know how to get the simple and explicit Haldane--Shastry wave functions directly by freezing. Luckily there are other approaches to derive this result.

\subsubsection{Frozen wedges} \label{sec:freezing_wedges}

Upon evaluation many single-particle basis vectors $u_k$ from \eqref{eq:wedge_single_particle} become the same because of \eqref{eq:wedge_boost}, and the wedges become linearly dependent. In order to evaluate $\widehat{u}_{\choice{k}}^{(M)}$ from \eqref{eq:wedgeM} we need to evaluate the Vandermonde determinants as well as the Schur polynomials. For the former we use that $\Delta(\vect{z})$ can be evaluated in two ways. On the one hand 
\begin{equation}
	\ev_\omega \, \Delta(\vect{z}) = (-1)^{(N+2)(N-1)/4} \, N^{N/2}
\end{equation}
is just a constant.%
\footnote{\ The value is found by noticing that $(-1)^{N(N-1)/2} \, \ev_\omega \, \Delta(\vect{z})^2 = \prod_{i=1}^N \omega^i \prod_{j(\neq i)}^N (1-\omega^{j-i}) = N^N \,\ev_\omega \,e_N(\vect{z}) = (-1)^{N-1}N^N$, where the middle equality uses \eqref{eq:prodroots}.}
On the other hand we can split the Vandermonde factor as
\begin{equation}
	\Delta(\vect{z}) = \Delta(\vect{z}_{I^\mathrm{c}}) \, \Delta(\vect{z}_I)^{-1} \prod_{\substack{i<j\\i\in I}}^N (z_i - z_j) \prod_{\substack{i<j \\ j \in I}}^N (z_i - z_j) \, ,
\end{equation}
where $\Delta(\vect{z}_I)^{-1}$ compensates for the spurious factors in the two products. The products can be explicitly evaluated. Indeed, since
\begin{equation} \label{eq:prodroots}
	\prod_{j=1}^{N-1}(1-\omega^j) = \lim_{u \to 1} \prod_{j=1}^{N-1}(u-\omega^j) = \lim_{u \to 1}\frac{u^N - 1}{u-1} = N
\end{equation}
we compute, setting $|I| \equiv \sum_{i \in I} i$,
\begin{equation}
	\begin{aligned}
		\prod_{\substack{i<j\\i\in I}}^N (\omega^i-\omega^j) \prod_{\substack{i<j\\j\in I}}^N (\omega^i-\omega^j) & = \prod_{\substack{i<j\\i\in I}}^N (\omega^i-\omega^j) \prod_{\substack{j<i\\i\in I}}^N (\omega^j-\omega^i) \\ 
		& = \prod_{i\in I} \biggl( (-1)^{i-1} \! \prod_{j (\neq i)}^N \!\! (\omega^i-\omega^j) \biggr) \\
		& = (-1)^{|I|-M} \prod_{i\in I} \biggl( \omega^{(N-1)i} \! \prod_{j (\neq i)}^N \! (1-\omega^{j-i}) \biggr) \\
		& = (-1)^{|I|-M} N^M \prod_{i\in I} \omega^{-i} \, .
	\end{aligned}
\end{equation}
Equating the two evaluations of the Vandermonde polynomial we obtain 
\begin{equation} \label{eq:delta_delta}
	\ev_\omega \, \Delta(\vect{z}_{I^\mathrm{c}}) = (-1)^{(N+2)(N-1)/4} \, N^{N/2 - M} \times (-1)^{|I|-M} \,\ev_\omega \, \Delta(\vect{z}_I) \, e_M(\vect{z}_I) \, ,
\end{equation}
where $e_M(\vect{z}_I) = \prod_{i\in I} z_i$.
Thus we may convert the factor $\Delta(\vect{z}_{I^\mathrm{c}})$ in \eqref{eq:wedgeM} into a constant times $(-1)^{|I|} \, \Delta(\vect{z}_I) \, e_M(\vect{z}_I)$ upon evaluation.
Regarding the Schur polynomials, by the evaluation \eqref{eq:p_r_ev} of power sums we have 
\begin{equation}
	\ev_\omega\, p_n(\vect{z}_I) = - \ev_\omega \, p_n(\vect{z}_{I^\mathrm{c}}) = \ev_\omega\, \Omega \,p_n(-\vect{z}_{I^\mathrm{c}}) \quad {\rm for} \ n<N \, ,
\end{equation}
where $\Omega \, p_n \equiv (-1)^{n-1} \, p_n$ is an involution that is known to transform the Schur polynomials according to $\Omega \, s_{\choice{\lambda}} =s_{\choice{\lambda}'}$, see \mbox{(2.13)} in \cite{Mac_95}. This implies that for partitions with $\lambda^{(\uparrow)}_1<N$ we have
\begin{equation} \label{eq:ev_schur}
	\ev_\omega \, s_{\choice{\lambda}^{(\uparrow)}}(\vect{z}_{I^\mathrm{c}}) = (-1)^{|\choice{\lambda}^{(\uparrow)}|} \; \ev_\omega \, s_{\choice{\lambda}^{(\uparrow)\mspace{2mu}\prime}}(\vect{z}_I)\, ,
\end{equation}
Upon evaluation the wedge~\eqref{eq:wedgeM} therefore becomes
\begin{equation} \label{eq:wedgepol}
	\ev_\omega \, \widehat{u}_{\choice{k}}^{(M)} = \mathrm{cst}_{N,M\mspace{-1mu},\mspace{1mu}\choice{\lambda}^{(\uparrow)}} \times \ev_\omega \!\!\! \sum_{\substack{I \subset \{1,\dots,N\} \\ \# I = M}} \!\!\!\! \Delta(\vect{z}_I)^2 \, e_M(\vect{z}_I) \, s_{\choice{\lambda}^{(\downarrow)}} (\vect{z}_{I}) \, s_{\choice{\lambda}^{(\uparrow)\mspace{2mu}\prime}}(\vect{z}_I) \, \cket{I} \, ,
\end{equation}
where the constant prefactor only depends on the partition~$\choice{\lambda}^{(\uparrow)}$ through the weight $|\choice{\lambda}^{(\uparrow)}|$, cf.~\eqref{eq:ev_schur}. The appearance of a symmetric polynomials in the $M$ variables associated to the coordinates of the magnons is in agreement with the discussion around \eqref{eq:p_r_ev}. We call \eqref{eq:wedgepol} a `frozen' wedge.

The product $s_{\choice{\lambda}^{(\downarrow)}} \, s_{\choice{\lambda}^{(\uparrow)\mspace{2mu}\prime}}$ in \eqref{eq:wedgepol} can be expanded in the Schur basis as $s_{\choice{\lambda}^{(\downarrow)} + \choice{\lambda}^{(\uparrow)\mspace{2mu}\prime}} + \mathrm{lower}$.
Therefore, to obtain all vectors after evaluation it suffices to take $\choice{\lambda}^{(\uparrow)} = \choice{0}$ empty, i.e.\ $\bar{\choice{k}}^{(\uparrow)} = \choice{\delta_{N-M}}$ by \eqref{eq:ktolambda}, and vary the partition $\choice{\lambda}^{(\downarrow)}$ associated to the magnons. Because of the evaluation we may furthermore take $\choice{\lambda}^{(\downarrow)}$ to be a partition with nonnegative parts.
Thus it suffices to consider the frozen wedges
\begin{equation} \label{eq:frozen_wedge}
	\begin{aligned}
		\ev_\omega \, \widehat{u}_{\choice{k}}^{(M)} & = \pm  \ev_\omega \; \widehat{u}_{\choice{k}^{(\downarrow)}} \wedge \widehat{u}_{\choice{\delta_{N-M}}} \\
		& = \mathrm{cst}_{N,M\mspace{-1mu},\mspace{1mu}\choice{\lambda}^{(\uparrow)} = \choice{0}} \times \ev_\omega \!\!\! \sum_{\substack{I \subset \{1,\dots,N\} \\ \# I = M}} \!\!\!\! \Delta(\vect{z}_I)^2 \, e_M(\vect{z}_I) \, s_{\choice{\lambda}^{(\downarrow)}} (\vect{z}_{I}) \, \cket{I} \, ,
	\end{aligned}
\end{equation}
which lie in the $M$-magnon sector of the spin-chain space $(\mathbb{C}^2)^{\otimes N}$. The wave function of \eqref{eq:frozen_wedge} is symmetric in the $i \in I$, matching the formal symmetry in magnon positions of the coordinate basis vectors~\eqref{eq:coord_basis}. It is completely determined by the symmetric polynomial
\begin{equation} \label{eq:pol_magnon}
	\begin{aligned}
		& \Delta(z_1,\dots,z_M)^2 \, e_M(z_1,\dots,z_M) \, s_{\choice{\lambda}^{(\downarrow)}}(z_1,\dots,z_M) && \qquad\mspace{2mu} \lambda^{(\downarrow)}_m = \bar{k}^{(\downarrow)}_m - M + m \, ,\\
		& \quad = \Delta(z_1,\dots,z_M)^2 \, s_{\choice{\nu}} (z_1,\dots,z_M) && \qquad\ \mspace{3mu} \nu_m \equiv \lambda^{(\downarrow)}_m + 1 \, , \\
		& \quad = s_{\choice{\mu}^+}(z_1,\dots,z_M) + \text{lower} && \mu_{M-m+1} \equiv \nu_m + 2\,(M-m) \, .
	\end{aligned}
\end{equation}
Here the we used the stability property~\eqref{eq:sym_jack_stability} for Schur polynomials to absorb the factor of $e_M$ in a shift of the partition, and $\choice{\mu}^+$ is the exponent of the highest monomial (symmetric polynomial). We have chosen $\choice{\mu}$ \emph{in}creasing (so not a partition) to match the literature. As we will see in \eqref{eq:hwcondk} below, it is closely related to the so-called `motifs' labelling the Haldane--Shastry eigenspaces. See Table~\ref{tb:motifs} (p.\,\pageref{tb:motifs}) for an example of \eqref{eq:pol_magnon}. In conclusion, (a suitable subset of) the frozen wedges \eqref{eq:frozen_wedge}--\eqref{eq:pol_magnon} span the whole spin-chain space, where we may take $\choice{\lambda}^{(\downarrow)}_M = 0$ by boost invariance.\,%
\footnote{\ Explicit examples suggests that in fact it suffices to restrict to partitions with $\bar{k}^{(\downarrow)}_1 < \bar{k}^{(\uparrow)}_1 = N - M -1$. This rule helps bounding the possible values of the energy~\eqref{eq:HSspectrum} that can occur for the Haldane--Shastry spin chain. We do not have a proof of this empirical observation.}

The Vandermonde factor in \eqref{eq:frozen_wedge} ensures that the wave function vanishes when two magnons are on top of each other, which fits with the fact that we cannot have two $\downarrow$s at same position. On the other hand, for that very reason the wave function at coinciding arguments is not physical, so these values may be chosen at will. The standard approaches to finding the Haldane--Shastry wave functions\,---\,see \cite{Hal_91a} (cf.~\textsection\ref{sec:lagrange}), \cite{BG+_93}\,---\,use the ansatz that the wave function vanishes at coinciding arguments, so that it is divisible by $\Delta(z_1,\dots,z_M)$. Symmetrisation then gives a second factor of $\Delta(z_1,\dots,z_M)$. The factor of $e_M(z_1,\dots,z_M)$ in \eqref{eq:pol_magnon} naturally leads to a (Yangian) highest-weight condition; we will return to this at the end of \textsection\ref{sec:connection}. As we have just seen, from the fermionic viewpoint these factors all come out automatically.

\bigskip
\begin{table}[h]
	\centering 
	\begin{tabular}{c|c|cc|cc|cccc} \toprule
		$M$ & $\bar{\choice{k}}$ & $\bar{\choice{k}}^{(\uparrow)}$ & $\bar{\choice{k}}^{(\downarrow)}$ & $\choice{\lambda}^{(\downarrow)}$ & $\choice{\nu}$ & $\choice{\mu}$ & $\choice{\mu}^+$ & $(\choice{\mu}^+)'$ & $\bigl((\choice{\mu}^+)'\bigr)^-$ \\ \midrule
		$0$ & $(3,2,1,0)$ & $(3,2,1,0)$ & $0$ & $0$ & $0$ & $0$ & $0$ & $0$ & $(0,0,0,0)$ \\
		$1$ & $(2,2,1,0)$ & $(2,1,0)$ & $(2)$ & $(2)$ & $(3)$ & $(3)$ & $(3)$ & $(1,1,1)$ & $(0,1,1,1)$ \\
		& $(2,1,1,0)$ & $(2,1,0)$ & $(1)$ & $(1)$ & $(2)$ & $(2)$ & $(2)$ & $(1,1)$ & $(0,0,1,1)$ \\
		& $(2,1,0,0)$ & $(2,1,0)$ & $(0)$ & $(0)$ & $(1)$ & $(1)$ & $(1)$ & $(1)$ & $(0,0,0,1)$ \\
		2 & $(1,1,0,0)$ & $(1,0)$ & $(1,0)$ & $(0,0)$ & $(1,1)$ & $(1,3)$ & $(3,1)$ & $(2,1,1)$ & $(0,1,1,2)$ \\
		\bottomrule
		\hline
	\end{tabular}
	\caption{An example of the partitions and motifs defined in \eqref{eq:splitk}, \eqref{eq:pol_magnon} and \eqref{eq:motif_sl2} for $N=4$ and $\mathfrak{sl}_2$. Note that \eqref{eq:k_to_motif} is satisfied.}
	\label{tb:motifs}
\end{table}

\subsubsection{Connection to scalar Calogero--Sutherland models} \label{sec:connection}

The Haldane--Shastry hamiltonian \eqref{eq:HS_AFM} acts on the frozen $M$-magnon wedges~\eqref{eq:frozen_wedge} as
\begin{equation} \label{eq:HS_wedge}
	H^\textsc{hs}_- \ \ev_\omega \, \widehat{u}_{\choice{k}}^{(M)} = \ev_\omega \, \Bigl( \partial_\beta \big|_{\beta=0} \widetilde{H}_-^{\prime\mspace{2mu}\textsc{cs}} \ \widehat{u}_{\choice{k}}^{(M)} \Bigr) \, ,
\end{equation}
where, due to \eqref{eq:hamwedge} and \eqref{eq:HSspectrum}, the linearised spin-Calogero--Sutherland hamiltonian acts by
\begin{equation} \label{eq:CS_wegde_linearised}
	\partial_\beta \big|_{\beta=0} \widetilde{H}_-^{\prime\mspace{2mu}\textsc{cs}} \; \widehat{u}_{\choice{k}}^{(M)} = E^\textsc{hs}_-\bigl(\bar{\choice{k}}\bigr) \, \widehat{u}_{\choice{k}} + \sum_{i<j}^N \widetilde{h}_{ij}\,\widehat{u}_{\choice{k}} \, .
\end{equation}
The squeezing operators $\widetilde{h}_{ij}$ defined in \eqref{eq:offdiag} kill the $\uparrow$s with `fully packed' momenta
\begin{equation} \label{eq:offdiag_reducing} 
	\widehat{u}_{\choice{k}} = \pm  \widehat{u}_{\choice{k}^{(\downarrow)}} \wedge \widehat{u}_{2\,\choice{\delta_{N-M}}} : \qquad
	\widetilde{h}_{ij}\,\widehat{u}_{\choice{k}} = \pm \Bigl(\widetilde{h}_{ij} \, \widehat{u}_{\choice{k}^{(\downarrow)}}\Bigr) \! \wedge \widehat{u}_{2\,\choice{\delta_{N-M}}} \, .
\end{equation}
Hence the $\uparrow$s are spectators in \eqref{eq:CS_wegde_linearised}, effectively reducing the rank of the wedges by one, that is, from spin-1/2 to the scalar case. 

In full detail, the reduction to the scalar case goes as follows. Recall, cf.~\eqref{eq:wedge_M=0}, that the $M$-magnon part of the wedges have the simple form $\widehat{u}_{\choice{k}^{(\downarrow)}} = a_{\bar{\choice{k}}^{(\downarrow)}}(z_1,\dots,z_M) \, \ket{\downarrow \cdots \downarrow}$. Using this along with the definition~\eqref{eq:offdiag} of the squeezing operators we find
\begin{equation}
	\begin{aligned}
		\widetilde{h}_{mn} \, \widehat{u}_{\choice{k}^{(\downarrow)}} & = \!\!\! \sum_{p=1}^{\bar{k}^{(\downarrow)}_m -\bar{k}^{(\downarrow)}_n -1} \!\!\! \bigl( \bar{k}^{(\downarrow)}_m - \bar{k}^{(\downarrow)}_n - p \bigr) \, \widehat{u}_{\choice{k}^{(\downarrow)} - 2 \, p \, (\choice{\varepsilon_m} - \choice{\varepsilon_n})} \\
		& = \bigl( h_{mn} \, a_{\bar{\choice{k}}^{(\downarrow)}}(z_1,\dots,z_M) \bigr) \, \ket{\downarrow \cdots \downarrow} \, ,
	\end{aligned}
\end{equation}
where we defined a scalar version of the squeezing operator,
\begin{equation} \label{eq:offdiag_scalar}
	h_{mn} \, a_{\bar{\choice{k}}^{(\downarrow)}}(z_1,\dots,z_M) \equiv \sum_{p=1}^{\bar{k}^{(\downarrow)}_m -\bar{k}^{(\downarrow)}_n -1} \!\!\! \bigl( \bar{k}^{(\downarrow)}_m - \bar{k}^{(\downarrow)}_n - p \bigr) \, a_{\bar{\choice{k}}^{(\downarrow)} - p \, (\choice{\varepsilon_m} - \choice{\varepsilon_n})} \, ,
\end{equation} 
which acts on antisymmetric polynomials in $M$ variables.
To get back to the spin case~\eqref{eq:offdiag_reducing} we just multiply \eqref{eq:offdiag_scalar} by $\ket{\downarrow\cdots\downarrow} = \cket{1,\dots,M}$ and take the wedge product with $\widehat{u}_{\vect{k}^{(\uparrow)} = 2\,\choice{\delta_{N-M}}} = a_{\choice{\delta_{N-M}}}(z_1,\dots,z_{N-M}) \, \ket{\uparrow \cdots \uparrow} = \Delta(z_1,\dots,z_{N-M}) \, \ket{\uparrow \cdots \uparrow}$, keeping in mind to shift the coordinates associated tot the $\uparrow$s as $z_j \mapsto z_{j+M}$ when taking the wedge product. 
Just as the squeezing operator~\eqref{eq:offdiag_reducing} is captured by the scalar action~\eqref{eq:offdiag_scalar}, the linearised spin-Calogero--Sutherland hamiltonian can be reduced by a scalar operator $H^\text{red}$ that acts on $\Delta(z_1,\dots,z_M) \, \mathbb{C}[z_1,\dots,z_M]^{S_M}$. Namely,
\begin{equation}
	\partial_\beta \big|_{\beta=0} \,  \widetilde{H}_-^{\prime\mspace{2mu}\textsc{cs}} \ \, \widehat{u}_{\choice{k}^{(\downarrow)}} \wedge \widehat{u}_{2\,\choice{\delta_{N-M}}} = \Bigl( \bigl( H^\text{red} \; a_{\bar{\choice{k}}^{(\downarrow)}}(z_1,\dots,z_M) \bigr) \, \ket{\downarrow \cdots \downarrow} \Bigr) \wedge \widehat{u}_{2\,\choice{\delta_{N-M}}} \, ,
\end{equation}
where
\begin{equation}
	H^\text{red} \; a_{\bar{\choice{k}}^{(\downarrow)}}(z_1,\dots,z_M) \equiv E^\textsc{hs}_-\bigl(\bar{\choice{k}}\bigr) \, a_{\bar{\choice{k}}^{(\downarrow)}}(z_1,\dots,z_M) + \sum_{m<n}^M h_{mn} \, a_{\bar{\choice{k}}^{(\downarrow)}}(z_1,\dots,z_M) \, .
\end{equation}
Since the $a_{\bar{\choice{k}}^{(\downarrow)}}(z_1,\dots,z_M)$ form a basis for $\Delta(z_1,\dots,z_M) \, \mathbb{C}[z_1,\dots,z_M]^{S_M}$ we obtain
\begin{equation}
	H^\text{red} = E^\textsc{hs}_-\bigl(\bar{\choice{k}}\bigr) + \sum_{m<n}^M h_{mn} \, .
\end{equation}
This starts to look like the effective hamiltonian of the Calogero--Sutherland model for $N^\star \equiv M$ spinless fermions. To verify that this is indeed the case and identify the parameter~$\beta$ it remains to massage the diagonal term into a more convenient form.

At this point we need to make an assumption about $\bar{\choice{k}}$, following \cite{Hal_91a} (cf.~Appendix~\ref{sec:lagrange}), \cite{BG+_93}, and \textsection3.2.3 of \cite{LPS_22}. The evaluations allow us, without loss of generality, to take $\choice{\lambda}^{(\downarrow)}$ to be a partition with nonnegative parts. Thus $\lambda^{(\downarrow)}_M \geqslant 0$, i.e.\ $\nu_M \geqslant 1$. Since $\omega^N = 1$ it furthermore seems reasonable to ask for the maximum degree of the polynomial in each variable to be less than $N$. By symmetry of \eqref{eq:pol_magnon} we may consider its degree in $z_1$. As the prefactor has degree $2\,(M-1)$ we thus restrict our attention to $\choice{\nu}$ with largest part bounded by $\nu_1 \leqslant N-2M + 1$.\,%
\footnote{\ We stress that due to the evaluation one can always take either $\nu_M \geqslant 1$ (as we do, provided one allows for parts $\geqslant N$) or $\nu_1 \leqslant N-2M + 1$ (allowing for negative parts). Since both conditions at the same time in particular requires $M \leqslant N/2$, at the very least we are missing all descendants `beyond the equator'. We will look at descendants in some more detail in \textsection\ref{sec:descendants}. }
The conditions $\nu_1 \leqslant N-2M + 1$ and $\nu_M \geqslant 1$ imply that the  integers $\choice{\mu}$ defined in \eqref{eq:pol_magnon} obey $0<\mu_m<N$ and are strictly increasing and non consecutive, i.e.\ 
\begin{equation} \label{eq:motif_sl2}
	0<\mu_m<N \, , \qquad \mu_{m+1} > \mu_m + 1 \, .
\end{equation}
Such an $M$-tuple $\choice{\mu} = (\mu_1,\dots,\mu_M)$ is called a \emph{motif}, or more precisely an $\mathfrak{sl}_2$-\emph{motif}.
In terms of the partitions, $\bar{\choice{k}}^{(\uparrow)}$ and $\bar{\choice{k}}^{(\downarrow)}$ the conditions $\choice{\lambda}^{(\uparrow)} = \choice{0}$ and $\nu_1 \leqslant N-2M+1$ mean that 
\begin{equation} \label{eq:hwcondk}
	\bar{k}_j^{(\uparrow)} = N-M-j \, , \qquad \bar{k}_1^{(\downarrow)} \leqslant N-M-1 \, .
\end{equation}
Hence the momenta associated with $\uparrow$s are fixed to their minimal values $\bar{\choice{k}}^{(\uparrow)} = \choice{\delta_{N-M}}$, while the momenta $\bar{\choice{k}}^{(\downarrow)}$ associated with the magnons~$\downarrow$ are a subset of $\choice{\delta_{N-M}}$.
The full $N$-tuple of coordinate integers $\bar{\choice{k}}$ and the motif are related by
\begin{equation} \label{eq:k_to_motif}
	\bar{k}_j = N-j -\sum_{m=1}^M \! \theta(N -j - \mu_m) \, ,
	\qquad \theta(x) \equiv \begin{cases} 0 & x<0\, , \\ 1 & x\geqslant 0 \, , \end{cases}
\end{equation} 
i.e.\ $\bar{\choice{k}} + (M^N) = \choice{\delta_N} +  \bigl((\choice{\mu}^+)')^-$, where $\bigl((\choice{\mu}^+)')^-$ is the reverse of the conjugate $(\choice{\mu}^+)'$ to the partition $\choice{\mu}^+$ that is itself obtained from the motif $\choice{\mu}$ by reversal. That is, $\choice{\mu}$ records positions of duplicates in $\bar{\choice{k}}$, counted from the right (since $\bar{\choice{k}}$ decreases). See Table~\ref{tb:motifs} for an example. Since the Haldane--Shastry energy~\eqref{eq:HSspectrum} is linear we have
\begin{equation}
	E^0 = E^\textsc{hs}_-(\choice{\delta_N}) = E^\textsc{hs}_-\bigl(\bar{\choice{k}} + \bigl((\choice{\mu}^+)'\bigr)^- \bigr) = E^\textsc{hs}_-\bigl(\bar{\choice{k}}\bigr) + E^\textsc{hs}_-\bigl(\bigl((\choice{\mu}^+)'\bigr)^-\bigr) \, . 
\end{equation}
The last term can be rewritten as 
\begin{equation}
	\begin{aligned} 
	E^\textsc{hs}_-\bigl(\bigl((\choice{\mu}^+)'\bigr)^-\bigr)  = E^\textsc{hs}_-\bigl((\choice{\mu}^+)'\bigr) & = \frac{1}{2} \sum_{i=1}^N \, (\mu^+)'_i \, (N-2\,i+1) \\
	& = \frac{1}{2} \sum_{i=1}^N \sum_{j=1}^{(\mu^+)'_i} (N-2\,i+1) \\
	& = \frac{1}{2} \sum_{m=1}^M \sum_{i=1}^{\mu_m} \, (N-2\,i+1) = \frac{1}{2} \sum_{m=1}^M \mu_m \, (N-\mu_m) \, ,
	\end{aligned}
\end{equation}
where in the third equality we use the standard trick of switching from summing row by row to column by column in the boxes of the Young diagram. Thus we can rewrite the Haldane--Shastry energy as
\begin{equation} \label{eq:energymotifs}
	E^\textsc{hs}_-(\bar{\choice{k}}) = E^0 - E^\textsc{hs}(\vect{\mu}) \, , \qquad E^\textsc{hs}(\vect{\mu}) \equiv \frac{1}{2} \sum_{m=1}^M \mu_m \, (N-\mu_m)\, .
\end{equation}
The summand $\tfrac{1}{2}\,\mu_m\,(N-\mu_m)$ of the difference $E^\textsc{hs}(\vect{\mu})$ with respect to the ferromagnetic energy~\eqref{eq:HSF} is the well-known quadratic dispersion relation of magnons for the Haldane--Shastry spin chain, and its sum gives their (strictly additive) correction to the energy \eqref{eq:HSF} of the ferromagnetic vector $\ket{\uparrow\cdots \uparrow}$.

From \eqref{eq:pol_magnon} we have $\mu_m = \bar{k}_{M-m+1}^{(\downarrow)} + m$. Plugging this into \eqref{eq:energymotifs} we get
\begin{equation}
	\begin{aligned}
	E^\textsc{hs}_-(\bar{\choice{k}}) 
	& = \Biggl( \frac{1}{2} \sum_{m=1}^M \bigl(\bar{k}_m^{(\downarrow)}\bigr)^2 + \frac{1}{2} \sum_{m=1}^M \bigl(M-2 \, m + 1 \bigr) \, \bar{k}_m^{(\downarrow)} \Biggr)  - \frac{1}{2} \, (N-M-1) \bigl| \bar{\choice{k}}^{(\downarrow)} \bigr| \\
	& \hphantom{ = } \ \, + E^0 - \frac{1}{6} \, M \, (M+1) \, (3\,N - 2 \, M -1)  \\
	& =  \bigl[E^{\prime\mspace{2mu}\textsc{cs}}\bigl(\bar{\choice{k}}^{(\downarrow)}\bigr) \bigr]_{N^\star = M, \, \beta^\star = 1} - \frac{1}{2} \, (N-M-1) \, \bigl| \bar{\choice{k}}^{(\downarrow)} \bigr| + \mathrm{cst}_{N,M} \, ,
	\end{aligned}
\end{equation}
where we recognised the Calogero--Sutherland energy~\eqref{eq:E_cs_eff} for $N^\star = M$ particles with effective coupling $\beta^\star = 1$, shifted by a multiple of the (total) momentum and a constant. 

Putting everything together, the action~\eqref{eq:CS_wegde_linearised} of the linearised spin-Calogero--Sutherland hamiltonian reduces to the scalar operator
\begin{equation} \label{eq:reduced_ham}
	H^\text{red} = \bigl[ H^{\prime\mspace{2mu}\textsc{cs}}_- \bigr]_{N^\star = M, \, \beta^\star = 1} - \frac{1}{2} \, (N-M-1) \sum_{m=1}^M \! z_m \, \partial_{z_m} + \mathrm{cst}_{N,M}
\end{equation}
defined on the space $\Delta(z_1,\dots,z_M) \, \mathbb{C}[z_1,\dots,z_M]^{S_M}$ of antisymmetric polynomials in $M$ variables. Here we recognised the action of the effective Calogero--Sutherland hamiltonian for spinless bosons [see e.g.\ \eqref{hamwedge_r}--\eqref{offdiag_r} with $\rk=1$] with $N^\star=M$ particles and coupling $\beta^\star=1$, up to an abelian symmetry proportional to the total momentum $\sum_{m=1}^M z_m \, \partial_m$ and a constant.
From \textsection\ref{sec:spinless}, see \eqref{eq:fermionic_wave_fns}, we know that the eigenfunctions \eqref{eq:reduced_ham} are given by
\begin{equation}
	\Delta(z_1,\dots,z_M) \, [P_{\choice{\lambda}}(z_1,\dots,z_M)]_{\beta = \beta^\star + 1} = \Delta(z_1,\dots,z_M) \, P_{\choice{\lambda}}^\star(z_1,\dots,z_M) \, , \quad P_{\choice{\lambda}}^\star \equiv P_{\choice{\lambda}}\big|_{\beta = 2} \, ,
\end{equation}
where we simplified $\choice{\lambda} \equiv \choice{\lambda}^{(\downarrow)}$.
The final step is to evaluate as in \eqref{eq:HS_wedge}, taking into account the contribution coming from $\Delta(z_{M+1},\dots,z_N)$ following \textsection\ref{sec:freezing_wedges}. The conclusion is that the Haldane--Shastry spin chain has wave functions $\Psi_{\choice{\mu}}^\textsc{hs}(i_1,\dots,i_M) = \ev_\omega \,\widetilde{\Psi}_{\choice{\mu}}^\textsc{hs}(z_{i_1},\dots,z_{i_M})$ with
\begin{equation} \label{eq:HS_wave_fn_result}
	\begin{aligned} 
	\widetilde{\Psi}^\textsc{hs}_{\choice{\mu}}(z_1,\dots,z_M) & = \Delta(z_1,\dots,z_M)^2 \, e_M(z_1,\dots,z_M) \, P_{\choice{\lambda}}^\star(z_1,\dots,z_M) \\
	& = \Delta(z_1,\dots,z_M)^2 \, P_{\choice{\nu}}^\star(z_1,\dots,z_M) \\
	& = m_{\choice{\mu}^+} + \text{lower} \, ,
	\end{aligned}
\end{equation}
which is of the form \eqref{eq:pol_magnon} as it should be. The corresponding motif is
\begin{equation}
	\mu_m = \lambda_{M-m+1} + 2\,m-1 = \nu_{M-m+1} + 2\,(m-1) \, ,
\end{equation}
in terms of which the energy is given by~\eqref{eq:energymotifs}. The motif conditions~\eqref{eq:motif_sl2} ensure that the degree of \eqref{eq:HS_wave_fn_result} in each variable is $<N$. Direct computation shows that the (lattice) momentum is $\tfrac{2\pi}{N}\,|\choice{\mu}|$. 
Due to the explicit factor of $e_M$, the polynomial~\eqref{eq:HS_wave_fn_result} vanishes when any argument is set to zero,
\begin{equation} \label{eq:hwcondition}
	\widetilde{\Psi}^\textsc{hs}_{\choice{\mu}}(z_1,\dots,z_M)\big|_{z_m=0} = 0 \, , \qquad 1\leqslant m \leqslant M \, .
\end{equation}
This is the Yangian highest-weight condition \cite{BPS_95a}, see \cite{LPS_22} for a proof in the $q$-deformed bosonic case. For each motif it is known to which Yangian irrep the highest-weight vector that we have constructed belongs~\cite{BG+_93}. In particular the dimension of the eigenspace labelled by $\choice{\mu}$ is known to equal
\begin{equation} \label{eq:degeneracy}
	\begin{cases} 
		N+1 & \text{if $\choice{\mu}$ is empty} \, , \\
		\displaystyle 
		(N-\mu_1) \, \mu_M \! \smash{\prod_{m=1}^{M-1}} (\mu_m - \mu_{m+1} - 1) \, , \quad & \text{else} \, .
	\end{cases}  
\end{equation}
A simple counting argument shows that we have obtained the complete spectrum; see Section~1.2.2 of \cite{LPS_22} for the full proof in the $q$-case, which has the exactly same combinatorics.%
\footnote{\ The $q$-deformed setting allows one to take the (crystal) limit $q \to 0,\infty$, where the spin chain simplifies drastically and the combinatorics is particularly simple (Section~1.2.5 of \cite{LPS_22}).}

Although the final result~\eqref{eq:HS_wave_fn_result} is the wave function \eqref{eq:bosonic_wave_fns} of the Calogero--Sutherland model for $N^\star = M$ spinless \emph{bosons} with coupling parameter $\beta = \beta^\star + 1 = 2$, we stress that it was obtained by passing through the fermionic model. The physical content of these results is that the magnons of the Haldane--Shastry spin chain behave like the particles of a discretised version of the scalar bosonic Calogero--Sutherland model with (bosonic) coupling parameter $\beta = \beta^\star + 1 = 2$. We emphasise that the value $\beta^\star = 1$ of the fermionic coupling parameter just rolls out of the computation.

\subsubsection{Comments about descendants} \label{sec:descendants}

In the preceding discussion we had to assume $\nu_1 \leqslant N-2M + 1$ (i.e.\ $\mu_M < N$) to make contact with the scalar Calogero--Sutherland model. The resulting eigenvectors have Yangian highest weight. Let us now investigate what happens for descendants. 

We start a wedge with $\bar{\choice{k}}^{(\uparrow)} = \choice{\delta_{N-M}}$ (i.e.\ $\choice{\lambda}^{(\uparrow)} = \choice{0}$) and $\bar{\choice{k}}^{(\downarrow)} = \choice{\lambda}^{(\downarrow)} + \choice{\delta_M}$ with parts contained in $\choice{\delta_{N-M}}$ as in \textsection\ref{sec:freezing_wedges}. Now we flip one spin from $\uparrow$ to $\downarrow$ without changing the coordinate part $\bar{\choice{k}}$. This amounts to deleting $\bar{k}_j^{(\uparrow),\,\mathrm{old}} = N-M-j$ for some $j\leqslant N-M$ such that $N-M-j$ is not among $\bar{\choice{k}}^{(\downarrow),\,\mathrm{old}}$ and introducing $\bar{k}_n^{(\downarrow),\,\mathrm{new}} \equiv N-M-j$ with some $n \leqslant M+1$ such that $\bar{\choice{k}}^{(\downarrow),\,\mathrm{new}}$ is sorted decreasingly. After freezing the leading wedge will have an expansion similar to \eqref{eq:wedgepol} with $\choice{\lambda}^{(\uparrow),\,\mathrm{old}}$ and $\choice{\lambda}^{(\downarrow),\,\mathrm{old}}$ replaced by $\choice{\lambda}^{(\uparrow),\,\mathrm{new}}$ and $\choice{\lambda}^{(\downarrow),\,\mathrm{new}}$, which are related to $\bar{\choice{k}}^{(\uparrow),\,\mathrm{new}}$ and $\bar{\choice{k}}^{(\downarrow),\,\mathrm{new}}$ as in \eqref{eq:ktolambda} with $M$ replaced by $M+1$. Explicitly we have $\choice{\lambda}^{(\uparrow),\,\mathrm{new}} = (1^{j-1})$, i.e.\ $\bar{\choice{k}}^{(\uparrow),\,\mathrm{new}} = \choice{\delta_{N-M-1}} + (1^{j-1})$, while the partition $\choice{\lambda}^{(\downarrow),\,\mathrm{new}}$ gets a part $\lambda^{(\downarrow)}_m$ inserted:
\begin{equation}
	\lambda^{(\downarrow),\,\mathrm{new}}_m = 
	\begin{cases}
		\lambda^{(\downarrow),\,\mathrm{old}}_m -1\, , & m<n \, , \\
		N-2M-1-j+m \, , \quad & m=n \, , \\
		\lambda^{(\downarrow),\,\mathrm{old}}_{m-1} \, , \quad & m> n \, .
	\end{cases}
\end{equation}

According to \eqref{eq:pol_magnon} the maximal degree in each variable of the associated polynomial \eqref{eq:wedgepol} is given by
\begin{equation}
	d_\mathrm{max} \equiv 2\,M + 1 + (\lambda^{(\uparrow),\,\mathrm{new}})'_1 + \lambda^{(\downarrow),\,\mathrm{new}}_1 \, .
\end{equation} 
If $n=1$ then $d_\mathrm{max} = 2M+1+(j-1)+(N-2M-j) = N$. If instead $n>1$ then $\bar{k}_j^{(\uparrow),\,\mathrm{old}} = \bar{k}_m^{(\downarrow),\,\mathrm{new}} < \bar{k}_1^{(\downarrow),\,\mathrm{new}} = \bar{k}_1^{(\downarrow),\,\mathrm{old}}$ since $m>1$ and the $\bar{k}$s are strictly decreasing, so $N-M-j < \lambda^{(\downarrow)}_1+M-1$, which leads to $d_\mathrm{max} = 2M+1+(j-1)+ \lambda^{(\downarrow)}_1 -1 > N$.

Thus, for descendants the polynomial~\eqref{eq:pol_magnon} is naturally of degree $\geqslant N$. To include these descendants the framework of \textsection\ref{sec:connection} would have to be modified. In addition, we have not been able to find a direct proof that the highest-weight condition~\eqref{eq:hwcondition} is violated for the descendants constructed above. Although we find this situation somewhat unsatisfactory, in view of the counting mentioned following \eqref{eq:hwcondition}, the Yangian highest-weight vectors derived in \textsection\ref{sec:connection} suffice to diagonalise the spin-1/2 Haldane--Shastry model.

\section{Higher rank} \label{sec:higer_rank}

In this section we treat the generalisation of the discussion about the spin-Calogero--Sutherland model and Haldane--Shastry spin chain in \textsection\ref{sec:rank_one}--\ref{sec:HS} to higher rank, with spins taking values in the (i.e.\ first fundamental, or vector) representation $V_{\,\Box} = \mathbb{C}^r$ of $\mathfrak{gl}_\rk$.
We will say that the basis vector $\ket{a}$ of $\mathbb{C}^r$ has `colour'~$a$; one could also interpret it as a `flavour' or particle `species'. For $\rk=2$ we retrieve the spin-1/2 setting of \textsection\ref{sec:rank_one} with $0 = \;\uparrow$ and $1=\;\downarrow$. 
The Hilbert space $(\mathbb{C}^\rk)^{\otimes N}$ for $N$ such spins has weight-space decomposition
\begin{equation} \label{eq:weight_decomp_higher_rk}
	(\mathbb{C}^\rk)^{\otimes N} = \bigoplus_{\vect{M}} \: (\mathbb{C}^\rk)^{\otimes N}[\vect{M}] \, ,
\end{equation}
where the sum runs over all $\rk$-tuples $\vect{M} = (M_0,\dots,M_{\rk-1})$ of nonnegative integers adding up to~$N$. The weight space $(\mathbb{C}^\rk)^{\otimes N}[\vect{M}]$ is spanned by all vectors $\ket{a_1 \dots a_N}$ with $M_a$ colours $a_j$ equal to $a$ for all $0\leqslant a<\rk$. Vectors with highest weight for $\mathfrak{gl}_\rk$ occur when $\vect{M} = (M_0\geqslant M_1\geqslant \ldots \geqslant M_{\rk-1})$ is a partition. (Of course the converse is not true: all weight spaces also contain descendants, except for the one-dimensional weight space with $\vect{M} = (N,0,\dots,0)$.)

\subsection{Fermionic spin-Calogero--Sutherland model}

Consider $N$ particles with coordinates $z_j$ as well as spins in $\mathbb{C}^\rk$. The \emph{fermionic space} is
\begin{equation} \label{eq:fermionic_space}
	\widetilde{\mathcal{F}}_{\rk} \equiv \Bigl\{ \ket{\widetilde{\Psi}} \in (\mathbb{C}^\rk)^{\otimes N} \otimes \mathbb{C}[\vect{z}, \vect{z}^{-1}]	: P_{ij} \, s_{ij}\, \ket{\widetilde{\Psi}} = 	-\ket{\widetilde{\Psi}} \Bigr\} \, .
\end{equation} 
so $\widetilde{\mathcal{F}}_1 \cong \Delta(\vect{z}) \, \mathbb{C}[\vect{z},\vect{z}^{-1}]$ gives the spinless fermions from \textsection\ref{sec:spinless}, while $\widetilde{\mathcal{F}}_2 = \widetilde{\mathcal{F}}$ is the fermionic space that we studied in \textsection\ref{sec:coord_basis}. From the spin space~\eqref{eq:weight_decomp_higher_rk}, the fermionic space~\eqref{eq:fermionic_space} inherits a weight decomposition determined by the spins,
\begin{equation}
	\widetilde{\mathcal{F}}_\rk = \bigoplus_{\vect{M}} \, \widetilde{\mathcal{F}}_\rk[\vect{M}] \, .
\end{equation}
For spin~1/2 we retrieve $\widetilde{\mathcal{F}}_2[N-M,M] = \widetilde{\mathcal{F}}[N-2M]$ from \eqref{eq:magnon_sectors}, where we identify the $\mathfrak{gl}_2$ weight $\vect{M} = (N-M,M)$ with the $\mathfrak{sl}_2$ weight $M_0 - M_1 = N-2M$.

\subsubsection{Coordinate basis}

Like in \textsection\ref{sec:coord_basis}, the vectors in each weight subspace~$\widetilde{\mathcal{F}}_\rk[\vect{M}]$ have an explicit form with respect to the coordinate basis. The generalisation of \eqref{eq:physvec} is a matter of bookkeeping, as follows. Consider the (Young) subgroup $S_{\vect{M}^-} \equiv S_{M_{\rk-1}} \times \cdots \times S_{M_0} \subset S_N$. The reversal of the parts of $\vect{M}$ ensures our notation is consistent with that of \textsection\ref{sec:coord_basis}: for $\rk=2$ with $\vect{M} = (N-M,M)$ we have $S_{\vect{M}^-} = S_M \times S_{N-M}$. Define the associated partial Vandermonde polynomial as
\begin{equation} \label{multiple_delta}
	\begin{aligned}
	\Delta_{\vect{M}}(\vect{z}) & \equiv \prod_{a=0}^{r-1} \! \Delta(z_{M_0 + \cdots + M_{a-1} + 1},\dots,z_{M_0 + \cdots + M_a}) \\
	& = \Delta(z_1,\dots,z_{M_{\rk-1}}) \, \Delta(z_{M_{\rk-1} + 1},\dots,z_{M_{\rk-1}+ M_{\rk-2}}) \cdots \Delta(z_{N - M_0 +1},\dots z_N) \, .
	\end{aligned}
\end{equation} 
Following the same steps as for $\rk =2$ one shows that the weight-$\vect{M}$ subspace is determined by partially (anti)symmetric polynomials via
\begin{equation}
	\begin{aligned}
	& \mathbb{C}[\vect{z}, \vect{z}^{-1}]^{S_{\vect{M}^-}} \!\!\!\!\! && \xrightarrow{\ \sim\ } \Delta_{\vect{M}}(\vect{z}) \, \mathbb{C}[\vect{z}, \vect{z}^{-1}]^{S_{\vect{M}^-}} \!\!\!\!\! && \xrightarrow{\ \sim\ } \widetilde{\mathcal{F}}_{\rk}[\vect{M}] \\
	& \ \ \, \widetilde{\Psi}(\vect{z}) && \xmapsto{\ \hphantom{\sim} \ } \ \ \, \Delta_{\vect{M}}(\vect{z}) \, \widetilde{\Psi}(\vect{z}) && \xmapsto{\ \hphantom{\sim} \ } \ \; \ket{\widetilde{\Psi}} \, .
	\end{aligned}
\end{equation}
Here
\begin{equation} \label{eq:physvec_higher_rk}
	\ket{\widetilde{\Psi}} = \!\! \sum_{\sigma \in S_N/S_{\vect{M}^-}} \mspace{-15mu} \mathrm{sgn}(\sigma) \, \Delta_{\vect{M}}(z_{\sigma(1)},\dots,z_{\sigma(N)}) \, \widetilde{\Psi}(z_{\sigma(1)},\dots,z_{\sigma(N)}) \, \cket{\sigma} \, ,
\end{equation}
and $\cket{\sigma} \equiv P_\sigma \, \ket{(\rk-1)^{M_{r-1}},\dots,0^{M_0}}$ denotes the vector with $M_{r-1}$ components $r-1$ at sites $\sigma(1),\dots,\sigma(M_{r-1})$, $M_{r-2}$ components $\rk-2$ at sites $\sigma(M_{r-1} + 1),\dots,\sigma(M_{r-1} + M_{r-2})$, etc.
For $\rk=2$ we recover \eqref{eq:physvec} by identifying the permutation $\sigma \in S_N/(S_M \times S_{N-M})$ with the subset $I=\{\sigma(1),\dots,\sigma(M)\} \subset \{1,\dots,N\}$. 

Here are again some examples. If the weight is $\vect{M} = (N,0,\dots,0)$ then \eqref{eq:physvec_higher_rk} boils down to
\begin{equation}
	\Delta(\vect{z}) \, \widetilde{\Psi}(\vect{z}) \, \ket{0\cdots0} \, ,
\end{equation}
which is \eqref{eq:physvecM=0} for $\rk=2$. We thus find a copy of the scalar case, and for spin-Calogero--Sutherland eigenvectors $\widetilde{\Psi}(\vect{z}) = [P_{\choice{\lambda}}(\vect{z})]_{\beta \mapsto \beta +1}$ are \emph{symmetric} Jack polynomials defined by \eqref{eq:nonsymm_to_asymm_Jack}. Similarly, at weight $\vect{M} = (N-M,M,0,\dots,0)$ we find a copy of the $M$-magnon sector of the spin-1/2 case, and with eigenvectors coming from \emph{partially symmetric} Jack polynomials as in \eqref{eq:partial_symm}. As the weight $\vect{M}$ decreases we get less and less symmetric versions of Jack polynomials. If $\rk \geqslant N$ is sufficiently high then at (central) weight $\vect{M} = (1^N) = (1,\dots,1)$ we finally get a copy of the nonsymmetric theory:  \eqref{eq:physvec_higher_rk} gives
\begin{equation}
	\ket{\widetilde{\Psi}} = \sum_{\sigma \in S_N} \! \mathrm{sgn}(\sigma) \, \widetilde{\Psi}(z_{\sigma(1)},\dots,z_{\sigma(N)}) \, \cket{\sigma} \, ,
\end{equation}
and for eigenvectors $\widetilde{\Psi}(\vect{z}) = E_{\choice{\lambda}}(\vect{z})$ is a \emph{nonsymmetric} Jack polynomial. That is, for $r\geqslant N$ the spin-Calogero--Sutherland model features symmetric, intermediate, and nonsymmetric Jack polynomials at the same time!

\subsubsection{Wedge basis}

The wedge basis of $\widetilde{\mathcal{F}}_{\rk}$ is obtained as a generalisation of \textsection\ref{sec:wedges}.  For each particle define the vectors\,%
\footnote{\ Note that our conventions differ from Uglov~\cite{Ugl_98} in that for us $k$ \emph{in}creases with $\bar{k}$.}
\begin{equation}
	u_k=z^{\bar{k}} \, \ket{\underbar{k}} \,, \quad k \in \mathbb{Z} \, , \quad \text{with $\bar{k},\underbar{k}$ now determined by} \quad k = \underbar{k}+\bar{k}\,\rk\,, \quad 0 \leqslant \underbar{k} \leqslant \rk-1 \, .
\end{equation}
Thus, \eqref{eq:wedge_diagram} generalises to
\begin{equation*}
	\begin{aligned}
		& \ \, \vdots && \qquad \ \, \vdots && && \qquad \quad \! \vdots \\[-.3\baselineskip]
		u_{2\rk} & = z^{2\hphantom{-}} \ket{0} \qquad && \mspace{-7mu} u_{2\rk+1} = z^{2\hphantom{-}} \ket{1} && \cdots && u_{3\rk-1} = z^{2\hphantom{-}} \ket{\rk-1}  \\
		u_{\rk} & = z^{\hphantom{-1}} \ket{0} && u_{\rk+1} = z^{\hphantom{-1}} \mspace{1mu} \ket{1} && \cdots && u_{2\rk-1} = z^{\hphantom{-1}} \ket{\rk-1} \\
		u_0 & = \hphantom{z^{-1}} \ket{0} && \mspace{18mu} u_1 = \hphantom{z^{-1}} \mspace{1mu} \ket{1} \qquad && \cdots \qquad && \mspace{7mu} u_{\rk-1} = \hphantom{z^{-1}} \ket{\rk-1} \\
		u_{-\rk} & = z^{-1} \ket{0} && u_{1-\rk} = z^{-1} \ket{1} && \cdots && \mspace{14mu} u_{-1} = z^{-1} \ket{\rk-1} \\[-.3\baselineskip]
		& \ \, \vdots && \qquad \ \, \vdots && && \qquad \quad \! \vdots
	\end{aligned}
\end{equation*}
Again, this is just a reordering of the natural basis of tensor product $\mathbb{C}^\rk \otimes \mathbb{C}[z]$.
From these vectors one builds a basis of $\widetilde{\mathcal{F}}_{\rk}$ to get wedges like in \eqref{eq:wedge},
\begin{equation} \label{asym_schur_gen}
	\begin{aligned}
		\widehat{u}_{\choice{k}} = u_{k_1}\wedge \cdots \wedge u_{k_N} & =  \sum_{\sigma \in S_N} \! \mathrm{sgn}(\sigma) \, u_1^{k_{\sigma(1)}} \otimes \cdots \otimes u_N^{k_{\sigma(N)}} \\
		& = \sum_{\sigma \in S_N} \! \mathrm{sgn}(\sigma) \, z_1^{\bar{k}_{\sigma(1)}} \cdots z_N^{\bar{k}_{\sigma(N)}} \ket{\underbar{k}_{\sigma(1)} \dots \underbar{k}_{\sigma(N)} } = \det_{i,j} \Bigl( z_i^{\bar{k}_j} \ket{\underbar{k}_j} \Bigr) \, .
	\end{aligned}
\end{equation}
For example, the ferromagnetic vacuum is
\begin{equation}
	\widehat{u}_{r\,\choice{\delta_N}} = u_{(N-1)\,r} \wedge \dots \wedge u_r \wedge u_0 = \Delta(\vect{z}) \, \ket{0\cdots 0} \, ,
\end{equation}
while for $\rk = N$ the antiferromagnetic vacuum equals
\begin{equation}
	\widehat{u}_{\choice{\delta_N}} = u_{N-1} \wedge \dots \wedge u_1 \wedge u_0 = \Pi^\text{sp}_- \, \ket{01\cdots N-1} \, .
\end{equation}

Consider a wedge in $\widetilde{\mathcal{F}}_{\rk}[\vect{M}]$, so that for each $0\leqslant a < \rk$ there are $M_a$ `spin integers' $\underbar{k}_j$ equal to $a$, with $\sum_a M_a = N$.
As in \eqref{eq:splitk} we group the `coordinate integers' $\bar{k}_j$ 
into $r$ strict `coordinate partitions'
\begin{equation} \label{splitkr}
	\bar{\choice{k}}^{(a)} = (\bar{k}_1^{(a)} > \ldots > \bar{k}_{M_a}^{(a)} ) \quad \text{such that the corresponding} \quad \underbar{k}_j = a \, , \qquad a=0,\ldots,\rk -1 \, .
\end{equation}
In this way, we split $\choice{k}$ into $r$ strict partitions
\begin{equation}
	\choice{k}^{(a)} \equiv \bigl( \rk \,\bar{k}_1^{(a)} + a, \ldots, \rk \,\bar{k}_{M_a}^{(a)} + a \bigr) \, , \qquad a=0,\ldots,\rk-1 
	\, .
\end{equation}
This generalises $\choice{k}^{(\uparrow)}$ and $\choice{k}^{(\downarrow)}$ of \eqref{eq:splitk} for the spin-1/2 case.
We denote by $\widehat{u}_{\choice{k}^{(a)}}$ the $M_a$-spin wedge corresponding to $\vect{k}^{(a)}$. It is a polynomial in $M_a$ variables times $\ket{a \dots a}$. By reordering some factors we bring the wedge product to the form
\begin{equation} \label{wedge_r}
	\widehat{u}_{\choice{k}}
	= \pm\, \widehat{u}_{\choice{k}^{(\rk-1)}}
	\wedge \dots \wedge \widehat{u}_{\choice{k}^{(0)}}
	\, .
\end{equation}
These wedges are decomposed in terms of Schur polynomials as
\begin{equation} \label{wedge_r_Schur}
	\begin{aligned} 
	\widehat{u}_{\choice{k}}
	= \pm \!\!\! \sum_{\sigma \in S_N/S_{\vect{M}^-}} \mspace{-15mu} \mathrm{sgn}(\sigma) \, & \Delta_{\vect{M}}(z_{\sigma(1)},\dots,z_{\sigma(N)}) \\
	\times {} & s_{\choice{\lambda}^{(r-1)}}(z_{\sigma(1)},\ldots,z_{\sigma(M_{r-1})}) \cdots s_{\choice{\lambda}^{(0)}}(z_{\sigma(N-M_0+1)},\ldots,z_{\sigma(N)})\, \cket{\sigma} \, ,
	\end{aligned}
\end{equation}
with $\Delta_{\vect{M}}$ from \eqref{multiple_delta}, and $\rk$ partitions $\choice{\lambda}^{(a)}$ defined by $\choice{\lambda}^{(a)} + \choice{\delta_{M_a}} \equiv \choice{k}^{(a)}$. This provides the link between wedges and the coordinate basis~\eqref{eq:physvec_higher_rk} for general rank.

On these vectors the Calogero--Sutherland hamiltonian acts as
\begin{equation} \label{hamwedge_r}
	\widetilde{H}^{\prime\mspace{2mu}\textsc{cs}}_- \, \widehat{u}_{\choice{k}}
	= E^{\prime\mspace{2mu}\textsc{cs}}(\bar{\choice{k}}) \, \widehat{u}_{\choice{k}}
	+ 2\,\beta \sum_{i<j}^N \widetilde{h}_{ij}\, \widehat{u}_{\choice{k}}
	\, .
\end{equation}
where the diagonal terms depend only on the coordinate part $\bar{\choice{k}}$ and are as in the scalar case, see \eqref{eq:E_cs_eff}, while the squeezing operators~\eqref{eq:offdiag} in the off-diagonal terms are generalised to
\begin{equation} \label{offdiag_r}
	\widetilde{h}_{ij} \, \widehat{u}_{\choice{k}}
	= \!\! \sum_{p=1}^{\bar{k}_i -\bar{k}_j-1} \!\!\! (\bar{k}_i-\bar{k}_j-p)
	\, \widehat{u}_{\choice{k} -\rk \, p \, (\choice{\varepsilon_i} - \choice{\varepsilon_j})} \, .
\end{equation}

\subsection{Haldane--Shastry spin chain}

\subsubsection{Frozen wedges}

Upon freezing we will restrict the maximum power $\bar k_1<N$. We will think of $\ket{0}$ as a hole, and $\ket{a}$ with $1\leqslant a < \rk$ as (different species of) particles.
Similarly to the case of $\mathfrak{gl}_2$, restricting to $\choice{\lambda}^{(0)} = \choice{0}$ corresponding to tightly-packed wedge momenta $\bar k_j^{(0)} = M_0 - j$ for the `colour' $0$, we reduce the analysis to $\mathfrak{gl}_\rk$ highest-weight states. To compute the evaluation $z_j \mapsto \omega^j$ we proceed as in \textsection\ref{sec:freezing_wedges}. We now have sets $I^{(1)},\dots,I^{(\rk-1)}$ for the positions of the different particle species, while $I^{(0)}$ plays the role of the set $I^\text{c}$ of empty sites. Using \eqref{eq:delta_delta} with $\vect{z}_{I^\text{c}}$ replaced by $(z_{N - M_0 +1},\dots, z_N)$ we find
\begin{equation} \label{wedge_r_Schur_freeze}
	\begin{aligned}
		\ev_\omega \,\widehat{u}_{\choice{k}}
		= \mathrm{cst}_{N,M_0\mspace{-1mu},\mspace{1mu}\choice{\lambda}^{(0)}} \times & \, \ev_\omega \!\! \sum_{\sigma \in S_N/S_{\vect{M}^-}} \!\!\!\!\!\!\! \mathrm{sgn}(\sigma) \, (-1)^{\sigma(1) + \dots + \sigma(M_0)} \\
		& \times \Delta(z_{\sigma(1)},\dots,z_{\sigma(M_{r-1})})^2 \cdots \Delta(z_{\sigma(N-M_0-M_1+1)},\dots,z_{\sigma(N - M_0)})^2 \\
		& \times s_{\choice{\lambda}^{(r-1)}}(z_{\sigma(1)},\dots,z_{\sigma(M_{r-1})}) \cdots s_{\choice{\lambda}^{(1)}}(z_{\sigma(N-M_0-M_1+1)},\dots,z_{\sigma(N - M_0)}) \\
		& \times e_{N-M_0}(z_{\sigma(1)},\dots,z_{\sigma(N-M_0)}) \, s_{(\choice{\lambda}^{(0)})'}(z_{\sigma(1)},\dots,z_{\sigma(N-M_0)}) \, \cket{\sigma} \, .
	\end{aligned}
\end{equation}
Like for spin~$1/2$ we may now take $\choice{\lambda}^{(0)} = \choice{0}$ to obtain the generalisation of \eqref{eq:frozen_wedge}. 
Prior to evaluation, the right-hand side of \eqref{wedge_r_Schur_freeze} is symmetric under simultaneous permutations of coordinates and spins. Indeed, permuting any two variables of particles with the same colour does not change anything because they are related by even powers  of Vandermonde determinants, while a permutation of variables from particle groups with different colours changes sign once because of the corresponding factor in  $\Delta_{M_1+\ldots +M_{r-1}}$ and once due to $\mathrm{sgn}(\sigma)$.

\subsubsection{Connection to rank $\rk-1$ Calogero--Sutherland models}

Notice that the discussion of freezing for the spin-1/2 Calogero--Sutherland hamiltonian is in fact independent of the rank, so it carries over verbatim to produce the $\mathfrak{gl}_\rk$-Haldane--Shastry hamiltonian.
The final step is to compute the action of this hamiltonian on the frozen wedges $\ev_\omega \,\widehat{u}_{\choice{k}}|_{\choice{\lambda}^{(0)} = \choice{0}}$ like in \textsection\ref{sec:connection}. 
The action of the antiferromagnetic Haldane--Shastry hamiltonian, again interpreted as the linear part in $\beta$ of the Calogero--Sutherland hamiltonian from \eqref{hamwedge_r}, yields
\begin{equation}
	\partial_\beta \big|_{\beta=0} \widetilde{H}_-^{\prime\mspace{2mu}\textsc{cs}} \ \widehat{u}_{\choice{k}^{(r-1)}} \wedge \dots \wedge \widehat{u}_{\choice{k}^{(1)}} \wedge \widehat{u}_{2\,\choice{\delta_{N-M_0}}} = \Bigl( \widetilde{H}^\text{red} \; \widehat{u}_{\choice{k}^{(r-1)}} \wedge \dots \wedge \widehat{u}_{\choice{k}^{(1)}} \Bigr) \wedge \widehat{u}_{2\,\choice{\delta_{N-M_0}}} \, ,
\end{equation}
where
\begin{equation}
	\widetilde{H}^\text{red} = E^\textsc{hs}_-\bigl(\bar{\choice{k}}\bigr) + \sum_{i<j}^{N-M_0} \bigl[\,\widetilde{h}_{ij}\bigr]_{r^\star = r - 1} \, .
\end{equation}
If we assume that the polynomials on which we act have degree $<N$ in each variable like in the text preceding \eqref{eq:motif_sl2} then we have
\begin{equation}
	E^\textsc{hs}_-\bigl(\bar{\choice{k}}\bigr) = E^0 - E^\textsc{hs}(\vect{\mu}) \, , \qquad E^\textsc{hs}(\vect{\mu}) \equiv \frac{1}{2} \sum_{m=1}^{N-M_0} \mu_m \, (N-\mu_m)\, ,
\end{equation}
where the integers
\begin{align}
	\mu_m \equiv \bar{k}_{N-M_0 - m+1} + m \, , \qquad m=1,\ldots,N-M_0 \, ,
\end{align} 
are \emph{$\mathfrak{gl}_r$-motifs} \cite{HH+_92}, obeying $1\leqslant \mu_m < N$, $\mu_m < \mu_{m+1}$ and $\mu_{m+\rk-1} > \mu_m + \rk-1$, where the last condition says that $\choice{\mu}$ does not contain any subsequence of $\rk$~successive integers.
In this way we obtain
\begin{equation} \label{eq:reduced_ham_r}
	H^\text{red} = \bigl[ H^{\prime\mspace{2mu}\textsc{cs}}_- \bigr]_{N^\star = N-M_0, \, \beta^\star = 1, \, \rk^\star = \rk-1} - \frac{1}{2} \, (M_0 - 1) \! \sum_{m=1}^{N-M_0} \! z_m \, \partial_{z_m} + \mathrm{cst}_{N,N-M_0}
\end{equation}
defined on the space 
\begin{equation}
	\Delta(z_1,\dots,z_{M_{\rk-1}}) \cdots \Delta(z_{N-M_0 - M_1 + 1},\dots,z_{N - M_0}) \, \mathbb{C}[z_1,\dots,z_{N-M_0}]^{S_{M_{\rk-1}} \times \cdots \times S_{M_1}} \, .
\end{equation}
In this way we relate the action of the $\mathfrak{gl}_\rk$ Haldane--Shastry spin chain to that of the Calogero--Sutherland hamiltonian with $\mathfrak{gl}_{\rk-1}$-spins, $N^\star= N - M_0$ particles, weight $\choice{M}^\star = (M_1,\dots,M_{r-1})$ a partition of $N^\star$, and (fermionic) coupling $\beta^\star = 1$ having the same value for any rank. 

This concludes our discussion of the generalisation of the material of \textsection\ref{sec:rank_one}--\ref{sec:HS} to the case of $\mathfrak{gl}_\rk$ spins for arbitrary $\rk$, proving a conjecture of \cite{BG+_93}.

\appendix

\section{Connection via Lagrange interpolation} \label{sec:lagrange}

In this appendix we show that the $N$-site Haldane--Shastry hamiltonian acts on $M$-magnon wave functions as a discretised version of the scalar bosonic Calogero--Sutherland hamiltonian with $M$ particles and (bosonic) coupling parameter $\beta^\star=2$. To this end we will use Lagrange interpolation and assume that the wave function (i) has a double zero at coinciding magnon positions, and (ii) is a polynomial of degree less than $N$ in each variable. This hypothesis is valid for Yangian highest-weight states; see \textsection\ref{sec:descendants} for a discussion about descendants.

\subsection{Lagrange interpolation}

Consider the space of polynomials in $w$ of degree at most $N-1$, which we denote by $\mathbb{C}[w]^{<N}$. One basis of this space is the monomial basis $w^n$. Another basis is given, for any choice of $N$ pairwise distinct points $z_i \in \mathbb{C}$, by
\begin{equation}
	\varphi_j(w) = \prod_{i(\neq j)}^N (w-z_i) \, , \qquad \varphi_j(z_i) = \delta_{ij} \prod_{k(\neq j)}^N (z_i - z_k) \, .
\end{equation}
Since only $\varphi_i$ is nonzero at $w=z_i$ these polynomials are algebraically independent, and their degree is $N-1$ so they lie in $\mathbb{C}[w]^{<N}$. Then any $f \in \mathbb{C}[w]^{<N}$ can be written as
\begin{equation}
	f(w) = \sum_{i=1}^N \frac{f(z_i)}{\varphi_i(z_i)} \, \varphi_i(w) = \sum_{i=1}^N \frac{\varphi(w)}{\varphi_i(z_i)} \, f(z_i) \, .
\end{equation}
This way of reconstructing the polynomial $f(w) \in \mathbb{C}[w]^{<N}$ from its values at $N$ independent points is known as Lagrange interpolation. 

We can get (kernel-type) expressions for various differential operators using Lagrange interpolation. We focus on $w \, \partial_w$. From
\begin{equation} \label{eq:dvarphi}
	\partial_w \, \varphi_j(w) 
	= \sum_{i(\neq j)}^N \frac{\varphi_j(w)}{w - z_i} 
	= \sum_{i(\neq j)}^N \prod_{k(\neq i,j)}^N \!\!\! (w - z_k ) 
\end{equation}
we obtain
\begin{equation}
	w\,\partial_w \, f(w) = \sum_{i \neq j}^N f(z_i)\, \frac{w}{z_{ij}} \, \prod_{k(\neq i,j)}^N \! \frac{w - z_k}{z_i - z_k} \, .
\end{equation}
At $w = z_l$ only the summands with $i=l$ or $j=l$ survive:
\begin{equation}
	w\,\partial_w \big|_{w = z_l} \, f(w) = f(z_l) \sum_{j(\neq l)}^N \frac{z_l}{z_{lj}} + \sum_{i (\neq l)}^N \frac{z_l}{z_{il}} \, \Biggl( \, \prod_{k(\neq i,l)}^N \! \frac{z_{lk}}{z_{ik}} \Biggr) \, f(z_i) \, .
\end{equation}
This gives an interpolation formula for the (multiplicative) derivative of $f$ at $z_i$ in terms of the value of $f$ at $z_i$ and the other $z_j$:
\begin{equation}
	w\,\partial_w \big|_{w = z_i} \, f(w) = \Biggl( \, \sum_{j(\neq i)}^N \frac{z_i}{z_{ij}} \Biggr) f(z_i) + \sum_{j (\neq i)}^N \frac{z_i}{z_{ij}} \, \frac{\varphi_i(z_i)}{\varphi_j(z_j)} \, f(z_j) \, .
\end{equation}

For the second derivative we likewise find
\begin{equation}
	\begin{aligned}
	w^2 \, \partial_w^2 \, f(w) & = \sum_{i=1}^N f(z_i) \sum_{\substack{j\neq j' \\ (\neq i)}}^N \frac{w}{z_{ij}} \, \frac{w}{z_{ij'}} \prod_{k(\neq i,j,j')}^N  \frac{w-z_k}{z_i - z_k} \, .
	\end{aligned}
\end{equation}
At $w = z_l$ the product requires one of $i,j,j'$ to equal $l$. By symmetry in $j,j'$ we get
\begin{equation}
	w^2 \, \partial_w^2 \big|_{w = z_l} \, f(w) = f(z_l) \sum_{\substack{j\neq j' \\ (\neq l)}}^N  \frac{z_l}{z_{lj}} \, \frac{z_l}{z_{lj'}} + 2 \sum_{i(\neq l)}^N f(z_i) \, \frac{z_l}{z_{il}} \sum_{j' (\neq i,l)}^N \frac{z_l}{z_{ij'}} \prod_{k(\neq i,j',l)}^N  \frac{z_{lk}}{z_{ik}} \, .
\end{equation}
Therefore
\begin{equation}
	w^2 \, \partial_w^2 \big|_{w = z_i} \, f(w) = f(z_i) \sum_{\substack{j\neq j' \\ (\neq i)}}^N  \frac{z_i}{z_{ij}} \, \frac{z_i}{z_{ij'}} + 2 \sum_{j(\neq i)}^N \Biggl( \, \sum_{j' (\neq i,j)}^N \frac{z_i}{z_{ij'}} \Biggr) \frac{\varphi_i(z_i)}{\varphi_j(z_j)} \, \frac{z_i}{z_{ij}} \, f(z_j) \, .
\end{equation}

Let us now shift our point of view a little and think of the $z_j$ as indeterminates rather than fixed numbers. Write $r_{ij}$ for the `replacement' $z_i \mapsto z_j$. Then we have derived the following identities:
\begin{equation}
	\begin{aligned}
	 z_i \,\partial_{z_i} & = \sum_{j(\neq i)}^N \frac{z_i}{z_{ij}} + \sum_{j (\neq i)}^N \frac{z_i}{z_{ij}} \, \frac{\varphi_i(z_i)}{\varphi_j(z_j)} \, r_{ij} \, ,\\ 
	z_i^2 \, \partial_{z_i}^2 & = \sum_{\substack{j\neq j' \\ (\neq i)}}^N  \frac{z_i}{z_{ij}} \, \frac{z_i}{z_{ij'}} + 2 \sum_{j(\neq i)}^N \Biggl( \, \sum_{j' (\neq i,j)}^N \frac{z_i}{z_{ij'}} \Biggr) \frac{z_i}{z_{ij}} \, \frac{\varphi_i(z_i)}{\varphi_j(z_j)} \, r_{ij} \, .
	\end{aligned}
\end{equation}
Together these further give
\begin{equation} \label{eq:Dsq}
	\begin{aligned}
	(z_i \, \partial_{z_i})^2 = z_i^2 \,\partial_{z_i}^2 + z_i \,\partial_{z_i} = {} & \sum_{j(\neq i)}^N \Biggl( 1+ \sum_{j' (\neq i,j)}^N \frac{z_i}{z_{ij'}} \Biggr) \frac{z_i}{z_{ij}} \\ 
	& + \sum_{j(\neq i)}^N \Biggl( 1+ 2 \sum_{j' (\neq i,j)}^N \frac{z_i}{z_{ij'}} \Biggr) \frac{z_i}{z_{ij}} \, \frac{\varphi_i(z_i)}{\varphi_j(z_j)} \, r_{ij} \, .
	\end{aligned}
\end{equation}
These expressions are in fact valid on the space $\mathbb{C}[z_1,\cdots\mspace{-1mu},z_N]^{<N}$ of multivariate polynomials of degree at most $N-1$ in each variable separately. Indeed, for the purpose of the preceding computations all other $z_j$ ($j\neq i$) are `spectators'; whether or not $f$ depended on them does not affect any of the calculations. (This is why we need $r_{ij}$ rather than $s_{ij}$.)

\subsection{Evaluation} 

The fun really starts when we evaluate the $z_j$ to roots of unity. Let us write
\begin{equation}
	X \overset{\ev}{=} Y \quad \text{to mean equality upon evaluation:} \quad \ev_\omega X = \ev_\omega Y \, .
\end{equation}
In \cite{LPS_22} we call these identities involving the evaluation map $z_j \longmapsto \omega^j$ \emph{on-shell identities}. 

Using the techniques from \textsection\ref{sec:strong_coupling_vs_class} one can show that for any $1\leqslant i \leqslant N$
\begin{equation} \label{eq:on-shell_sums}
	\begin{gathered}
	z_i \prod_{k(\neq i)}^N \!\! z_{ik} \;\overset{\ev}{=}\; N 
	\quad \text{whence} \quad 
	\varphi_i(z_i) \;\overset{\ev}{=}\; \frac{N}{\omega^i} \, , \\
	\sum_{j(\neq i)}^N \frac{z_i}{z_{ij}} \;\overset{\ev}{=}\; - \sum_{j(\neq i)}^N \frac{z_j}{z_{ij}} \;\overset{\ev}{=}\; \frac{1}{2}\,(N-1) \, , \\
	\sum_{\substack{ j \neq j' \\ (\neq i)}}^N \frac{z_i}{z_{ij}} \, \frac{z_i}{z_{ij'}} \;\overset{\ev}{=}\; \frac{1}{3} \, (N-1)(N-2) \, .
	\end{gathered}
\end{equation}
Therefore we obtain the elegant formula
\begin{equation}
	z_i \, \partial_{z_i} \;\overset{\ev}{=}\; \sum_{j (\neq i)}^N \frac{z_j}{z_{ij}} \, r_{ij} + \frac{1}{2} \, (N-1) \, ,
\end{equation}
where we stress that $r_{ij}$ is to be interpreted as acting \emph{prior} to evaluation.
For the second derivative \eqref{eq:Dsq} we first compute
\begin{equation}
	\frac{z_i}{z_{jj'}} \prod_{k(\neq i,j,j')}^N  \frac{z_{ik}}{z_{jk}} \;\overset{\ev}{=}\; \frac{z_j}{z_{ij'}} \, , \qquad 
	\sum_{j' (\neq i,j)}^N \, \frac{1}{z_{ij'}} \;\overset{\ev}{=}\; \frac{1}{2}\,(N-1) \, \frac{1}{z_i} - \frac{1}{z_{ij}} \, .
\end{equation}
Therefore
\begin{equation}
	\begin{aligned}
	z_i^2 \, \partial_{z_i}^2 \;\overset{\ev}{=}\; 2 \sum_{j(\neq i)}^N \frac{z_i\,z_j}{z_{ij} \, z_{ji}} \, r_{ij} + (N-1) \sum_{j(\neq i)}^N \frac{z_j}{z_{ij}} \, r_{ij} + \frac{1}{3} \, (N-1)(N-2) \, .
	\end{aligned}
\end{equation}
Altogether we obtain
\begin{equation} \label{eq:Dsq_ev}
	\begin{aligned}
	(z_i \, \partial_{z_i})^2 \;\overset{\ev}{=}\; {} & 2 \sum_{j(\neq i)}^N \frac{z_i\,z_j}{z_{ij} \, z_{ji}} \, r_{ij} + N \sum_{j(\neq i)}^N \frac{z_j}{z_{ij}} \, r_{ij} + \frac{1}{6} \, (2N-1)(N-1) \\
	\;\overset{\ev}{=}\; {} & 2 \sum_{j(\neq i)}^N \frac{z_i\,z_j}{z_{ij}\,z_{ji}} \, r_{ij} + N \, z_i \, \partial_{z_i}- \frac{1}{6} \, (N^2 - 1) \, .
	\end{aligned}
\end{equation}
Armed with these identities we return to the spin chain.

\subsection{The \textit{M}-particle difference equation}

We find it convenient to work with the antiferromagnetic hamiltonian $H^\textsc{hs}_-$, where
\begin{equation} \label{eq:pairwise}
	H^\textsc{hs}_\pm = \sum_{i<j}^N V(i-j) \, (1 \pm P_{ij}) \, ,
\end{equation}
which differs from $H^\textsc{hs}_-$ from \eqref{eq:HS_AFM} by an additive constant. Here we abbreviate the inverse-square potential as
\begin{equation}
	V(i-j) \equiv \mathrm{ev} \, \frac{z_i \, z_j}{z_{ij} \, z_{ji}} = \frac{1}{4\,\sin^2 [\pi(i-j)/N]} \, .
\end{equation}
With respect to the coordinate basis of the spin-chain space $(\mathbb{C}^2)^{\otimes N}$, any vector in the $M$-magnon sector can be written as
\begin{equation}
	\ket{\Psi} = \! \sum_{i_1<\cdots<i_M}^N \!\!\!\!\! \Psi(i_1,\cdots\mspace{-1mu},i_M) \, \cket{i_1,\cdots\mspace{-1mu},i_M} \, .
\end{equation}
Since the spin lowering matrix $\sigma^-_{i_m}$ commute we may take the wave function $\Psi(\vect{i})$ to be symmetric without loss of generality.

The projection of the eigenvalue equation $H^\textsc{hs}_\pm \, \ket{\Psi} = E \, \ket{\Psi}$ on the coordinate basis is sometimes called the \emph{$M$-particle difference equation}. For any translationally invariant and isotropic long-range spin chain with pairwise interactions, i.e.\ with hamiltonian of the form~\eqref{eq:pairwise}, it can be written as 
\cite{Ino_90,KL_22} 
\begin{equation} \label{eq:M_particle_diff_eq}
	\begin{aligned}
		\sum_{m=1}^M \sum_{j(\notin \vect{i})}^N \! V(i_m - j) \, [\Psi(\vect{i})]_{i_m \mapsto j} = \Biggl( E + M \sum_{j=1}^{N-1} \! V(j) - \sum_{m\neq m'}^M \!\! V(i_m - i_{m'}) \Biggr) \, \Psi(\vect{i}) \, .
	\end{aligned}
\end{equation}
As shown in \eqref{eq:cont_int_1}--\eqref{eq:cont_int_2} we have
\begin{equation} \label{eq:sum_trig}
	\sum_{j=1}^{N-1} V(j) = \frac{1}{12} \, (N^2 - 1) \, .
\end{equation}
We further recall the expression \eqref{eq:CS_scalar} for the scalar bosonic Calogero--Sutherland hamiltonian for $N$ particles,
\begin{equation}
	{H}_+^\textsc{cs}
	=  \frac12 \sum_{i=1}^N \, (z_i \, \partial_{z_i})^2  + \beta\,(\beta-1)  \sum_{i<j}^N \frac{z_i\,z_j}{z_{ij} \, z_{ji}}  \, .
\end{equation}
and the total momentum operator ${P}^\textsc{cs} = \sum_{i=1}^N \, z_i \, \partial_{z_i}$ from \eqref{eq:momentum_operator}.

\subsection{Into the difference equation}
As a warm up we start with the $M=1$ case of \eqref{eq:M_particle_diff_eq},
\begin{equation} \label{eq:preceding}
	\sum_{j(\neq i)}^N \! V(i - j) \, \Psi(j) = \Biggl( E + \sum_{j=1}^{N-1} V(j) \Biggr) \, \Psi(i) \, .
\end{equation}
We make the ansatz that the wave function depends on the coordinate as $\Psi(i) = \mathrm{ev} \, \widetilde{\Psi}(z_i)$ for some polynomial $\widetilde{\Psi}(w) \in \mathbb{C}[w]^{<N}$. Then \eqref{eq:preceding} is
\begin{equation} 
	\sum_{j(\neq i)}^N \frac{z_i \, z_j}{z_{ij} \, z_{ji}} \, \widetilde{\Psi}(z_j) \;\overset{\ev}{=}\; \Biggl( E + \frac{1}{12}\,(N^2 - 1)  \Biggr) \, \widetilde{\Psi}(z_i) \, .
\end{equation}
By \eqref{eq:Dsq_ev} the operator on the left-hand side is nothing but
\begin{equation} \label{eq:sum}
	\sum_{j(\neq i)}^N \frac{z_i \, z_j}{z_{ij} \, z_{ji}} \, r_{ij} \;\overset{\ev}{=}\; \frac{1}{2} \, (z_i\, \partial_{z_i})^2  - \frac{N}{2} \, z_i \, \partial_{z_i} + \frac{1}{6} \, (N^2 - 1) \, .
\end{equation}
Therefore $\widetilde{\Psi}$ must obey
\begin{equation} 
	\biggr( \frac{1}{2} \, (z_i \, \partial_{z_i})^2 - \frac{N}{2} \, z_i \, \partial_{z_i} \biggl) \, \widetilde{\Psi}(z_i) = E \, \widetilde{\Psi}(z_i) \, .
\end{equation}
This is a very simple instance of a Calogero--Sutherland-type model: the operator on the left-hand side is just a linear combination of the free hamiltonian $(z_i\,\partial_{z_i} )^2 \propto -\partial_{x_i}^2$ (in additive notation) and the momentum operator $z_i\,\partial_{z_i} \propto -\I\, \partial_{x_i}$ for $N=1$ particle with coordinate that we happen to denote by $z_i$, restricted to the $N$ sites of the spin chain. A basis of eigenfunctions is given by the plane waves $\widetilde{\Psi}_n = \E^{2\pi \I k x_i/N} = z_i^k$, with eigenvalue
\begin{equation}
	E^\textsc{hs}(k) = \frac{1}{2} \, k^2 - \frac{N}{2} \, k = \frac{1}{2} \, k \, (N-k) \, .
\end{equation}
This is the quadratic dispersion relation of Haldane--Shastry. The sign comes from \eqref{eq:pairwise}.

On to general $M$. The ansatz is $\Psi(i_1,\cdots\mspace{-1mu},i_M) = \mathrm{ev}_\omega\, \widetilde{\Psi}(z_{i_1},\cdots\mspace{-1mu},z_{i_M})$ for $\widetilde{\Psi}(z_1,\cdots\mspace{-1mu},z_M) \in \mathbb{C}[z_1,\cdots\mspace{-1mu},z_M]^{<N}$ a polynomial in the $z_m$ of degree at most $N-1$ in each variable separately. As we observed at the start we may take 
\begin{equation}
	\widetilde{\Psi}(z_1,\cdots\mspace{-1mu},z_M) \in \Bigl(\mathbb{C}[z_1,\cdots\mspace{-1mu},z_M]^{S_M}\Bigr)^{\! <N} = \bigl(\mathbb{C}[z_1,\cdots\mspace{-1mu},z_M]^{<N})^{S_M}
\end{equation}
to be symmetric without loss of generality. The $M$-particle difference equation is the application to $\widetilde{\Psi}(z_{i_1},\cdots\mspace{-1mu},z_{i_M})$ of the eigenvalue equation
\begin{equation} \label{eq:eig_eqn}
	\begin{aligned}
	\sum_{m=1}^M \sum_{j(\notin \vect{i})}^N \frac{z_{i_m} \, z_j}{z_{i_m j} \, z_{j i_m}} \, r_{i_m,j} \;\overset{\ev}{=}\; \Biggl( E + \frac{1}{12} \, M \,(N^2 - 1) - \sum_{m\neq m'}^M  \frac{z_{i_m} \, z_{i_{m'}}}{z_{i_m i_{m'}} \, z_{i_{m'}i_m }} \Biggr) \, .
	\end{aligned}
\end{equation}
Focus on
\begin{equation} \label{eq:partial_sum}
	\begin{aligned}
	\sum_{m=1}^M \sum_{j(\notin \vect{i})}^N \frac{z_{i_m} \, z_j}{z_{i_m j} \, z_{j i_m}} \, r_{i_m,j} + \sum_{m\neq m'}^M \frac{z_{i_m} \, z_{i_{m'}}}{z_{i_m i_{m'}} \, z_{i_{m'}i_m }} \, .
	\end{aligned}
\end{equation}
We would like to use \eqref{eq:sum}, but the terms with $j \in \vect{i} \setminus \{ i_m \}$ are missing from the sum. This is remedied if we make one more assumption: if $\widetilde{\Psi}(z_{i_1},\cdots\mspace{-1mu},z_{i_M})$ vanishes whenever two arguments coincide\,---\,which is very natural from the point of view of the coordinate basis, where two excitations~$\downarrow$ cannot occupy the same site\,---\,then $r_{i_m,j} \,\widetilde{\Psi}(z_{i_1},\cdots\mspace{-1mu},z_{i_M}) = 0$ for $j \in \vect{i} \setminus \{ i_m \}$. That is, we assume that $\widetilde{\Psi}$ is divisible by the Vandermode polynomial $\Delta(w_1,\cdots\mspace{-1mu},w_M)$. 
Taking into account that $\widetilde{\Psi}$ should be symmetric we thus seek wave functions in
\begin{equation}
	\Bigl( \Delta(z_1,\cdots\mspace{-1mu},z_M)^2 \ \mathbb{C}[z_1,\cdots\mspace{-1mu},z_M]^{S_M} \Bigr)^{\!<N} \subset\, \Bigl(\mathbb{C}[z_1,\cdots\mspace{-1mu},z_M]^{S_M}\Bigr)^{\! <N} \, .
\end{equation}

On polynomials with such zeroes the first term in \eqref{eq:partial_sum} can be completed to
\begin{equation} \label{eq:sum_completed}
	\sum_{m=1}^M \sum_{j(\neq i_m)}^N \frac{z_{i_m} \, z_j}{z_{i_m j} \, z_{j i_m}} \, r_{i_m,j} \,= \frac{1}{2} \, \sum_{m=1}^M (z_{i_m}\, \partial_{z_{i_m}})^2- \frac{N}{2} \sum_{m=1}^M z_{i_m}\, \partial_{z_{i_m}}  + \frac{1}{12} \, M \, (N^2 - 1) \, ,
\end{equation}
while the second term is
\begin{equation} \label{eq:factor_two}
	\sum_{m\neq m'}^M \sum_{m\neq m'}^M \frac{z_{i_m} \, z_{i_{m'}}}{z_{i_m i_{m'}} \, z_{i_{m'}i_m }}
	= 2 \sum_{m < m'}^M  \sum_{m\neq m'}^M \frac{z_{i_m} \, z_{i_{m'}}}{z_{i_m i_{m'}} \, z_{i_{m'}i_m }} \, .
\end{equation}
Again the constants cancel in the eigenvalue equation \eqref{eq:eig_eqn}\
which on shell becomes
\begin{equation}
	\frac{1}{2} \, \sum_{m=1}^M (z_{i_m}\, \partial_{z_{i_m}})^2 + 2 \, \sum_{m < m'}^M \sum_{m\neq m'}^M \frac{z_{i_m} \, z_{i_{m'}}}{z_{i_m i_{m'}} \, z_{i_{m'}i_m }} - \frac{N}{2} \sum_{m=1}^M z_{i_m}\, \partial_{z_{i_m}} \;\overset{\ev}{=}\; E
\end{equation}
This is nothing but
\begin{equation} \label{eq:eig_eqn_2}
	E^\textsc{hs} \;\overset{\ev}{=}\; \Bigl[H^\textsc{cs}_+ - \frac{N}{2} \, {P}^\textsc{cs} \Bigr]_{\beta^\star = 2,N^\star =M, z_m^\star = \ev_\omega z_{i_m}} \, .
\end{equation}
This shows that eigenfunctions for $M$-particle sector of the spin chain can be obtained from those of $M$-particle $\beta=2$ Calogero--Sutherland model by `freezing' the coordinates. The actual eigenfunctions and eigenvalues then follow from the analysis of the Calogero--Sutherland model from \textsection\ref{sec:spinless}.

\section{Spin-Calogero--Sutherland models revisited}

\subsection{Bosonic vs fermionic case} \label{sec:bosonic}

To highlight the differences with the fermionic setting from \textsection\ref{sec:rank_one} we treat both cases simultaneously.
The bosonic and fermionic physical spaces
\begin{equation} 
	\widetilde{\mathcal{H}}_{\rk,\pm} \equiv \Bigl\{ \ket{\widetilde{\Psi}} \in (\mathbb{C}^\rk)^{\otimes N} \otimes \mathbb{C}[\vect{z}, \vect{z}^{-1}] : P_{ij} \, s_{ij}\, \ket{\widetilde{\Psi}} = \pm\ket{\widetilde{\Psi}} \Bigr\} 
\end{equation} 
are the images in \eqref{eq:big_space} of the total (anti)symmetrisers
\begin{equation} 
	\Pi^\mathrm{tot}_\pm \equiv \sum_{\sigma \in S_N} \! (\pm1)^{\ell(\sigma)} \, P_{\sigma} \, s_{\sigma} \, , \qquad \bigl(\Pi^\mathrm{tot}_\pm\bigr)^2 = N! \; \Pi^\mathrm{tot}_\pm \, .
\end{equation}
In \textsection\ref{sec:rank_one} we write $\widetilde{\mathcal{F}} = 	\widetilde{\mathcal{H}}_-$ for $\rk=2$.

The abelian symmetries include the momentum operator~\eqref{eq:momentum_operator_eff} and the effective hamiltonian~\eqref{eq:CS_eff}
\begin{equation} \label{eq:CS_eff_bosferm_spin}
	\widetilde{H}_\pm^{\prime\mspace{2mu}\textsc{cs}} = \frac{1}{2} \sum_{j=1}^N (z_j \, \partial_{z_j})^2 + \frac{\beta}{2} \sum_{i<j}^N \frac{z_i + z_j}{z_i - z_j} \, \bigl(z_i \, \partial_{z_i} - z_j \, \partial_{z_j}\bigr) + \beta \sum_{i<j}^N \frac{z_i \, z_j}{z_{ij} \, z_{ji}} \, (1 \mp P_{ij}) \, .
\end{equation}
A gauge transformation~\eqref{eq:Ham_gauge} gives the spin-Calogero--Sutherland hamiltonian~\eqref{eq:CS_spin}. The energy is the same as in the scalar case, see \eqref{eq:E_cs_eff}--\eqref{eq:E_cs}; the difference is what the allowed momenta (quantum numbers) are.

For the spin-1/2 case the physical space decomposes into magnon sectors
\begin{equation}
	\widetilde{\mathcal{H}}_\pm = \bigoplus_{M=0}^N \widetilde{\mathcal{H}}_{\pm}[N-2M] \, , \qquad \widetilde{\mathcal{H}}_{\pm}[N-2M] \equiv \ker\bigl[S^z - \tfrac12(N-2M) \bigr] \subset \widetilde{\mathcal{H}}_\pm \, .
\end{equation}
The bosonic analogues of the two bases given in the main text for spin-1/2 fermions are as follows. With respect to the coordinate basis~\eqref{eq:coord_basis} any $\ket{\widetilde{\Psi}} \in \widetilde{\mathcal{H}}_{+}[N-2M]$ has the form
\begin{equation} 
	\ket{\widetilde{\Psi}} = \sum_{\substack{ I \subset \{1,\dots,N\} \\ \# I = M }}^N \!\!\!\! \widetilde{\Psi}(\vect{z}_I \, ; \vect{z}_{I^\text{c}}) \, \cket{I\mspace{1mu}} \, ,
\end{equation}
completely determined by the simple component $\cbraket{1,\dots,M}{\widetilde{\Psi}} = \widetilde{\Psi}(\vect{z}) \in \mathbb{C}[\vect{z}]^{S_M \times S_{N-M}}$. This is the $q\to 1$ limit of the coordinate basis of \cite{LPS_22}, and should be compared with its fermionic counterpart \eqref{eq:physvec}.
A bosonic analogue of the wedges from \textsection\ref{sec:wedges} is
\begin{equation} 
	u_{k_1} \odot \cdots \odot u_{k_N} \equiv \sum_{\sigma\in S_N} \! u_{k_{\sigma(1)}} \otimes \cdots \otimes u_{k_{\sigma(N)}} \, ,
\end{equation}
which can be viewed as a spin-generalisation of the Schur polynomials $s_{\choice{k}}$ just like wedges $\widehat{u}_{\choice{k}}$ are for $a_{\choice{k}}$. For example, $u_{2k_1} \odot \cdots \odot u_{2k_N} = s_{\choice{k}}(\vect{z}) \, \ket{\uparrow\cdots \uparrow}$. We do not use this basis in this work.

\subsection{Freezing} \label{sec:freezing_symm_vs_asymm}

By freezing as in \textsection\ref{sec:freezing}, see especially \eqref{eq:HS_AFM}, the effective hamiltonians~\eqref{eq:CS_eff_bosferm_spin} give
\begin{equation} 
	H^\textsc{hs}_\pm = \ev_\omega \, \partial_\beta \big|_{\beta=0} \, \widetilde{H}_\pm^{\prime\mspace{2mu}\textsc{cs}} = \sum_{i < j}^N \ev_\omega \, \frac{z_i \, z_j}{z_{ij} \, z_{ji}} \, (1 \mp P_{ij}) \, .
\end{equation}
In the bosonic (fermionic) case we get the (anti)ferromagnetic Haldane--Shastry spin chain. In view of \eqref{eq:class_equilibrium_energy} the two are related by
\begin{align}
	H^\textsc{hs}_+ + H^\textsc{hs}_- = E^0 \, .
\end{align}
Since the bosonic and fermionic spin-Calogero--Sutherland models have the same energy, the ferromagnetic hamiltonian has energy
\begin{equation} 
	E^\textsc{hs}_+(\choice{\lambda}) = \partial_\beta\big|_{\beta = 0} \, E^{\prime\mspace{2mu}\textsc{cs}}(\choice{\lambda}) = \frac{1}{2}\sum_{i=1}^N \, (N-2\mspace{1mu}i+1)\, \lambda_i \, .
\end{equation}
The only difference with the energy $E^\textsc{hs}_-(\bar{\choice{k}})$ found in \eqref{eq:HSspectrum} is that the partition $\choice{\lambda}$ need not be strict. To match the ferromagnetic and antiferromagnetic momenta we need
\begin{equation} 
	E^\textsc{hs}_+(\choice{\lambda}) + E^\textsc{hs}_-(\bar{\choice{k}}) = E^0 
\end{equation}
From \textsection\ref{sec:connection}, cf.~in particular \eqref{eq:energymotifs}, we find the solution
\begin{equation}
	\lambda_j = N - j - \bar{k}_j \, .
\end{equation}
By \eqref{eq:k_to_motif} the $\lambda_j$ are thus related to the motifs as
\begin{equation}
	\lambda_j = \sum_{m=1}^M \theta(N - j - \mu_m) \, ,
\end{equation}
i.e.\ $\choice{\lambda}' = \choice{\mu}^+$. This matches the definition of motifs in Section 4.3 of \cite{Ugl_95u}.

\section{Yangian symmetry} \label{sec:nonabelian}

\subsection{Spin-Calogero--Sutherland model}

We continue to use upper (lower) signs for the bosonic (fermionic) case as in \textsection\ref{sec:bosonic}. From the \textit{R}-matrix we build the monodromy matrix \cite{BG+_93, Ugl_96} 
\begin{equation} \label{eq:monodromy}
	\widetilde{L}_0(u) \equiv R_{0N}(u \pm d_N) \cdots R_{01}(u \pm d_1) \, , \qquad R(u) \equiv 1 + u^{-1} P \, .
\end{equation}
The proper algebraic meaning of the Dunkl operators instead of `quantised inhomogeneities' stems from affine Schur--Weyl duality~\cite{Dri_86}. The relations~\eqref{eq:dAHA} guarantee that \eqref{eq:monodromy} acts on $\widetilde{\mathcal{H}}_\pm$, obeys the \textit{RLL}-relations defining the Yangian
\begin{equation}
	R_{0 0'}(u-v) \, \widetilde{L}_0(u) \, \widetilde{L}_{0'}(v) = \widetilde{L}_{0'}(v) \, \widetilde{L}_0(u) \, R_{0 0'}(u - v) \, ,
\end{equation}
and commutes with the abelian symmetries.\,%
\footnote{\ The quantum determinant of the monodromy matrix \eqref{eq:monodromy} reproduces the family of abelian symmetries containing~\eqref{eq:CS_eff_ferm_spin}, in agreement with the role of \eqref{eq:monodromy} as nonabelian \emph{symmetry}.} 
The spin-Calogero--Sutherland model thus has a large amount of nonabelian symmetries. These include usual (global) spin operators $S^+ = \sum_i \sigma^+_i$, $S^- = \sum_i \sigma^-_i$, $S^z = \sum_i \sigma^z_i/2$ that obey the $\mathfrak{sl}_2$ relations
\begin{equation}
	\begin{aligned} 
		& [S^z,S^+] = \hphantom{+}S^+ \, , \\
		& [S^z,S^-] = -S^- \, , 
	\end{aligned} 
	\qquad [S^+,S^-] = 2\,S^z \, ,
\end{equation}
plus the `affine generators' of the Yangian (in Drinfeld's first presentation) given by \cite{Dri_86, Ugl_96}\,%
\footnote{\ One finds $S^\pm, S^z$ in the first nontrivial coefficient in the expansion of $\log \widetilde{L}_0(u)$ in $u^{-1}$. The affine generators appear as the next coefficient in this expansion.}%
\begin{equation} \label{eq:Q^pmz_dual}
	\begin{aligned}
		\widetilde{Q}^+ & \equiv \hphantom{+}\frac{1}{2}\sum_{i<j}^N  \bigl(\sigma^z_i \, \sigma^+_j - \sigma^+_i \, \sigma^z_j \bigr) \mp \hphantom{2} \sum_{i=1}^N \sigma^+_i \, d_i \,  , \\
		\widetilde{Q}^- & \equiv -\frac{1}{2}\sum_{i<j}^N  \bigl(\sigma^z_i \, \sigma^-_j - \sigma^-_i \, \sigma^z_j \bigr) \mp \hphantom{2} \sum_{i=1}^N \sigma^-_i \, d_i \,  , \\
		\widetilde{Q}^z & \equiv \hphantom{\pm}\frac{1}{2} \sum_{i<j}^N \bigl(\sigma^+_i \sigma^-_j - \sigma^-_i \sigma^+_j \bigr) \mp \frac{1}{2} \sum_{i=1}^N \sigma^z_i \, d_i \, .
	\end{aligned}
\end{equation}
These live in the adjoint representation of $\mathfrak{sl}_2$,
\begin{equation}
	\begin{aligned}
	& [S^z, \widetilde{Q}^+] = \hphantom{+}\widetilde{Q}^+ \, , \qquad [S^+, \widetilde{Q}^-] = \hphantom{+}2\, \widetilde{Q}^z \, , \qquad [S^+, \widetilde{Q}^z] = -\widetilde{Q}^+ \, , \\
	& [S^z, \widetilde{Q}^-] = - \widetilde{Q}^- \, , \qquad [S^-, \widetilde{Q}^+] = -2\, \widetilde{Q}^z \, , \qquad [S^-, \widetilde{Q}^z] = \hphantom{+}\widetilde{Q}^- \, ,
	\end{aligned}
\end{equation}
and obey the Serre-like relation
\begin{equation} \label{eq:Serre}
	[\widetilde{Q}^z,[\widetilde{Q}^+,\widetilde{Q}^-]] = \mp(S^+ \,\widetilde{Q}^- - \widetilde{Q}^+ \, S^-) \, S^z \, .
\end{equation}

Using the explicit expression~\eqref{eq:Dunkl} of the Dunkl operators together with the fermionic condition we can rewrite \eqref{eq:Q^pmz_dual} in the compact form
\begin{equation} \label{eq:Q^pmz_phys_compact}
	\begin{aligned}
		\widetilde{Q}^+ & = \hphantom{2} \sum_{i=1}^N \sigma^+_i \, \biggl(\frac{1}{2} \sum_{j(\neq i)}^N \frac{z_i+z_j}{z_i - z_j} \, (1 \mp P_{ij}) \mp\frac{1}{\beta}\, z_i \, \partial_{z_i} \biggr) \, , \\
		\widetilde{Q}^- & = \hphantom{2} \sum_{i=1}^N \sigma^-_i \, \biggl(\frac{1}{2} \sum_{j(\neq i)}^N \frac{z_i+z_j}{z_i - z_j} \, (1 \mp P_{ij}) \mp\frac{1}{\beta}\, z_i \, \partial_{z_i} \biggr) \, , \\
		\widetilde{Q}^z & = \frac{1}{2} \sum_{i=1}^N \sigma^z_i \, \biggl( \frac{1}{2} \sum_{j(\neq i)}^N \frac{z_i+z_j}{z_i - z_j} \, (1 \mp P_{ij}) \mp\frac{1}{\beta}\, z_i \, \partial_{z_i} \biggr) \, ,
	\end{aligned}
\end{equation}
or, more explicitly, as [cf.\ (4.6) in \cite{BPS_94}]
\begin{equation} \label{eq:Q^pmz_phys}
	\begin{aligned}
		\widetilde{Q}^+ & = \hphantom{+}\frac{1}{2}\sum_{i<j}^N \frac{z_i + z_j}{z_i - z_j} \, \bigl(\sigma^z_i \, \sigma^+_j - \sigma^+_i \, \sigma^z_j \bigr) \mspace{2mu} \mp \hphantom{2} \sum_{i=1}^N \, \sigma^+_i \, \biggl(\frac{1}{\beta}\, z_i \, \partial_{z_i} - \frac{1}{2} \sum_{j(\neq i)}^N \frac{z_i+z_j}{z_i - z_j} \biggr) \, , \\
		\widetilde{Q}^- & = -\frac{1}{2}\sum_{i<j}^N \frac{z_i + z_j}{z_i - z_j} \, \bigl(\sigma^z_i \, \sigma^-_j - \sigma^-_i \, \sigma^z_j \bigr) \mspace{2mu} \mp \hphantom{2} \sum_{i=1}^N \, \sigma^-_i \, \biggl(\frac{1}{\beta}\, z_i \, \partial_{z_i} - \frac{1}{2} \sum_{j(\neq i)}^N \frac{z_i+z_j}{z_i - z_j} \biggr) \, , \\
		\widetilde{Q}^z & = \hphantom{+}\frac{1}{2}\sum_{i<j}^N \frac{z_i + z_j}{z_i - z_j} \, \bigl(\sigma^+_i \sigma^-_j - \sigma^-_i \sigma^+_j \bigr) \mp \frac{1}{2} \sum_{i=1}^N \, \sigma^z_i \, \biggl(\frac{1}{\beta} \, z_i \, \partial_{z_i} - \frac{1}{2} \sum_{j(\neq i)}^N \frac{z_i+z_j}{z_i - z_j} \biggr) \, .
	\end{aligned}
\end{equation}

\subsection{Haldane--Shastry spin chain} 

The identities~\eqref{eq:ev_sum_cot} make it easy to freeze \eqref{eq:Q^pmz_phys} and obtain (both for bosons and fermions) the affine generators that were proposed in \cite{HH+_92}, 
\begin{equation}
	\begin{aligned}
		Q^x + \mathrm{i} \, Q^y = Q^+ = \ev_\omega \lim_{\beta \to \infty} \widetilde{Q}^+ 
		& = \hphantom{+} \frac{\mathrm{i}}{2}\sum_{i<j}^N \cot\bigl[\tfrac{\pi}{N}(i-j)\bigr] \, \bigl(\sigma^+_i \, \sigma^z_j - \sigma^z_i \, \sigma^+_j \bigr) \, , \\
		Q^x - \mathrm{i} \, Q^y = Q^- = \ev_\omega \lim_{\beta \to \infty} \widetilde{Q}^- 
		& = - \frac{\mathrm{i}}{2}\sum_{i<j}^N \cot\bigl[\tfrac{\pi}{N}(i-j)\bigr] \, \bigl(\sigma^-_i \, \sigma^z_j - \sigma^z_i \, \sigma^-_j \bigr) \, , \\
		Q^z = \ev_\omega \lim_{\beta \to \infty} \widetilde{Q}^z \,
		& = \hphantom{+}\frac{\mathrm{i}}{2}\sum_{i<j}^N \cot\bigl[\tfrac{\pi}{N}(i-j)\bigr] \, \bigl(\sigma^+_i \sigma^-_j - \sigma^-_i \sigma^+_j \bigr) \, .
	\end{aligned}
\end{equation}
Although not hard to establish, to the best of our knowledge this explicit link between these affine generators and the monodromy matrix~\eqref{eq:monodromy} from \cite{BG+_93} was missing in the literature.

\end{document}